\documentclass[useAMS,usenatbib,letterpaper]{mn2e}
\usepackage[pass]{geometry}
\usepackage{natbib}
\usepackage{amssymb,amsmath}
\usepackage{graphicx}
\usepackage{verbatim}
\usepackage{color,hyperref}
\usepackage{lastpage}
\definecolor{linkcolor}{rgb}{0,0,0.25}
\hypersetup{
  colorlinks=true,        
  linkcolor=linkcolor,    
  citecolor=linkcolor,    
  filecolor=linkcolor,    
  urlcolor=linkcolor      
}
\newcommand{\ie}{i.e.}

\newcommand{\dd}{\mathrm{d}}
\newcommand{\eg}{e.g.}
\newcommand{\eqnname}{equation}
\newcommand{\Eqnname}{Equation}
\newcommand{\equationname}{\Eqnname}

\newcommand{\figurename}{Figure}

\newcommand{\sectionname}{$\mathsection$}
\newcommand{\normal}{\ensuremath{\mathcal{N}}}
\newcommand{\erf}{\ensuremath{\mathrm{erf}}}

\renewcommand{\vec}[1]{\ensuremath{\mathbf{#1}}}

\newcommand{\vecx}{\ensuremath{\vec{x}}}
\newcommand{\vecv}{\ensuremath{\vec{v}}}
\newcommand{\vecw}{\ensuremath{\vec{w}}}

\newcommand{\vecj}{\ensuremath{\vec{J}}}
\newcommand{\veco}{\ensuremath{\vec{\Omega}}}
\newcommand{\veca}{\ensuremath{\boldsymbol\theta}}
\newcommand{\opar}{\ensuremath{\Omega_\parallel}}
\newcommand{\apar}{\ensuremath{\theta_\parallel}}
\newcommand{\dopar}{\ensuremath{\Delta \opar}}
\newcommand{\mopar}{\ensuremath{\langle\Delta \opar\rangle}}
\newcommand{\dapar}{\ensuremath{\Delta \apar}}
\newcommand{\operp}{\ensuremath{\Omega_\perp}}
\newcommand{\aperp}{\ensuremath{\theta_\perp}}
\newcommand{\doperp}{\ensuremath{\Delta \operp}}
\newcommand{\daperp}{\ensuremath{\Delta \aperp}}
\newcommand{\vecvkick}{\ensuremath{\delta \vec{v}^g}}
\newcommand{\doparb}{\ensuremath{\Delta \Omega_{\parallel,b}}}
\newcommand{\doparbpone}{\ensuremath{\Delta \Omega_{\parallel,b+1}}}
\newcommand{\daparb}{\ensuremath{\Delta \theta_{\parallel,b}}}

\newcommand{\dopartb}{\ensuremath{\Delta \Omega_{\parallel,bb'}}}
\newcommand{\dopartbtilde}{\ensuremath{\Delta \tilde{\Omega}_{\parallel,bb'}}}

\newcommand{\doparbprime}{\ensuremath{\Delta \Omega_{\parallel,b'}}}
\newcommand{\doparbprimeprime}{\ensuremath{\Delta \Omega_{\parallel,b''}}}
\newcommand{\doparbprimepone}{\ensuremath{\Delta \Omega_{\parallel,b'+1}}}
\newcommand{\doparbprimeprimepone}{\ensuremath{\Delta \Omega_{\parallel,b''+1}}}
\newcommand{\dapari}{\ensuremath{\Delta \theta_{\parallel,i}}}

\newcommand{\Myr}{\ensuremath{\,\mathrm{Myr}}}
\newcommand{\Gyr}{\ensuremath{\,\mathrm{Gyr}}}
\newcommand{\kpc}{\ensuremath{\,\mathrm{kpc}}}
\newcommand{\pc}{\ensuremath{\,\mathrm{pc}}}
\newcommand{\kms}{\ensuremath{\,\mathrm{km\ s}^{-1}}}
\newcommand{\msun}{\ensuremath{\,\mathrm{M}_{\odot}}}

\newcommand{\nsims}{500,000}

\title[Linear Perturbation Theory for Tidal Streams]{Linear Perturbation Theory for Tidal Streams and the Small-scale CDM Power Spectrum}
\author[Bovy, Erkal, \& Sanders]{Jo Bovy$^1$\thanks{E-mail: bovy@astro.utoronto.ca}, Denis Erkal$^2$, and Jason L. Sanders$^2$\\
$^1$Department of Astronomy and Astrophysics, University of Toronto, 50 St. George Street, Toronto, ON M5S 3H4, Canada\\
$^2$Institute of Astronomy, Madingley Road, Cambridge, CB3 0HA, United Kingdom}
\pagerange{\pageref{firstpage}--\pageref{LastPage}} \pubyear{2016}
\date{23 November 2016}

\begin{document}
\maketitle
\label{firstpage}
\begin{abstract}
  Tidal streams in the Milky Way are sensitive probes of the
  population of low-mass dark--matter subhalos predicted in
  cold-dark-matter (CDM) simulations. We present a new calculus for
  computing the effect of subhalo fly-bys on cold streams based on the
  action--angle representation of streams. The heart of this calculus
  is a line-of-parallel-angle approach that calculates the perturbed
  distribution function of a stream segment by undoing the effect of
  all relevant impacts. This approach allows one to compute the
  perturbed stream density and track in any coordinate system in
  minutes for realizations of the subhalo distribution down to
  $10^5\msun$, accounting for the stream's internal dispersion and
  overlapping impacts. We study the statistical properties of density
  and track fluctuations with large suites of simulations of the
  effect of subhalo fly-bys. The one-dimensional density and track
  power spectra along the stream trace the subhalo mass function, with
  higher-mass subhalos producing power only on large scales, while
  lower mass subhalos cause structure on smaller scales. We also find
  significant density and track bispectra that are observationally
  accessible.  We further demonstrate that different projections of
  the track all reflect the same pattern of perturbations,
  facilitating their observational measurement. We apply this
  formalism to data for the Pal 5 stream and make a first rigorous
  determination of $10^{+11}_{-6}$ dark--matter subhalos with masses
  between $10^{6.5}\msun$ and $10^9\msun$ within $20\kpc$ from the
  Galactic center (corresponding to $1.4^{+1.6}_{-0.9}$ times the
  number predicted by CDM-only simulations or to
  $f_{\mathrm{sub}}(r<20\kpc)\approx0.2\,\%$) assuming that the Pal 5
  stream is $5\Gyr$ old. Improved data will allow measurements of the
  subhalo mass function down to $10^5\msun$, thus definitively testing
  whether dark matter is clumpy on the smallest scales relevant for
  galaxy formation.
\end{abstract}


\section{Introduction}

One of the fundamental predictions of the cold-dark-matter (CDM)
cosmological model is that the halos of galaxies like the Milky Way
should be filled with abundant substructure in the form of bound
dark--matter subhalos. These subhalos are predicted to make up
$\approx10\,\%$ of the mass of the parent halo and to follow a mass
spectrum that is approximately $\dd n /\dd M \propto M^{-2}$
\citep[\eg,][]{Klypin99a,Moore99a,Springel08a,Diemand08a} between the
mass scale of the parent halo and the free-streaming scale
\citep{Schmid99a,Hofmann01a,Profumo06a}. Yet this mass spectrum has so
far eluded detection, except at the massive end where dark--matter
subhalos are expected to host luminous satellite galaxies ($M \gtrsim
10^{8.5}\msun$) \citep{Strigari08a,Koposov09a} and where measurements
of the small-scale power spectrum from the Lyman-$\alpha$ forest show
the expected clustering ($M > 3\times10^8\msun$; \eg,
\citealt{Viel13a}). Whether or not dark matter clusters on smaller
scales is a question that is of fundamental importance for the nature
of dark matter. A clear detection of a CDM-like population of
$M_{\mathrm{lim}} = 3\times10^6\msun$ dark--matter subhalos would, for
example, improve constraints on the mass $m_{\mathrm{WDM}}$ of the
particle in thermal-relic warm dark matter models by a factor of
$\approx4$ to $m_{\mathrm{WDM}}\gtrsim 13$ keV as the lower limit
scales as $m_{\mathrm{WDM}} >
3.3\,\mathrm{keV}\,\left(3\times10^8\msun/M_{\mathrm{lim}}\right)^{0.3}$
following \citet{Viel05a,Viel13a}.

Various techniques have been proposed to search for dark subhalos. In
external galaxies, flux anomalies in strong gravitational lenses can
reveal the presence of massive substructures
\citep{Mao98a,Chiba02a,Dalal02a,Vegetti12a} and statistically probe
the subhalo mass spectrum down to $\approx10^7\msun$
\citep{Hezaveh16a}. In our own galaxy, massive dark--matter clumps
affect the structure of dynamically cold objects such as the disk
\citep{Lacey85a} and tidal streams
\citep{Johnston98a,Johnston02a,Ibata02a}. The latter is the cleaner
and more sensitive method, because of the number of other heating
mechanisms that affect the structure and dynamics of the Galactic
disk.

Early work on the interaction between dark--matter subhalos and tidal
streams investigated the cumulative heating induced by the massive end
of the subhalo mass spectrum using various $N$-body techniques
\citep{Johnston02a,Ibata02a,Carlberg09a}. These papers predicted that
the thin stellar streams formed by disrupting globular clusters would
experience significant heating over a few Gyr and would therefore
quickly disperse \citep{Ibata02a,Carlberg09a}. The very existence of
long, thin tidal streams such as GD-1 \citep{Grillmair06a} would
therefore rule out a CDM-like population of dark subhalos. This strong
effect is not seen in more recent simulations that contain a more
realistic subhalo population (\eg, \citealt{Yoon11a};
\citealt{Carlberg16a}) and long, narrow stellar streams appear
compatible with a CDM-like subhalo population in a realistic halo
\citep{Ngan16a}.

Recent work has focused on the density perturbations in tidal streams
that are induced by the dynamical effect of subhalo
fly-bys. \citet{SiegalGaskins08a} demonstrated with $N$-body
simulations of a massive disrupting satellite in the presence of a
CDM-like subhalo population that its tidal tails exhibit a large
amount of clumpiness compared to simulations without subhalos. Much of
the work in the past few years has specifically looked at gaps induced
by subhalo fly-bys (\eg, \citealt{Yoon11a}; \citealt{Carlberg12a};
\citealt{Carlberg13a}; \citealt{Erkal15a}, \citealt{Erkal15b}) as the
easiest detectable signal. Simplified dynamical modeling of increasing
sophistication has elucidated the physical reason for gap formation
due to subhalo encounters and allowed for analytic solutions for
induced gap profiles and their evolution in time for circular orbits
and in the absence of internal stream dispersion
\citep{Yoon11a,Carlberg13a,Erkal15a}. These methods have made it
plausible that near-future measurements of the density and phase-space
structure of streams will allow sensitive measurements of subhalo
impacts down to at least $M \approx 10^7\msun$ \citep{Erkal15b}.

A number of cold streams have been found in the Milky Way's stellar
halo as overdensities of stars in color and magnitude (\eg, Pal 5;
\citealt{Odenkirchen01a}; GD-1; \citealt{Grillmair06a};
\citealt{Grillmair09a}). However, despite improving measurements of
their densities and increasingly sophisticated modeling of the effect
of subhalo encounters, only vague statements regarding the consistency
with the observed structure have been made so far (\eg,
\citealt{Yoon11a}, \citealt{Carlberg12b}; \citealt{Carlberg13b};
\citealt{Carlberg16a}). While the level of density structure---often
determined using the number of density gaps of a certain
size---appears similar to that expected in CDM-like simulations, no
clear measurement or constraint has been made so far. We believe that
the problem is two-fold. Firstly, modeling the effect of subhalo
impacts either assumes circular stream orbits and vanishingly small
stream dispersion or uses expensive $N$-body modeling to run a small
number of simulations. Thus, it has so far remained impossible to
generate a large and realistic statistical sampling of the expected
stream structure for a given model of the subhalo mass spectrum that
could be compared with the data. Secondly, $N$-body simulations only
very approximately return a model for an observed stream. Depending on
the initial conditions and the perturbation history of the stream, the
resulting stream today typically does not lie in the same location as
where a stream is observed. Comparisons between the models and the
data are therefore largely qualitative: no direct, rigorous comparison
between the model and the data is possible.

In this paper we address both of these problems. Over the past few
year, extensive progress has been made in modeling tidal streams
\citep[\eg,][]{Eyre11a,Varghese11a,Bonaca14a,Bovy14a,Gibbons14a,PriceWhelan14a,Sanders14a,Amorisco15a,Bowden15a,Fardal15a,Kuepper15a}. Here,
we specifically build on the advancements in modeling tidal stream in
action--angle coordinates. While it has long been clear that
action--angle coordinates provide the simplest description of the
dynamics of tidal streams \citep{Helmi99a,Tremaine99a}, advances in
computing the transformation between configuration space and
action--angle space for realistic potentials to arbitrary accuracy in
the last few years \citep{Bovy14a,Sanders14b,Binney16a} has made
action--angle modeling of tidal streams practical. In particular,
\citet{Bovy14a} and \citet{Sanders14a} have demonstrated that the
structure of unperturbed tidal streams---disrupting clusters or
low-mass dwarf galaxies in a smooth background potential---can be
accurately and efficiently modeled using simple prescriptions in the
space of orbital frequencies and angles. \citet{Sanders15a} recently
showed that this modeling framework can be extended to model the
effect of a single dark--matter impact on a stellar stream: the
impulsive velocity kicks from the impact can be transformed to
frequency--angle space where they produce a instantaneous change in
the frequencies and angles of all stream stars that gets added to
their otherwise linear frequency--angle dynamics.

Here, we extend this approach into a new calculus for computing the
present-day structure of a tidal stream perturbed by many dark--matter
subhalo encounters. As in \citet{Sanders15a}, we employ simple models
for the distribution of stars in the unperturbed stream and combine it
with impulsive kicks transformed to frequency--angle space. We
demonstrate how we can then compute the present-day stream structure
to high accuracy in the linear regime where all kicks are calculated
based on the location of the stream in the absence of
impacts. However, we apply the kicks to the perturbed streams and take
the dispersion in the stream and the overlapping effects of multiple
impacts into account. We test that this linear perturbation theory
applies using a set of $N$-body simulations that include the full
non-linear interactions. In the linear regime, we then develop a new
``line-of-parallel-angle'' algorithm for the fast computation of the
present-day stream structure by efficiently undoing the effect of all
impacts that affect a given position along the stream. Like in
\citet{Sanders15a}, these simulations can be tailor-made to an
observed tidal stream, returning predictions that may be directly and
quantitatively compared to observed streams.

We use this fast and accurate algorithm to run a large suite of
simulations of the effect of subhalo impacts on cold stellar
streams. Rather than looking for gaps of different sizes, we directly
consider the power spectrum of density fluctuations and demonstrate
that the effect of a given subhalo mass range leaves a clear imprint
on a specific scale that can be used to robustly infer the number of
subhalos in different mass ranges down to $M = 10^5\msun$. We also
investigate power spectra of fluctuations in the position and velocity
of the stream track, cross power spectra of track and density
fluctuations, and bispectra of the density and track. All of these are
within reach of observations that will occur in the next decade and a
detection in multiple of these channels would constitute a clear proof
that the observed stream structure is due to low-mass dark--matter
subhalos.

In total, the results in this paper are based on $\approx\nsims$
simulations of subhalo impacts on tidal streams, many with hundreds of
impacts per realization. This can be compared to the number of
realizations performed in state-of-the-art $N$-body simulations of
subhalo--stream interactions, which only number in the hundreds
\citep{Carlberg16a}. The statistical sampling that is opened up by our
fast method clarifies many of the effects that are normally buried
within the realization-to-realization scatter.

To aid the reader in navigating this long paper, we provide a brief
overview of all sections here. Section \ref{sec:model} contains
generalities about the manner in which we model tidal streams in
frequency--angle space (\sectionname~\ref{sec:smooth}), how we apply
subhalo kicks in this space (\sectionname~\ref{sec:sanders}), and how
we sample the number of impacts and the fly-by parameters
(\sectionname~\ref{sec:sample}). Section \ref{sec:perturb} discusses
various methods for evaluating the phase--space structure of tidal
streams perturbed by a large number of subhalo impacts:
\sectionname~\ref{sec:mock} describes how to do this by Monte Carlo
simulations of tracer particles in the perturbed stream and
\sectionname~\ref{sec:pOparapar} gives a general method for evaluating
the phase--space distribution function of a perturbed stream
analytically. Section \ref{sec:lineofpar} presents the fast
line-of-parallel-angle method for evaluating the phase--space
distribution function and its moments that allows for the full density
and phase--space structure of a perturbed stream to be computed in
minutes. The last subsection of Section \ref{sec:perturb},
\sectionname~\ref{sec:convert_obs}, discusses how to convert the
computed structure in frequency--angle space to configuration space,
where it can more easily be compared to
observations. Section~\ref{sec:powspec} employs this fast method to
investigate the density and track power spectra of tidal streams
perturbed by a variety of subhalo
populations. Section~\ref{sec:bispec} goes beyond the power spectrum
and demonstrates that perturbed tidal streams also have significant
bispectra, that is, third order moments. We discuss the power spectra
and bispectra in configuration space (as opposed to frequency--angle
space as in \sectionname\sectionname~\ref{sec:powspec} and
\ref{sec:bispec}) in Section~\ref{sec:obsspace}. We apply our
formalism to density data for the Pal 5 stream in
\sectionname~\ref{sec:pal5} and perform a first measurement of the
number of subhalos in the inner Milky Way based on this
data. Section~\ref{sec:discussion} discusses various aspects of the
work presented here and we conclude with some final remarks in
\sectionname~\ref{sec:conclusion}.

A collection of appendices contains various further technical aspects
of our work. Appendix~\ref{sec:nbody} describes a suite of tests of
our formalism using full $N$-body simulations of the interaction
between dark--matter subhalos and tidal
streams. Appendix~\ref{sec:detail} holds a detailed derivation of the
fast line-of-parallel angle algorithm presented in
\sectionname~\ref{sec:lineofpar} for the case of multiple
impacts. Various convergence tests for the power spectra from
\sectionname~\ref{sec:powspec} are presented in
Appendix~\ref{sec:convtests}. Finally, Appendix~\ref{sec:mockpal5}
tests our analysis of the Pal 5 data by repeating it for a few
$N$-body simulations of the Pal 5 stream perturbed by varying levels
of substructure. All calculations in this paper make heavy use of the
\texttt{galpy} galactic dynamics code \citep{BovyGalpy} (see
\sectionname~\ref{sec:conclusion} for full details on code
availability).

\section{Stream and impact modeling}\label{sec:model}

\begin{figure*}
\includegraphics[width=0.8\textwidth]{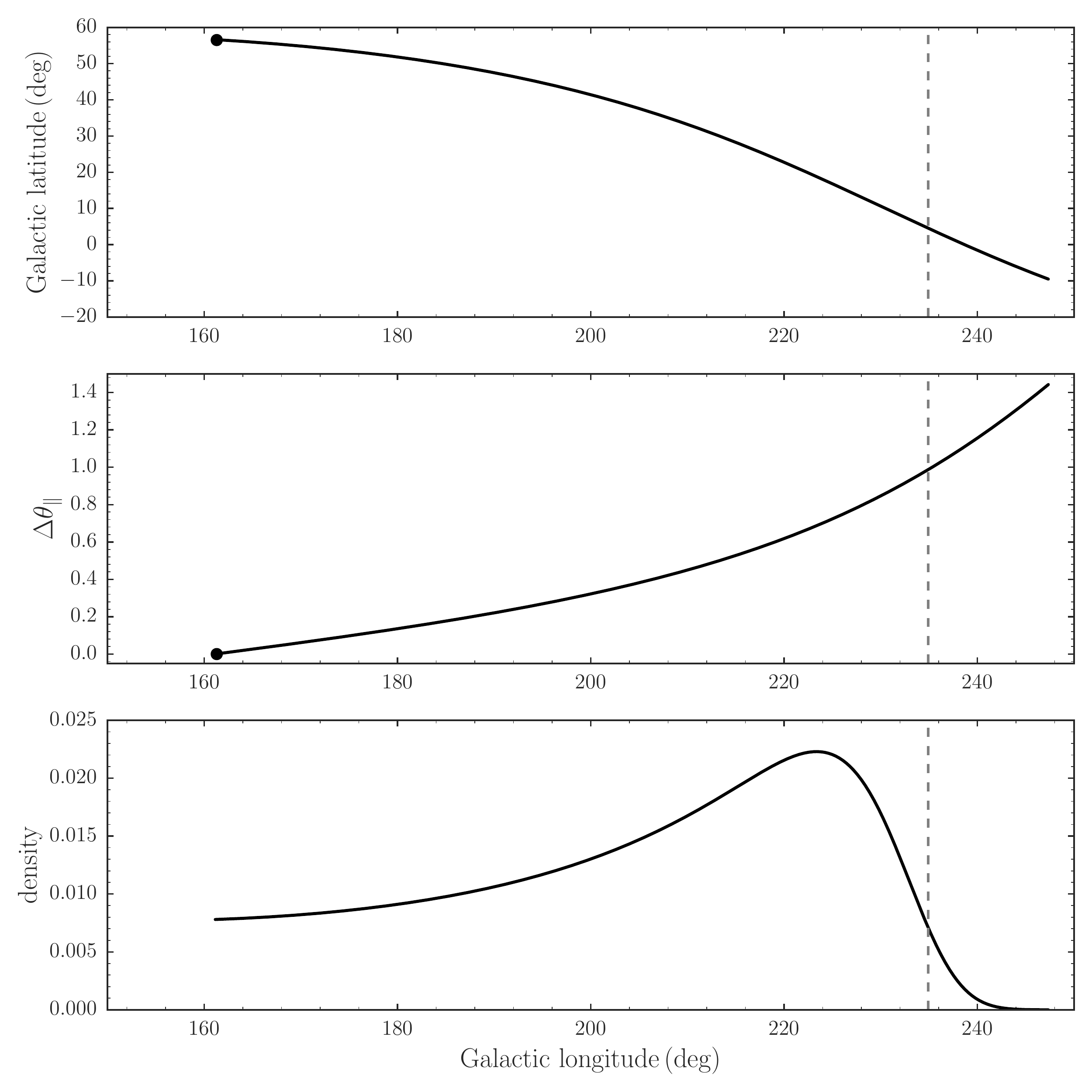}
\caption{Properties of the mock GD-1-like stream used throughout this
  paper. The top panel displays the position on the sky of the leading
  arm (the trailing arm is not used here), the middle panel gives the
  relation between the Galactic longitude and the parallel angle along
  the stream, and the bottom panel gives the density of the
  stream. All of these properties are for the smooth, unperturbed
  stream. The position of the progenitor cluster is indicated with a
  black dot in the top two panels. The dashed line is the location of
  the end of the stream for the purposes of this
  paper.\label{fig:gd1props}}
\end{figure*}

\subsection{Smooth stream model}\label{sec:smooth}

In this paper, we employ a mock stellar stream similar to that used by
\citet{Bovy14a} to investigate the effect of multiple dark-matter-halo
impacts on tidal streams. In particular, we make use of the simple
model in frequency--angle space from \citet{Bovy14a} as the basis of
all our calculations of the effect of impacts. The model stream in
frequency--angle space that we use to illustrate our computations is
the same as that from sections 3.2 and 3.3 of \citet{Bovy14a}, except
that we make the stream twice as old ($9\Gyr$ vs. $4.5\Gyr$) and twice
as cold (a model velocity dispersion $\sigma_v = 0.1825\kms$ rather
than $\sigma_v = 0.365\kms$) to create a stellar stream that is older,
but has the same length as the stream from \citet{Bovy14a}. The
current progenitor position and past orbit in a flattened logarithmic
potential with a circular velocity of $220\kms$ and a potential
flattening of $q=0.9$ are kept the same and the resulting stream
resembles the GD-1 stream \citep{Grillmair06a,Koposov10a}. The
progenitor's orbit has a pericenter of $13.7\kpc$, an eccentricity of
$0.31$, and is close to pericenter at the current time ($r =
14.4\kpc$). The radial period is approximately $400\Myr$. The details
of the Milky-Way-like potential are unimportant for the forecasting
that we perform using the GD-1-like stream as long as it realistically
describes the divergence of nearby orbits due to kicks from
dark-matter-halo fly-bys, which this flattened potential does.

The model of \citet{Bovy14a}---as well as the similar one of
\citet{Sanders14a}---fundamentally lives in frequency--angle
space. Briefly, based on a model host potential, current progenitor
position, velocity-dispersion parameter $\sigma_v$, and a time $t_d$
at which disruption started, a model for the leading or trailing tail
of the stream is created in frequency--angle space as follows. The
properties of the progenitor orbit are employed to create an
approximate Gaussian action $\vecj$ distribution for the tidal debris
\citep{Eyre11a}, which is then transformed to frequency $\veco$ space
using the Hessian $\partial \veco/\partial \vecj$ evaluated at the
progenitor's actions. The principal eigenvector of the resulting
variance tensor in frequency space is the parallel-frequency
direction, that is, the direction in frequency space along which the
stream spreads. The eigenvalues of the eigenvectors perpendicular to
the parallel frequency are typically a factor of 30 or more smaller
than the largest eigenvalue \citep{Sanders13a}. The dispersion of the
debris in frequency space is scaled by the velocity-dispersion
parameter $\sigma_v$ and the relative eigenvalues; the full frequency
distribution is modeled as a Gaussian with a mean that is offset from
the progenitor's frequency and the variance tensor resulting from
propagating the Gaussian action distribution to frequency space,
multiplied by the magnitude of the parallel frequency (the latter
because this simplifies analytical calculations). For the purposes of
this paper, the initial spread in angle of the tidal debris is
zero. After debris is tidally stripped from the progenitor with the
frequency distribution given above, it evolves in the host potential
only and the future location at all times can be computed from the
linear evolution in angle space, with the frequency remaining
constant. In angle space, the stream spreads primarily in the
direction parallel to the parallel-frequency direction and the angle
offset from the progenitor in this direction is denoted as $\dapar$;
the corresponding parallel frequency is denoted $\dopar$. Debris is
generated with a distribution of stripping times, assumed constant up
to a maximum time $t_d$---the time since disruption commenced---in the
simplest model. As discussed below, none of the specific assumptions
of (close-to) Gaussian frequency distribution or uniform stripping
rate is crucial to the fast method presented below for computing the
effect of impacts.

As discussed by \citet{Bovy14a}, the above analytic model in
frequency--angle space can be efficiently transformed to configuration
space by linearizing the transformation between configuration space
and frequency--angle space near the track of the stream. Starting from
the progenitor orbit, the mean path of the stream in configuration
space can be iteratively computed starting from the progenitor's
orbit, with convergence typically attained after one iteration. This
generates a continuous model for the phase-space structure of a
stellar stream (as opposed to discrete, $N$-body models) and in
particular, a smooth, continuous representation of the stream
track---the average phase-space location as a function of distance
from the progenitor. In this paper, this allows us to transform
determinations of the stream structure due to dark--matter-halo
impacts from frequency--angle space---where they are most easily
computed---to configuration space.

The model of \citet{Bovy14a} and the computation of the effects of
impacts on a stellar stream below both consider the leading and
trailing arms of a stream separately. Therefore---and because it
creates a long stream on its own---we only consider the leading arm of
the stream. For the purpose of modeling dark-matter-halo kicks,
whether a tail is leading or trailing does not matter greatly, because
the high surface-brightness part of a tidal tail has approximately
constant actions \citep{Bovy14a}. Some of the properties of the
leading arm (without perturbations due to dark--matter-halo impacts)
are displayed in \figurename~\ref{fig:gd1props}; other projections of
the current phase--space position are similar to those in Figs. 7 and
8 of \citet{Bovy14a}. The top panel demonstrates the path on the
celestial sphere traced by the smooth stream. The leading arm spans
approximately $93^\circ$ on the sky, has a physical length of
$\approx12.4\kpc$, and an angular Gaussian width of $\approx14'$,
similar to the GD-1 stream \citep{Koposov10a,Carlberg13b}. The middle
panel shows the relation between Galactic longitude $l$ and the
parallel angle $\dapar$. Most of the calculations in this paper are
performed as a function of parallel angle $\dapar$ and we will discuss
the structure of the stream induced by impacts as a function of this
angle. The middle panel demonstrates that $\dapar$ is a smooth
transformation of the observable coordinate along the stream, in this
case Galactic longitude. The bottom panel of
\figurename~\ref{fig:gd1props} displays the density of the smooth
stream, computed analytically from the smooth stream model. The mock
stream has a smooth, close-to-constant density profile all the way
until the edge of the stream, defined here as the location along the
stream where the density (as a function of $\dapar$, not longitude)
drops below $20\,\%$ of that near the progenitor; this location is
indicated by the dashed vertical line. We use the density as a
function of $\dapar$ because this is easy to compute and it returns a
length that increases linearly with time; we demonstrate in
Appendix~\ref{sec:convtests} that the exact definition of the length
is unimportant for predicting the statistical structure of perturbed
streams. The density enhancement around $l \approx 220^\circ$ is due
to a factor of two variation in the Jacobian of the transformation
between $\dapar$ and $l$ in the smooth logarithmic potential.

\begin{figure}
\includegraphics[width=0.48\textwidth]{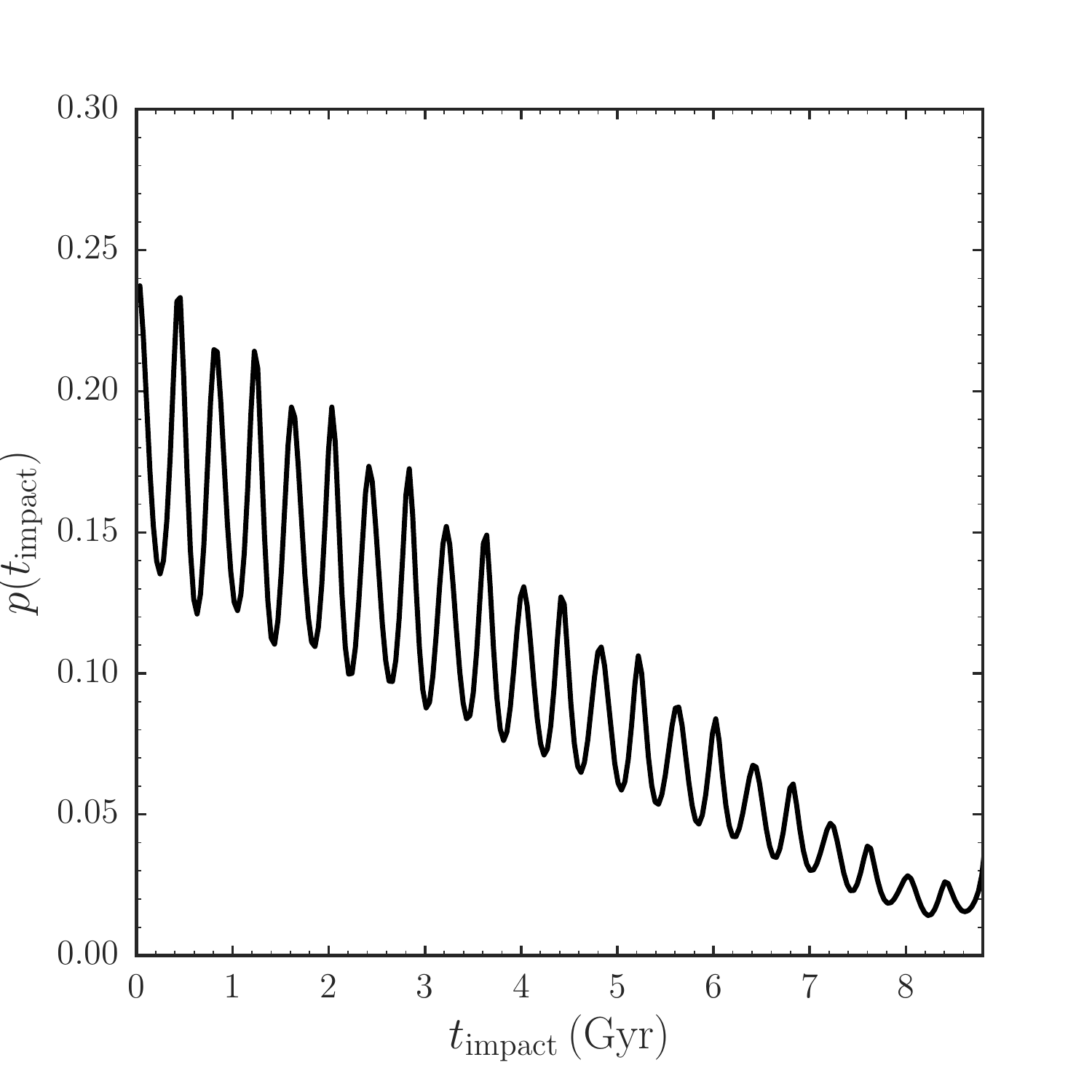}
\caption{Relative probability of impacts at different times in the
  past ($t_{\mathrm{impact}}$) for the GD-1-like stream from
  \figurename~\ref{fig:gd1props}. The relative probability is
  proportional to the length of the stream in configuration space at
  different times and normalized such that it integrates to one. The
  oscillations are due to the eccentric nature of the progenitor
  orbit, with the stream being longer near pericentric passages and
  shorter near apocentric passages. The radial period for the
  progenitor cluster is $\approx400\Myr$.\label{fig:gd1ptimpact}}
\end{figure}

\subsection{Angle--frequency perturbations due to subhalo impacts}\label{sec:sanders}

\citet{Sanders15a} demonstrated that the interaction between a stellar
stream and a dark--matter halo can be efficiently modeled in
frequency--angle space by transforming the impulsive kicks resulting
from this interaction to frequency--angle space. Because this
formalism forms the basis of the calculations in this paper, we
briefly review its main ingredients here. 

A dark--matter halo having a close encounter with a stellar stream
imparts a perturbation to the orbits of stars in the stream that can
be accurately computed in the impulse approximation
\citep{Yoon11a,Carlberg13a,Erkal15a}. Thus, the effect of the
dark--matter halo is approximated as resulting in an instantaneous
velocity kick $\vecvkick$ imparted at the time of closest approach
between the dark--matter halo and the stream. This velocity kick
$\vecvkick$ can be transformed to frequency--angle space using the
Jacobians $\partial \veco / \partial \vecv$ and $\partial \veca /
\partial \vecv$, resulting in frequency and angle kicks $\delta
\veco^g$ and $\delta \veca^g$. Before the impact, a star with
coordinates $(\veco,\veca)$ has the equation of motion $\veco =
\veco_0 = $ constant, $\veca = \veco_0\,t+\veca_0$, with
$(\veco_0,+\veca_0)$ the initial frequency--angle position at the time
of stripping. After the impact at time $t^g$, this star has the
equation of motion $\veco = \veco_0+\delta \veco^g = $ constant,
$\veca = \veco_0\,t+\delta \veco^g\,(t-t^g)+\delta
\veca^g+\veca_0$. As in \citet{Sanders15a}, we will assume that
observations happen at $t=0$ and specify impact times as happening
$t_i$ in the past (\ie, $t_i$ is positive for an impact in the past
and is zero for the current time).

\citet{Sanders15a} demonstrated that the following approximate form of
the above formalism accurately models the effect of a dark--matter
halo fly-by at all times in the future and in both frequency--angle
and configuration space. Rather than computing the kicks $\vecvkick$
over the full six-dimensional phase--space volume of the stream, we
can compute it for the mean stream track $(\vecx,\vecv)(\dapar)$ and
apply the kicks based on the nearest $\dapar$ to any phase--space
location populated by the stream. That is, we only need to compute
$\vecvkick$---and the resulting $\delta \veco^g$ and $\delta
\veca^g$---along the one-dimensional mean locus of the stream at the
time of impact (which can be efficiently computed for the
\citet{Bovy14a} stream model as discussed above). Furthermore,
\citet{Sanders15a} showed that the angle kicks $\delta \veca^g$ are
small compared to the frequency kicks $\delta \veco^g$ after
approximately one period. Because the orbital periods of old, long
stellar streams are much shorter than their age, we therefore ignore
the angle kicks in the calculations in this paper. These
approximations are extensively tested in the following sections and in
Appendix~\ref{sec:nbody}.

\subsection{Statistical sampling of multiple impacts}\label{sec:sample}

If the Milky Way's halo is populated by a population of dark--matter
halos orbiting within it, tidal streams are expected to have
interactions with multiple halos over their lifetimes. In this
section, we describe how we sample the distribution of possible
impacts to generate a realization of a tidal stream perturbed by
multiple encounters. In the model of \citet{Sanders15a} summarized
above, for a given underlying smooth stream model, a single impact is
fully described by (a) the time $t_i$ at which the impact occurs, (b)
the angular offset $\dapar^i$ from the progenitor of the closest
approach between the dark--matter halo and the stellar stream, (c) the
fly-by velocity $\vecw$ of the dark--matter halo, (d) the mass $M$ and
internal structure of the perturber (specified using a single scale
parameter $r_s$), and (e) the impact parameter $b$. We sample the
parameters in this order, \ie, later parameters in this sequence are
sampled conditional on the value sampled earlier in the sequence.

We sample impact times from a distribution $p(t_i)$ that corresponds
to the relative length of the stream at different times. That is, the
probability that the stream is hit at time $t_i$ is proportional to
the physical length of the stream at that time. In practice, we
compute the length of the stream as the path length in position space
between the progenitor and the point along the stream where the
density as a function of $\dapar$ drops below 20\,\% (see
\sectionname~\ref{sec:smooth}). The probability distribution $p(t_i)$
computed in this way is shown in
\figurename~\ref{fig:gd1ptimpact}. The probability $p(t_i)$ displays a
general increase toward more recent times (smaller $t_i$), because of
the increasing length of the stream, but it also takes into account
the relative shortening and lengthening of the stream near apo- and
pericenter. The latter gives rise to the oscillatory behavior in
\figurename~\ref{fig:gd1ptimpact}. 

To sample the angular offset $\dapar^i$ of the closest approach for a
given $t_i$, we similarly compute the relative physical lengths of
different parts of the stream at $t_i$ and sample $\dapar^i$
proportional to this (in practice we do this by discretizing the
stream at each time using $\approx1000$ segments). This $t_i$ sampling
assumes that the subhalo population does not evolve over time. The
subhalo population is expected to change as the accretion rate of
subhalos varies with time or because of mass-loss from tidal stripping
\citep{Dieman07a}. In the inner Milky Way, subhalos may also get
depleted through their interaction with the Milky Way's disk
\citep{DOnghia10a}, in which case the incidence of substructure is
higher at earlier times. The $p(t_i)$ distribution computed based on
the relative length of the stream could be multiplied by a function to
account for the time evolution of the number of subhalos to account
for these effects.

As discussed by \citet{Erkal16a} (see also \citealt{Carlberg12a}), for
a Gaussian distribution of velocities of the population of
dark--matter halos characterized by a velocity dispersion $\sigma_h$,
the distribution of velocities $\vecw$ that enter a cylinder around
the stream is as follows. In the cylindrical coordinate frame centered
on the position of the stream at the point of closest approach
(specified by $\dapar^i$), with the $z$ axis pointing along the
stream, and at rest with respect to the Galactic center (thus, not
co-moving with the stream), the $z$ and tangential velocities are
normally distributed with dispersion $\sigma_h$, while the negative
radial velocity $w_r$ has a Rayleigh distribution with the same
dispersion (positive radial velocities have zero probability, because
these move away from the stream): $p(w_r) =
|w_r|/\sigma_h^2\,\exp\left(-w_r^2/[2\,\sigma_h^2]\right)$. In the
impulse approximation, the velocity at closest approach is the same as
when the dark--matter subhalo enters this
``cylinder-of-influence''. Therefore, we sample the fly-by velocity in
the cylindrical frame from these distributions and rotate it to the
Galactocentric frame using our knowledge of the stream track at the
time of impact.

The mass $M$ and internal profile of the dark--matter perturber and
the impact parameter are sampled last. In this paper, we specify the
internal profile in terms of a scale radius $r_s$ parameter and we
consider both Hernquist and Plummer profiles for the perturbers. Thus,
the internal profile is sampled by drawing $r_s$ values. As discussed
by \citet{Erkal16a} (see also \citealt{Yoon11a,Carlberg13a}), the
impact parameter $b$ for a subhalo entering the roughly cylindrical
volume of a stellar stream is uniformly distributed. We thus sample
$b$ from a uniform distribution between $-b_{\mathrm{max}}$ and
$b_{\mathrm{max}}$. Smaller, lower-mass dark--matter halos need to
pass more closely to a stellar stream to have an observable
effect. Below, we find that setting $b_{\mathrm{max}}$ equal to a
multiple $X$ of the scale radius of the dark--matter halo works well
to take the decreasing volume of the interaction with decreasing mass
into account. Therefore, we need to sample $M$ and $r_s$ of the
dark--matter halo first, from the marginalized distribution $p(M,r_s)
= \int \dd b\,p(M,r_s,b)$. We further make use of a deterministic
$r_s(M)$ relation (\ie, we assume that the scale radius is exactly
specified by the mass) of the form $r_s \propto M^{0.5}$
\citep{Diemand08a}. Then, if the population of dark--matter halos has
a spectrum $\dd n / \dd M$, we sample $M$ from a distribution $\propto
M^{0.5}\,\dd n / \dd M$. We use a fiducial $\dd n / \dd M \propto
M^{-2}$, for which $p(M) \propto M^{-1.5}$. After sampling $M$ in this
way, we determine $r_s$ and sample $b$ uniformly between $-X r_s$ and
$X r_s$. Our fiducial dark--matter subhalo model is that of a
Hernquist sphere with $r_s(M) =
1.05\kpc\,\left(M/10^8\msun\right)^{0.5}$, obtained by fitting
Hernquist profiles to the circular-velocity--mass relation in the Via
Lactea II simulation \citep{Diemand08a}.

Finally, we need to sample the number of impacts from a Poisson
distribution for the expected number of impacts. We refer to this
interchangeably as the ``expected number'' or as the ``rate'' below;
the rate is always per stream age. For this we follow \citet{Yoon11a}
in writing the number of impacts in an interval $\dd t$ and integrate
that over the lifetime of the stream, but using the modified formalism
of \citet{Erkal16a}, that uses the correct distribution of fly-by
velocities (see above). We could use the same approach as in
\figurename~\ref{fig:gd1ptimpact} to compute the length of the stream
at all times, but for simplicity we approximate the stream as
increasing its length linearly in time as $\Delta \Omega^m$, where
$\Delta \Omega^m$ is the mean-parallel-frequency parameter of the
smooth stream (see \citealt{Bovy14a}). The expression for the expected
number of impacts for either the leading or trailing arm is then
\begin{equation}\label{eq:nenc}
  N_{\mathrm{enc}} =
  \sqrt{\frac{\pi}{2}}\,r_{\mathrm{avg}}\,\sigma_h\,t_d^2\,\Delta \Omega^m\,b_{\mathrm{max}}\,n_h\,,
\end{equation}
where $r_{\mathrm{avg}}$ is the average spherical radius of the stream
and $n_h$ is the number density of dark--matter halos in the mass
range considered. For the GD-1-like stream, this rate is approximately
62.7 when considering impacts between $10^5\msun$ and $10^9\msun$ out
to $b_{\mathrm{max}} = 5\,r_s(M)$ for $n_h$ corresponding to 425.79
dark--matter subhalos in this mass range within $25\kpc$ from the
Galactic center. The latter number is obtained by considering 38.35
subhalos within $25\kpc$ from the Galactic center with masses between
$10^6\msun$ and $10^7\msun$ and scaling this number by a factor of ten
more or less for each higher or lower mass decade ($\dd n / \dd
\log_{10} M \propto M^{-1}$; similar to the numbers obtained from the
Via Lactea II simulation by \citealt{Yoon11a}).

\section{The phase--space structure of a perturbed stream}\label{sec:perturb}

In this section, we discuss methods for evaluating the phase--space
structure of perturbed tidal streams. Our modeling approach is that of
a generative model for a tidal stream in frequency--angle space. This
model has two main ingredients: (a) the prescription for how stars are
released from the progenitor cluster or satellite and how they orbit
in the smooth background potential and (b) the set of
dark-matter--halo kicks that perturbs the orbits of stream
stars. Ingredient (a) on its own produces the smooth stream and we use
the model of \citet{Bovy14a} (see \sectionname~\ref{sec:smooth}
above). Part (b) applies the perturbations from subhalo
impacts.

For a given set of subhalo perturbations, a star in a tidal stream
with a given initial phase--space position has a deterministic path
that leads to its observed position today. Because of the conservation
of phase--space volume, the phase--space distribution function
$p(\veco,\veca)$ of tidal debris today at a point $(\veco,\veca)$ ago
is therefore equal to the distribution of the debris just after
release (assuming that $|\partial \veco / \partial \vecj|$ does not
vary substantially, which is a good assumption for a cold stellar
stream). Therefore, evaluating the phase--space structure of a
perturbed stream can be performed in two ways. We can start from the
distribution of debris at the time that it is stripped and apply
perturbations while evolving the debris forward to arrive at the
present-day distribution of debris. A simple implementation of this is
to generate mock perturbed--stream data. We start by describing this
in \sectionname~\ref{sec:mock}, because this is the simplest manner in
which to determine the perturbed structure of a stream. While
straightforward, it is difficult to compute the phase--space
distribution at a given \emph{current} location in this manner,
because one does not know where one will end up when starting at an
initial debris location.

\begin{figure}
\includegraphics[width=0.5\textwidth]{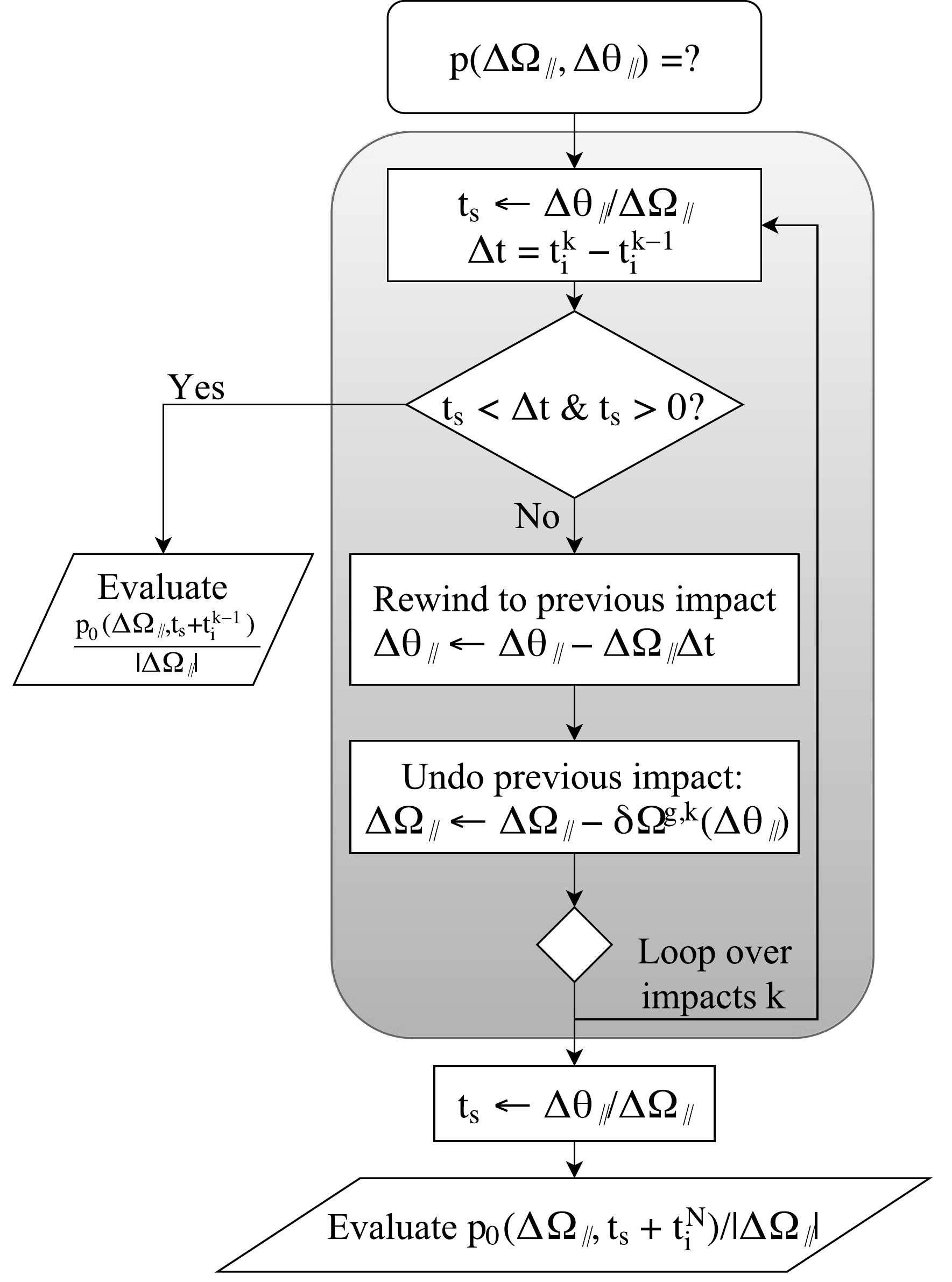}
\caption{Flowchart for the direct determination of $p(\dopar,\dapar)$
  by undoing the effect of all impacts up to the time of
  stripping. This procedure is illustrated graphically in
  \figurename~\ref{fig:illustrate_pOa}. In this flowchart, $t_i^0 =
  0$.\label{fig:pOparapar}}
\end{figure}

To evaluate the phase--space distribution at a given $(\veco,\veca)$,
we need to specify the time at which this point was stripped (a time
$t_s$ in the past) and to run the stream backward in time while
undoing the effect of all impacts. Then we arrive at the initial
$(\veco_0,\veca_0)$ a time $t_s$ in the past. We can then evaluate the
probability distribution for the initial distribution of the debris
$p_0(\veco_0,\veca_0,t_s)$.  This backward-evolution method allows for
faster and more accurate ways of computing the perturbed stream
structure today. Various forms of the backward-evolution method are
described in \sectionname\sectionname~\ref{sec:pOparapar} and
\ref{sec:lineofpar}.

Readers uninterested in the details of how we compute the perturbed
structure of tidal streams can safely skip ahead to
\sectionname\sectionname~\ref{sec:powspec} and beyond, which do not
require a detailed understanding of the calculation.

\subsection{Sampling mock perturbed--stream data}\label{sec:mock}

To sample mock data, we start with the model of the smooth stream,
which provides a model for the orbits (given by $\Delta\veco$ relative
to the progenitor) onto which stars are released from the progenitor
at different times $t_s$ in the past. We denote this model by the
probability distribution $p_0(\Delta \veco,t_s)$. In the simple model
of \citet{Bovy14a} described in \sectionname~\ref{sec:smooth}, this
model has a uniform distribution of $t_s$ up to a time $t_d$---which
defines the start of tidal disruption---and a stationary distribution
of $\Delta \veco$: $p_0(\Delta \veco,t_s) \propto p_0(\Delta \veco)$
for $t_s < t_d$. We sample $\Delta \veco$ and $t_s$ from this model to
generate the leading or trailing arm of a tidal stream.

For a given mock star $(\Delta \veco,t_s)$, we compute the parallel
angle this star reaches at the first impact that occurs after it was
stripped ($\dapar = \dopar \times \max_{k; t_i^k < t_s} t^k_i$, for
impacts $k$ at times $t_i^k$). We add the kick $\delta \veco^{g,k}$ at
this $\dapar$ to $\Delta \veco$ and similarly add the kick $\delta
\veca^{g,k}$ to $\Delta \veca$. We then repeat this procedure for each
subsequent impact. After the final (most recent) impact, the star
reaches its present $(\Delta \veco,\Delta \veca)$. At this point, we
can use the linearized frequency--angle transformation in the vicinity
of the stream to compute the configuration-space coordinates of this
point: $(\Delta \veco,\Delta \veca) \rightarrow (\vecx,\vecv)$.

This procedure is very simple to apply, but becomes quite
computationally expensive when many impacts are involved (streams have
$\mathcal{O}(100)$ impacts when considering dark--matter subhalos down
to $10^5\msun$, see \sectionname~\ref{sec:sample} above). However, it
does allow one to fully apply the kicks imparted by dark--matter
subhalos in the linear regime and thus can act as a check on the
approximations that we perform below.

\subsection{Direct evaluation of $p(\Delta \veco,\Delta \veca)$}\label{sec:pOparapar}

\begin{figure}
\includegraphics[width=0.48\textwidth]{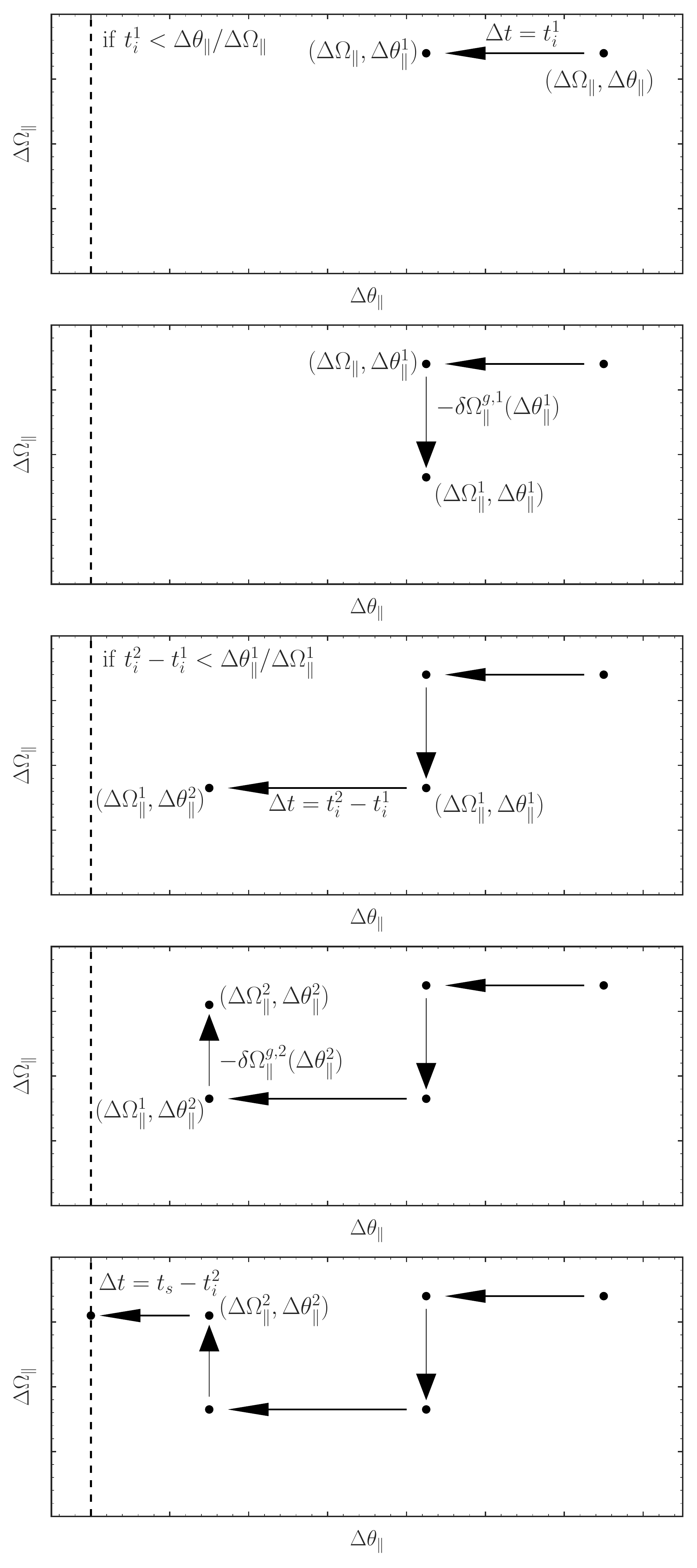}
\caption{Illustration of the straightforward algorithm to evaluate the
  phase--space probability $p(\dopar,\dapar)$. For a given point
  $(\dopar,\dapar)$, we rewind the angle $\dapar$ using its frequency
  $\dopar$ to a previous impact, starting with the first one in the
  top panel. We then subtract the frequency kick from that impact to
  obtain the pre-impact frequency (second panel from the top). This
  process is repeated for each impact until zero $\dapar$ is reached
  (bottom three panels), at which point the unperturbed release
  probability $p_0(\dopar,t_s)$ can be evaluated (see
  \eqnname~[\ref{eq:p0}]).\label{fig:illustrate_pOa}}
\end{figure}

\begin{figure*}
\includegraphics[width=\textwidth]{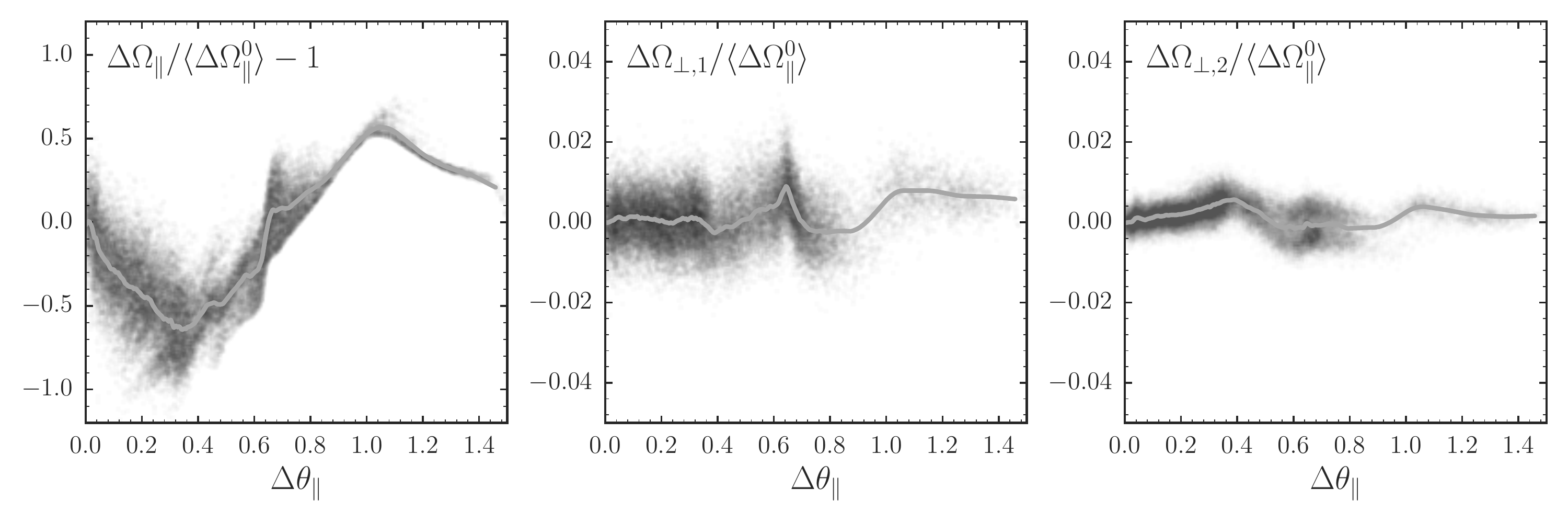}
\caption{Mock-data simulation of the present-day frequency
  distribution of the GD-1-like stream from
  \sectionname~\ref{sec:smooth} with 61 dark--matter subhalo impacts
  between $10^5\msun$ and $10^9\msun$ sampled as described in
  \sectionname~\ref{sec:sample}. Mock data are sampled using the
  algorithm given in \sectionname~\ref{sec:mock}. The left panel
  displays the parallel frequency as a function of $\dapar$, the
  middle and right panels show the frequency in the two perpendicular
  directions. The gray line is a lowess trendline (a locally linear
  weighted regression). All frequencies are normalized by the mean
  frequency $\langle \dopar^0\rangle$ as a function of $\dapar$ of the
  mock stream in order to demonstrate the deviations from the smooth
  stream due to impacts. Because of the structure of phase--space near
  the tidal stream, the ratio of the typical deviations in the
  parallel and perpendicular directions is that of the eigenvalues of
  $\partial \veco/\partial\vecj$, which is $\approx30$. Because of
  this, the effect of impacts in the perpendicular directions is much
  less than that in the parallel direction for any thin tidal
  stream.\label{fig:meanOparOperp}}
\end{figure*}

Rather than relying on mock-data sampling to determine the current
phase--space structure of a perturbed tidal stream, we can directly
evaluate the present-day phase--space distribution function $p(\Delta
\veco,\Delta \veca)$. Because we make the approximation that the kicks
$\delta \veco^g$ are only a function of $\dapar$, it is expedient to
write this as $p(\Delta \veco,\Delta \veca) =
p(\dopar,\dapar)\,p(\doperp,\daperp|\dopar,\dapar)$; we will primarily
focus on the first factor on the right-hand side in the remainder of
this paper.

{We assume that the initial distribution of angles $\Delta \veca$ is so
narrow that we can approximate it as a delta function. The evaluation
of $p(\dopar,\dapar)$ then becomes
\allowdisplaybreaks
\begin{align}
p(\dopar,\dapar) & = \int \dd t_s\, p(\dopar,\dapar,t_s)\,,\nonumber\\
& = \int \dd t_s \,p(\dapar|\dopar,t_s)\,p(\dopar,t_s)\,,\nonumber\\
& = \int \dd t_s \,p(\dapar^0|\dopar^0,t_s)\,p_0(\dopar^0,t_i^l+t_s)\,,\nonumber\\
& \approx \int \dd t_s \,\delta(\dapar^0-\dopar^0 t_s)\,p_0(\dopar^0,t_i^l+t_s)\,,\nonumber\\
& = \frac{1}{|\dopar^0|}\,p_0\left(\dopar^0,t=t_i^l+\frac{\dapar^0}{\dopar^0}\right)\,,\label{eq:p0}
\end{align}
where $(\dopar^0,\dapar^0)$ are the frequency and angle offset
\emph{before} the first impact that the point at $(\dopar,\dapar)$
experiences; this impact occurs at $t_i^l$ (second to third line in
the derivation above). The probability $p_0(\cdot,t=\cdot)$ is that of
the initial frequency offset $\dopar^0$ being generated at the time
$t$. The release time has to be equal to $t_i^l + \dapar^0/\dopar^0$
to place the initial point $(\dopar^0,\dapar=0)$ at
$(\dopar^0,\dapar^0)$ at time $t_i^l$ when it experiences its first
impact and starts on an exciting journey of kicks and evolution in the
smooth potential that ends up at $(\dopar,\dapar)$ today. We have
assumed here that there is only a single initial $\dopar^0$ and time
$t$ that lead to the current ($\dopar,\dapar$). In principle it is
possible that a star gets kicked from the leading to the trailing arm
and vice versa and that after two such kicks it crosses $\dapar = 0$
again, but this is highly unlikely for the low number of impacts
expected for a CDM-like population of subhalos.}

The probability $p(\dopar,\dapar)$ can then be evaluated by rewinding
the phase-space position to all previous impacts, starting with the
most recent one, and undoing their effect on $\dopar$ to determine
$(\dopar^0,\dapar^0,t_i^l)$.  We do this by rewinding until $\dapar =
0$ between two of the impacts or until the effect of all impacts
during the history of the stream has been undone. We then evaluate the
release probability as in \equationname~(\ref{eq:p0}). This algorithm
is presented in detail in the flowchart in
\figurename~\ref{fig:pOparapar} and illustrated in
\figurename~\ref{fig:illustrate_pOa}. We could similarly undo the
impact of angle kicks $\delta \apar^g$, but we ignore angle kicks
because they are small (see discussion above and below).

The probability $p(\doperp,\daperp|\dopar,\dapar)$ can be evaluated in
a similar manner and ideally in parallel with the computation of
$p(\dopar,\dapar)$. That is, we again determine $\dapar$ at all
impacts in between the present time and when the point $(\Delta
\veco,\Delta \veca)$ was stripped, starting with the most recent
impact, and undo the kicks $\delta \Omega^g_\perp$ in the
perpendicular direction in the same manner as for $\dopar$. At the
time of stripping determined from the $(\dopar,\dapar)$ history in the
previous paragraph, we then evaluate the release distribution
$p_0(\doperp^0,\daperp^0|\dopar,\dapar)$, which may also depend on
time.

Being able to evaluate $p(\Delta \veco,\Delta \veca)$, we can evaluate
moments of this probability distribution function by direct numerical
integration. This includes the density $p(\dapar)$ as a function of
$\dapar$ and the mean parallel frequency $\langle
\dopar\rangle(\dapar)$. These are the two main moments that we focus
on in the rest of this paper. We compute these as
\begin{align}
  p(\dapar) & = \int \dd\dopar \,p(\dopar,\dapar)\,,\\
  \langle \dopar\rangle(\dapar) & = \frac{1}{p(\dapar)}\,\int\dd\dopar\,\dopar\,p(\dopar,\dapar)\,,
\end{align}
that is, we use that $\int
\dd\doperp\dd\daperp p(\doperp,\daperp|\dopar,\dapar) = 1$, because for
a given $(\dopar,\dapar)$ all $(\doperp,\daperp)$ get shifted by the
same amount. Other moments that are important for the mean track
$(\vecx,\vecv)(\dapar)$ of the stream in configuration space are
$\langle \doperp\rangle(\dapar)$ and $\langle
\daperp\rangle(\dapar)$. Kicks in the perpendicular direction are
always much smaller than those in the parallel direction, because the
ratio between the perpendicular and parallel frequency kicks is
essentially the ratio of the eigenvalues of $\partial \veco/\partial
\vecj$ again and this ratio is small for tidal streams. Unlike kicks
in the parallel direction, which always push the stream material away
from the point of impact \citep{Erkal15a}, kicks in the perpendicular
direction can be positive or negative depending on the relative
orientation of the stream and the fly-by velocity. Thus, we both
expect kicks in the perpendicular direction to be small and to cancel
out if many impacts occur.

We test this explicitly using mock--data simulations of the GD-1-like
stream described in \sectionname~\ref{sec:smooth}. We generate mock
$(\Delta \veco,\Delta \veca)$ data with the method from
\sectionname~\ref{sec:mock} for a set of 61 impacts with masses
between $10^5\msun$ and $10^9\msun$ sampled using the procedure in
\sectionname~\ref{sec:sample}. The frequencies as a function of
$\dapar$ are displayed in \figurename~\ref{fig:meanOparOperp}
normalized by the frequency of the smooth stream as a function of
$\dapar$ to focus on the deviations induced by impacts. It is clear
that the deviations in the perpendicular direction are indeed much
smaller than those in the parallel direction, with the ratio
approximately equal to that of the eigenvalues of $\partial
\veco/\partial \vecj$. In Appendix~\ref{sec:nbody} we test this
assumption further by comparing mock streams generated by only
applying kicks in the parallel direction to full $N$-body simulations
of impacts. We find good agreement between these $N$-body simulations
and the formalism that only applies parallel frequency kicks (that is,
no perpendicular frequency kicks and no angle kicks). In what follows,
we will therefore focus on the two-dimensional phase--space
distribution $p(\dopar,\dapar)$ and we only apply the kicks to the
parallel frequency, as these dominate the observed structure of tidal
streams.

\begin{figure}
\includegraphics[width=0.45\textwidth]{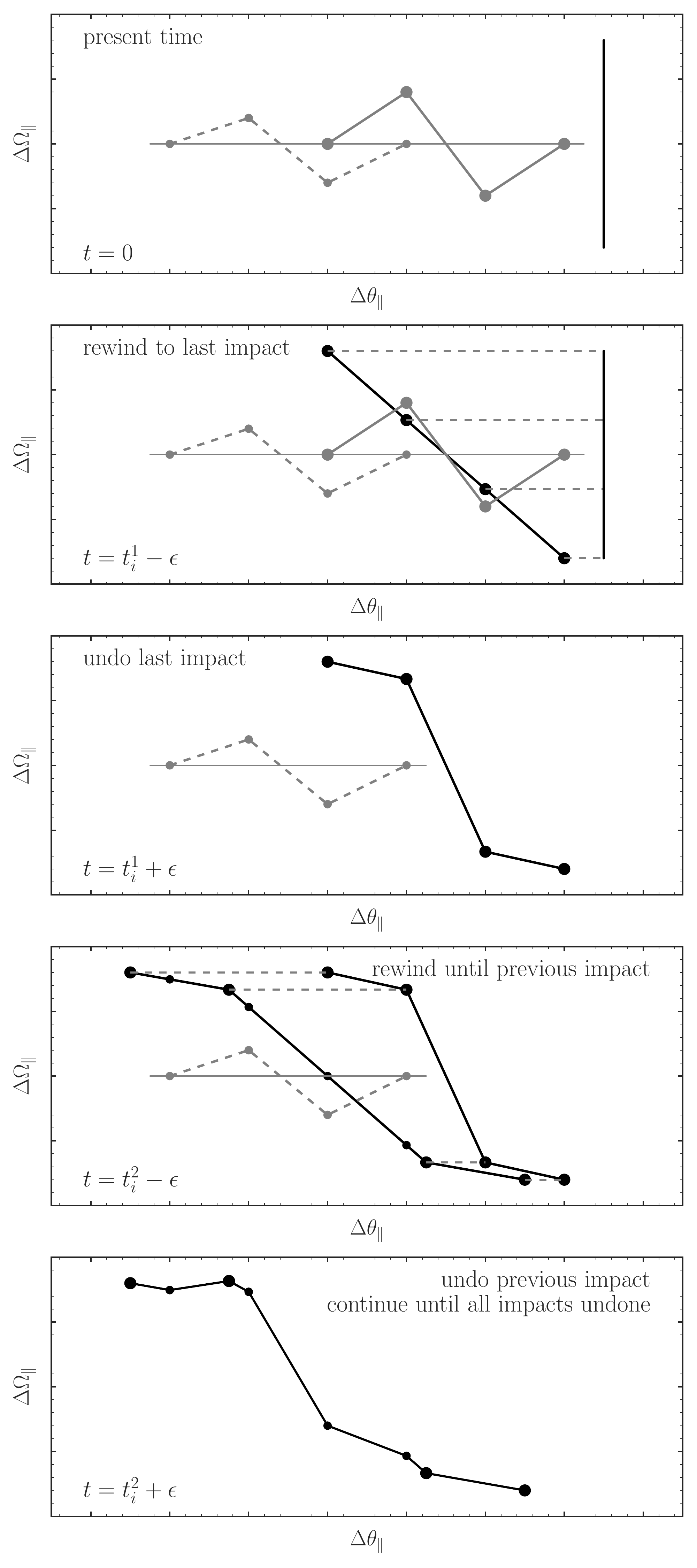}
\caption{Illustration of the line-of-parallel-angle algorithm to
  efficiently compute $p(\dopar,\dapar)$ for a set of $\dopar$ at
  fixed $\dapar$. The initially vertical line $\dopar$ at fixed
  $\dapar$ is propagated backwards through two impacts (solid and
  dashed gray lines). The propagation starts with the most recent
  impact (top three panels), using a piecewise-linear representation
  of minus the kick (solid gray line in the top two panels; zero kick
  is given by the horizontal gray line; we show minus the kick,
  because the effect of the kick is \emph{removed} in the backward
  propagation). This creates a piecewise-linear transformation of the
  initially vertical $(\dapar,\dopar)$ line (black line in the middle
  panel) that can then be propagated through the previous impact in
  the same manner (bottom three panels). After the line has been
  propagated through all impacts, we know the release times and
  initial frequencies that make up the vertical line at the final
  time. The unperturbed release probability can then be evaluated to
  give the probability everywhere along the line. As in
  \figurename~\ref{fig:illustrate_pOa}, parts of the line that reach
  $\dapar= 0$ at an intermediate stage (between impacts before the
  first one) are not propagated through earlier
  impacts.\label{fig:illustrate_perturbations}}
\end{figure}

\subsection{A line-of-parallel-angle approach to computing $p(\dopar,\dapar)$}\label{sec:lineofpar}

\begin{figure*}
\includegraphics[width=0.48\textwidth]{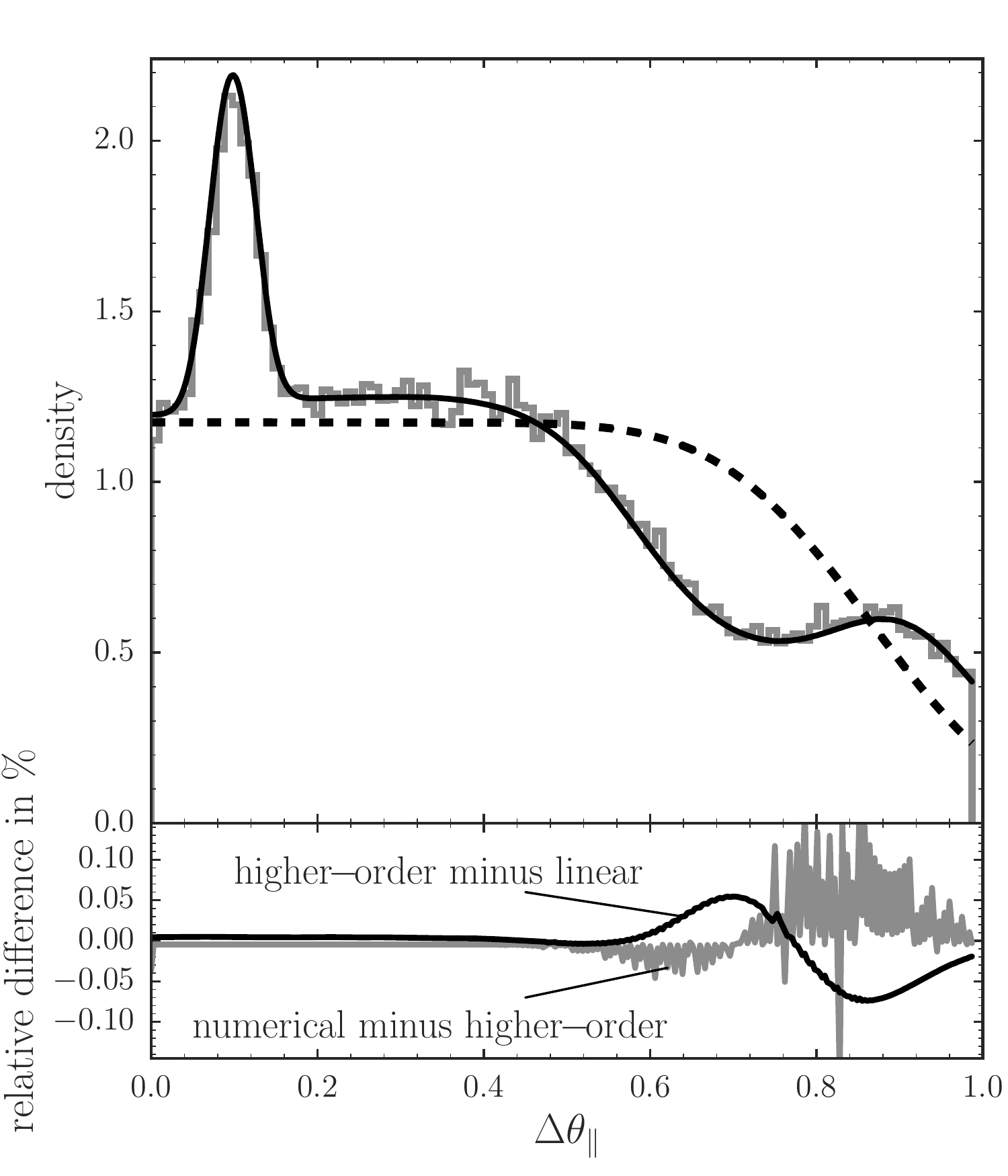}
\includegraphics[width=0.48\textwidth]{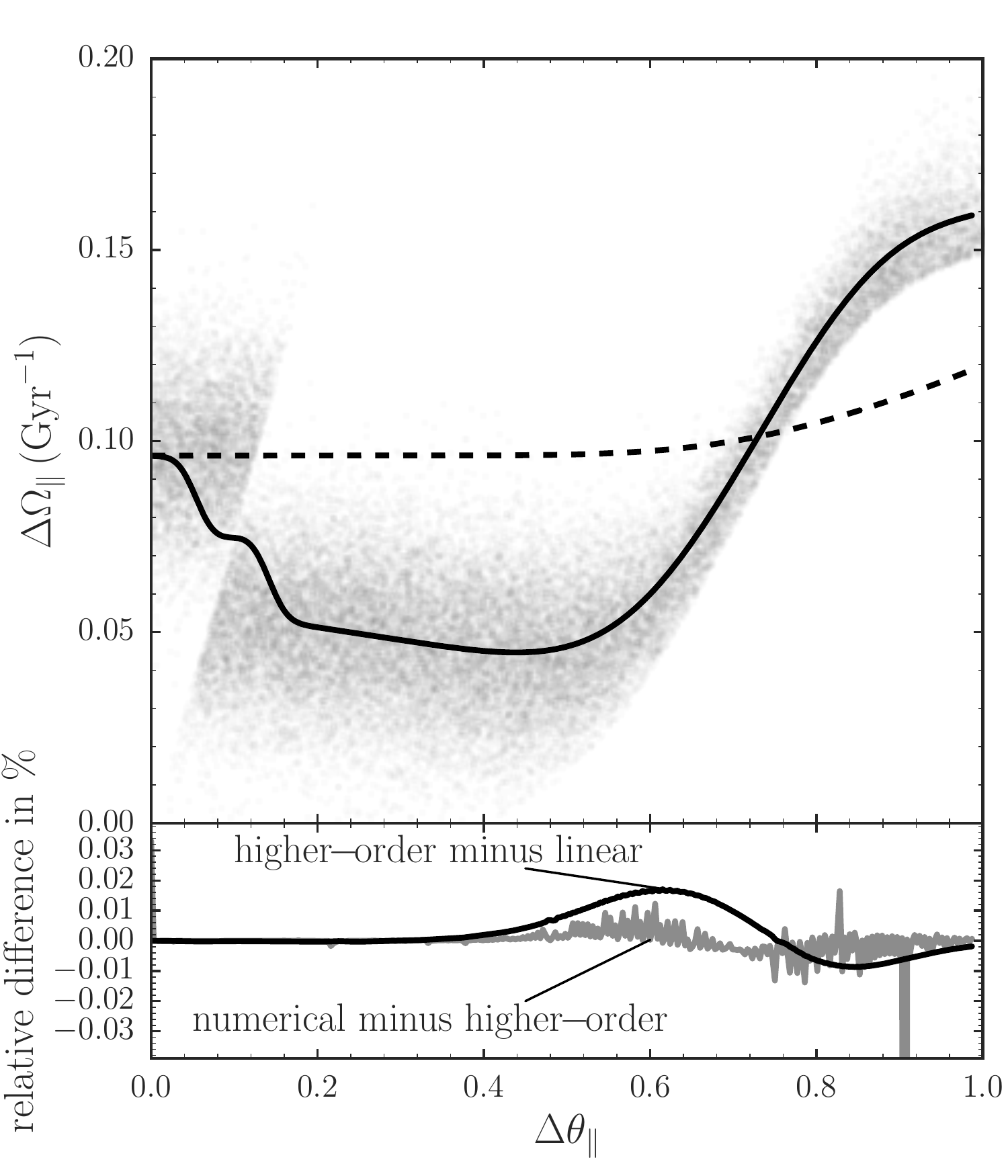}
\caption{Density $p(\dapar)$ and mean parallel frequency
  $\mopar(\dapar)$ for a single impact. The left panel displays the
  density for a single encounter $1.3\Gyr$ ago of the GD-1-like mock
  stream with a $10^8\msun$ dark--matter subhalo with an impact
  parameter of $525\pc$, a fly-by velocity of $160\kms$, and with a
  closest approach at $\dapar = 0.6$. The solid line in the top, left
  panel shows the perturbed density, the dashed line gives the density
  in the absence of an encounter, and the histogram displays the
  density obtained from mock-stream data sampled as in
  \sectionname~\ref{sec:mock}. The bottom, left panel compares
  different methods for computing the density: the black curve shows
  the difference between the density computed using only the linear
  part of the piecewise-polynomial representation of the $\delta
  \Omega_\parallel^{g}$ kick (\equationname~[\ref{eq:pdenssingle}])
  and including the higher-order parts
  (\equationname~[\ref{eq:higherorderdens}]); the gray line displays
  the difference between the higher-order calculation and a direct
  numerical integration of $p(\dopar,\dapar)$ using a generic adaptive
  integrator. The right panels show the same for $\mopar(\dapar)$,
  with the mock-stream data presented as a grayscale background in the
  top, right panel. The break around $\dapar \approx0.1$ is due to the
  part of the stream released after the impact (that is therefore
  unperturbed) and the perturbed part meeting at this angle; this
  break is unphysically hard here because the impacts occur
  instantaneously in our calculation. The piecewise-linear
  approximation of the kick allows one to compute the density and
  $\mopar(\dapar)$ to much better than $1\,\%$ and is about three
  orders of magnitude faster than using direct numerical
  integration.\label{fig:densmOparApprox}}
\end{figure*}

While it is possible and straightforward to compute moments of
$p(\dopar,\dapar)$ by direct numerical integration, for the
$\mathcal{O}(100)$ number of impacts that is expected for old tidal
streams within a few tens of kpc from the Galactic center this becomes
prohibitively expensive and numerically difficult. As an example, the
direct numerical evaluation of $\mopar(\dapar)$ for 30 impacts of
$10^6\msun$ dark--matter halos happening all at the same time for 200
values of $\dapar$ takes approximately a half hour on a single
cpu. Full simulations with impacts happening at a range of times and a
wide range of masses would therefore take prohibitively long. The
approximate method that we describe in this section reduces the
computational time for this example to 0.1 s---a speed-up of about a
factor of 20,000---and allows us to compute $\mopar(\dapar)$ and the
stream density simultaneously for full, realistic simulations in a few
minutes on a single cpu.

The approximate method that we develop in this section works similarly
to the method for the direct evaluation of $p(\dopar,\dapar)$, but
rather than running a single $(\dopar,\dapar)$ point through all
impacts between the present time and the time that this point was
released from the progenitor, we run all values of $\dopar$ at a given
$\dapar$ through all of the past impacts that they have experienced
since their release. We call this approach the
``line-of-parallel-angle'' approach to computing $p(\dopar,\dapar)$ as
it considers the history of a line of $\dopar$ at constant
(present-day) $\dapar$. The essence of this approach is that when we
approximate each impact as a piecewise-linear function of $\dapar$, an
initially straight line in $(\dopar,\dapar)$ remains piecewise linear
throughout all past impacts and to fully characterize it we only need
to track the breakpoints of this piecewise-linear representation and
the linear dependence between each two breakpoints. Eventually we end
up with a piecewise-linear representation of the phase--space
coordinates $(\dopar^0,\dapar^0,t_i^l)$ before the first encountered
impact at $t_i^l$ as a function of the present-day
$(\dopar,\dapar)$. This procedure is illustrated in
\figurename~\ref{fig:illustrate_perturbations}.

We first discuss how this is done in detail for a single impact. In
this case, we can also approximately account for higher-order
components in a piecewise-polynomial representation of the frequency
kicks, allowing us to determine the requisite $\dapar$ spacing at
which to compute the kicks such that the piecewise-linear
representation is sufficient. We then describe the algorithm for the
general case of multiple impacts occurring at different times.

\subsubsection{Single impacts}\label{sec:single}

In the line-of-parallel-angle approach we consider all $\dopar$ at a
given $\dapar$ and we want to determine how to evaluate
$p(\dopar,\dapar)$ in terms of the release probability
$p_0(\dopar^0,t_s)$. For a single impact giving a kick $\delta
\opar^{g,1}(\dapar)$ a time $t_i^1$ in the past we first determine the
maximum $\dopar$ for which the point $(\dopar,\dapar)$ was released
before the impact: $\dopar^{\mathrm{max}} = \dapar/t_i^1$. Any $\dopar
> \dopar^{\mathrm{max}}$ must have been released after the impact and
for these $\dopar$ we have that $p(\dopar,\dapar) = p_0(\dopar,\dapar)
= p_0(\dopar,t_s)/|\dopar|$. For $\dopar \leq \dopar^{\mathrm{max}}$
we need to determine the release $\dopar$ by undoing the kick due to
the impact.

We build a piecewise-polynomial representation of
$\delta \opar^{g,1}(\dapar)$ by computing this function at a set of $\dapar$
and constructing a spline function of a given order that goes through
these data. This representation has the form
\begin{equation}
\begin{split}
  \delta \opar^{g,1}(\dapar) = \sum_k c_{bk}\left(\dapar \right.&\left. -x_b\right)^k\,,\\
  & x_b \leq \dapar < x_{b+1}\,,
\end{split}
\end{equation}
for a set of breakpoints $\{x_b\}$. For this set, we can compute the
present $\dopar$ as
\begin{equation}\label{eq:breakSingle}
  \doparb = \frac{\dapar-x_b}{t_i^1}\,.
\end{equation}
The sequence of breakpoints is reversed in this way, because $\doparb$
increases for decreasing $x_b$. The set $\{\doparb\}$ represents the
$\dopar$ at the present time at which the kicks $\delta
\opar^{g,1}(\dapar)$ at time $t_i^1$ have a breakpoint for the
line-of-parallel-angle at fixed present $\dapar$. The coefficients of
the piecewise-polynomial representation of the kicks for a value
$\dopar$ are those of the smallest $\doparb$ with $\doparb \geq
\dopar$. The breakpoints $\{x_b\}$ are the gray dots in the top panel
of \figurename~\ref{fig:illustrate_perturbations} (for the solid gray
curve) and the corresponding $\{\doparb\}$ are the black dots on the
black line in the second panel.

We can then compute $p(\dopar,\dapar)$ for $\dopar \leq
\dopar^{\mathrm{max}}$ as
{\footnotesize
\allowdisplaybreaks
\begin{align}\label{eq:p0single}
  p&(\dopar,\dapar) = p_0\left(\dopar-\delta \opar^{g,1}\left(\dapar-\dopar\,t_i^1\right),\dapar-\dopar\,t_i^1\right)\,,\nonumber\\
  & = p_0\left(\dopar-\sum_k c_{bk}\left(\dapar-\dopar\,t_i^1-x_b\right)^k,\dapar-\dopar\,t_i^1\right)\,\nonumber\\
  & \qquad \qquad \qquad x_b \leq \dapar < x_{b+1}\,,\nonumber\\
  & = p_0\left(\dopar-\sum_k c_{bk}\,[t_i^1]^k\left(\doparb-\dopar\right)^k,\dapar-\dopar\,t_i^1\right)\,\nonumber\\
  & \qquad \qquad \qquad \doparb < \dopar \leq \doparbpone\,.
\end{align}}
The final right-hand side can be easily evaluated in terms of the
release probability as $p_0(\tilde{\dopar},t_s)/|\tilde{\dopar}|$ with
$\tilde{\dopar} = \dopar-\sum_k
c_{bk}\,[t_i^1]^k\left(\doparb-\dopar\right)^k$ and $t_s =
(\dapar-\dopar\,t_i^1)/\tilde{\dopar}$. Thus, we have a direct
piecewise representation of $p(\dopar,\dapar)$ for all $\dopar$ at
fixed $\dapar$.

{To compute moments of $p(\dopar,\dapar)$ we proceed as follows. The
density is
\allowdisplaybreaks
\begin{equation}
  p(\dapar) = \int\dd \dopar\,p(\dopar,\dapar)\,,
\end{equation}
which can be written in terms of the piecewise representation of
$p(\dopar,\dapar)$ as
\begin{equation}
\begin{split}
   p&(\dapar) = \int_{\dopar^{\mathrm{max}}}^\infty\dd\dopar\,p_0(\dopar,\dapar)\\
&  +\sum_b \int_0^{\doparbpone-\doparb}\dd w\,p_0(\doparbpone-w-\sum_k c_{bk}\,[t_i^1]^k w^k,\\
   & \quad \qquad \qquad \qquad \qquad \qquad \qquad \dapar-\dopar\,t_i^1)\,.
\end{split}
\end{equation}
This remains a fully general, exact expression that can be evaluated
for any release probability (including, for example, release of bursts
of stars at pericentric passages).}

{For the fiducial model of \citet{Bovy14a}, $p_0(\dopar,\dapar) =
p_0(\dopar,t_s)/|\dopar| = \normal(\dopar|\Delta
\Omega^m,\sigma^2_{\Omega,1})$ if $t_s < t_d$; this is a Gaussian
distribution with mean $\Delta \Omega^m$ and variance
$\sigma^2_{\Omega,1}$. In this case, we can write the density as
\allowdisplaybreaks
\begin{align}\label{eq:taylor}
p&(\dapar)  = \int_{\dopar^{\mathrm{max}}}^\infty\dd\dopar\,\normal(\dopar|\Delta \Omega^m,\sigma^2_{\Omega,1})\nonumber\\
&\quad +\sum_{b\geq b_0} \int_0^{\doparbpone-\doparb}\!\!\!\!\!\!\!\!\!\!\!\!\!\dd w \,\normal(\doparbpone-w-\sum_k c_{bk}\,[t_i^1]^k w^k)\nonumber\\
& \approx \int_{\dopar^{\mathrm{max}}}^\infty\dd\dopar\,\normal(\dopar|\Delta \Omega^m,\sigma^2_{\Omega,1})\nonumber\\
& \quad +\sum_{b\geq b_0} \int_0^{\doparbpone-\doparb}\dd w \,\normal(\doparbpone-c_{b0}-(1+c_{b1}\,t_i^1)w)\nonumber\\
& \qquad \left[1+\frac{\doparbpone-c_{b0}-(1+c_{b1}t_i^1)w-\Delta \Omega^m}{\sigma^2_{\Omega,1}}\,\sum_{k>1} c_{bk}\,[t_i^1]^k w^k\right]\,,
\end{align}
where we have Taylor-expanded the integrand of the second integral
around $w=0$ up to first order and the sum is only over
frequency-ranges that were released between $t_d$ and $t_i^1$ (denoted
as $b\geq b_0$; see below). These integrals can be performed
analytically. The first term in the square brackets returns the
density up to linear order in the piecewise-polynomial kick
representation: {\footnotesize
\begin{align}
\label{eq:pdenssingle}
p&(\dapar)  \approx \frac{1}{2}\left[1+\erf\left(\frac{1}{\sqrt{2}\sigma_{\Omega,1}}\left[\Delta \Omega^m-\dopar^{\mathrm{max}}\right]\right)\right]\nonumber\\
& + \sum_{b\geq b_0}\frac{1}{2(1+c_{b1} t_i^1)}\,\left[\erf\left(\frac{1}{\sqrt{2}\sigma_{\Omega,1}}\left[\doparbpone-c_{b0}-\Delta \Omega^m\right]\right)\right.-\nonumber\\
  & \left.\erf\left(\frac{1}{\sqrt{2}\sigma_{\Omega,1}}\left[\doparb-c_{b0}-\Delta \Omega^m-c_{b1}\,t_i^1(\doparbpone-\doparb)\right]\right)\right]
\end{align}}
The second term adds
{\footnotesize
\begin{equation}\label{eq:higherorderdens}
\begin{split}
  & \sum_{b\geq b_0}-\frac{1}{2}\sum_{j=0}^{\mathrm{order}+1}\,E(u,l;j+1)\times \Bigg[(-\sqrt{2})^{j+1}\,\sigma_{\Omega,1}^{j-1}\,\\
    & \ \ \sum_{k>1}\frac{c_{bk}\,[t_i^1]^k}{(1+c_{b1}\,t_i^1)^{k+1}}\binom{k}{j}\left(\doparbpone-c_{b0}-\Delta \Omega^m\right)^{k-j}\Bigg]\,,
\end{split}
\end{equation}}
where $u$ is the argument of the first error function in the sum over
$b$ in \equationname~(\ref{eq:pdenssingle}) and $l$ is the argument of
the second error function in the same equation; the sum over $j$ runs
up to the order of the piecewise polynomial plus one. The factor
$E(u,l;n)$ is defined as
\begin{equation}
  E(u,l;n) = \int_l^u\,\dd t\,\frac{2}{\sqrt{\pi}}\,e^{-t^2}\,t^n\,,
\end{equation}
which can be efficiently computed using the recurrence relations
\begin{align}\label{eq:recurse1}
  E(u,l;0) & = \erf(u)-\erf(l)\,,\\
  E(u,l;1) & = -\frac{1}{\sqrt{\pi}}\left(e^{-u^2}-e^{-l^2}\right)\,,\\
  E(u,l;n) & = -\frac{1}{\sqrt{\pi}}\left(e^{-u^2}\,u^{n-1}-e^{-l^2}\,l^{n-1}\right)\label{eq:recurse2}\\
  & \ \ +\frac{n-1}{2}\,E(u,l;n-2)\,.\nonumber
\end{align}}

To finish our discussion of how to evaluate $p(\dapar)$ approximately
using the line-of-parallel-angle approach, we need to determine the
lower limit $b_0$ of the integration. This lower limit is obtained by
finding the segment where the line-of-parallel-angle switches from
being released after $t_d$ to being released before $t_d$ (which is
impossible). In the linear approximation, this interval is that for
which the equation
\begin{equation}\label{eq:lowerlimit}
\begin{split}
  \left[\doparbpone-c_{b0}-(1+c_{b1}\,t_i^1)x\right]& (t_d-t_i^1) = \\
  & \dapar-(\doparbpone-x)\,t_i^1\,,
\end{split}
\end{equation}
has a solution within the interval, \ie, $0 \leq x <
\doparbpone-\doparb$. To obtain a better approximation of the
interval, we then adjust the lower limit of this interval such that it
starts from the value of $x_0$ that solves this equation: $\Delta
\Omega_{\parallel,b_0} \rightarrow \Delta
\Omega_{\parallel,b_0+1}-x_0$.

The calculation of other moments of $p(\dopar,\dapar)$ is similar and
we only discuss the first moment of $\dopar$ such that we can compute
the mean parallel frequency $\mopar(\dapar)$. We find that the linear
terms of the kicks produce
{\footnotesize
\allowdisplaybreaks
\begin{align}\label{eq:mOsingle}
  & \mopar\,p(\dapar) = \nonumber\\
  & \quad \frac{1}{2}\left[\sigma_{\Omega,1}\sqrt{\frac{2}{\pi}}\,\exp\left(-\frac{1}{2\sigma^2_{\Omega,1}}\left[\Delta \Omega^m-\dopar^{\mathrm{max}}\right]^2\right)\right.\nonumber\\
& \left.\quad \quad\ \   +\Delta \Omega^m\left(1+\erf\left[\frac{1}{\sqrt{2}\sigma_{\Omega,1}}\left(\Delta \Omega^m-\dopar^{\mathrm{max}}\right)\right]\right)\right]\nonumber\\
& +\sum_{b\geq b_0}
  \left(\doparbpone+\frac{\Delta \Omega^m+c_{b0}-\doparbpone}{1+c_{b1}\,t_i^1}\right)p_b(\dapar)\nonumber\\
  & \quad 
 +\frac{\sigma_{\Omega,1}}{\sqrt{2\pi}(1+c_{b1}\,t_i^1)^2}\times\nonumber\\
& \ \   \left[\exp\left(-\frac{1}{2\sigma^2_{\Omega,1}}\left[\doparb-c_{b0}-\Delta \Omega^m-c_{b1}\,t_i^1(\doparbpone-\doparb)\right]^2\right)\right.\nonumber\\
& \ \ \left.   -\exp\left(-\frac{1}{2\sigma^2_{\Omega,1}}\left[\doparbpone-c_{b0}-\Delta \Omega^m\right]^2\right)\right]\,,
\end{align}}
where $p_b(\dapar)$ is the summand in
\equationname~(\ref{eq:pdenssingle}). 
The higher-order polynomial coefficients in the first-order Taylor
expansion of \equationname~(\ref{eq:taylor}) add
{\footnotesize
\begin{equation}\label{eq:mOhigherorder}
\begin{split}
  \sum_{b\geq b_0}& \doparbpone\,p^h_b(\dapar)\\
  &+\frac{1}{2}\sum_{j=0}^{\mathrm{order}+2}\,E(u,l;j+1)\,\times\Bigg[(-\sqrt{2})^{j+1}\,\sigma^{j-1}_{\Omega,1}\\
    & \ \sum_{k>1}\frac{c_{bk}\,[t_i^1]^k}{(1+c_{b1}\,t_i^1)^{k+2}}\,\binom{k+1}{j}\,(\doparbpone-c_{b0}-\Delta \Omega^m)^{k-j+1}\Bigg]\,,
\end{split}
\end{equation}}
where $p^h_b(\dapar)$ is the summand in
\equationname~(\ref{eq:higherorderdens}).

As an example of these calculations,
\figurename~\ref{fig:densmOparApprox} displays the density and mean
parallel frequency $\mopar(\dapar)$ of the GD-1-like mock stream
after an encounter with a single $10^8\msun$ dark--matter halo. In
this figure, we compare the approximations above with (a) the
structure obtained using mock data generated as described in
\sectionname~\ref{sec:mock} above and (b) direct numerical
integrations of $p(\dopar,\dapar)$---evaluated using the method of
\sectionname~\ref{sec:pOparapar}. All four different methods agree
very well. The bottom panels of \figurename~\ref{fig:densmOparApprox}
compare the computed density and $\mopar(\dapar)$ for the direct
numerical integration and using the approximate expressions
above. These all agree to less than $0.1\,\%$. The differences between
only including the linear part of the piecewise-polynomial
representation of the impact kick (\equationname
s~[\ref{eq:pdenssingle}] and [\ref{eq:mOsingle}]) and including the
higher-order parts to first order (\equationname
s~[\ref{eq:higherorderdens}] and [\ref{eq:mOhigherorder}] are small
and only substantial near the impact point. The numerical integration
has significant noise at the $0.1\,\%$ level due to the breakpoints in
the kick representation and the sharp lower edge in $\dopar$ that
separates regions of phase--space that can and cannot be populated by
the stream. These latter issues are much more significant when
considering multiple impacts.

Using the approximation in \equationname s~(\ref{eq:pdenssingle})
allows one to compute $p(\dapar)$ at a single $\dapar$ in
approximately 0.1 ms on a single cpu in a pure Python
implementation. Including the higher-order terms in \equationname
s~(\ref{eq:higherorderdens}) takes about ten times as long, because of
the recursion necessary in \equationname
s~(\ref{eq:recurse1})-(\ref{eq:recurse2}). Computing the density at a
single $\dapar$ using direct numerical integration of
$p(\dopar,\dapar)$ takes approximately $115$ ms, or more than 1300
times longer than using the piecewise-linear approximation of the
kicks. The computation times for $\mopar(\dapar)$ are similar.

When considering multiple impacts below, we will only allow these to
occur on a grid of times. We presented the procedure in this section
for a single impact, but it can in practice be applied to all impacts
that occur at the same time. This can be done by combining all of the
individual-kick $\delta \Omega_\parallel^{g,j}$ that occur at a given
impact time into a single $\delta \Omega_\parallel^{g}$ for that time
and working with the piecewise-polynomial representation of the
combined kick.

\subsubsection{Multiple impacts}\label{sec:multiple}

\begin{figure*}
\includegraphics[width=0.48\textwidth]{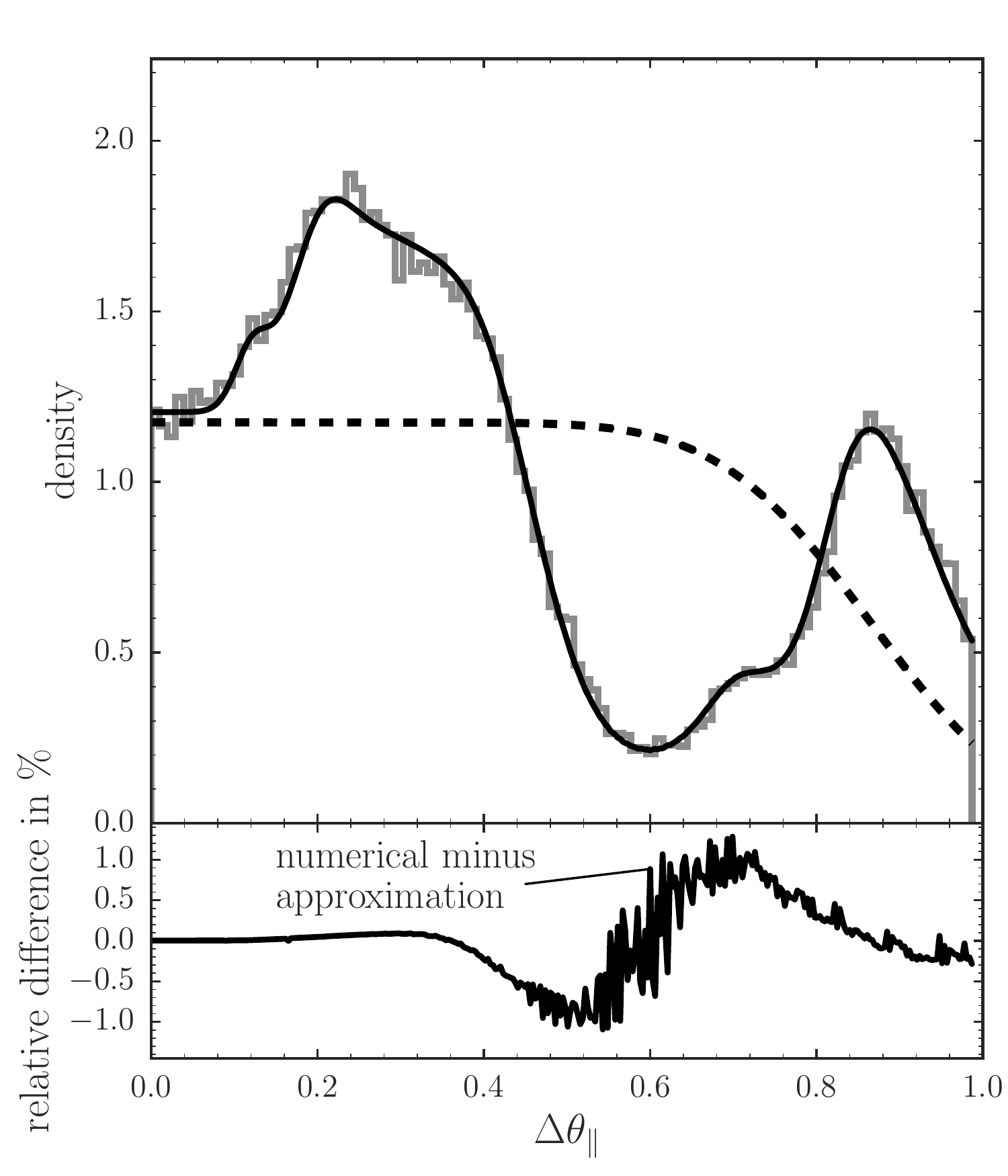}
\includegraphics[width=0.48\textwidth]{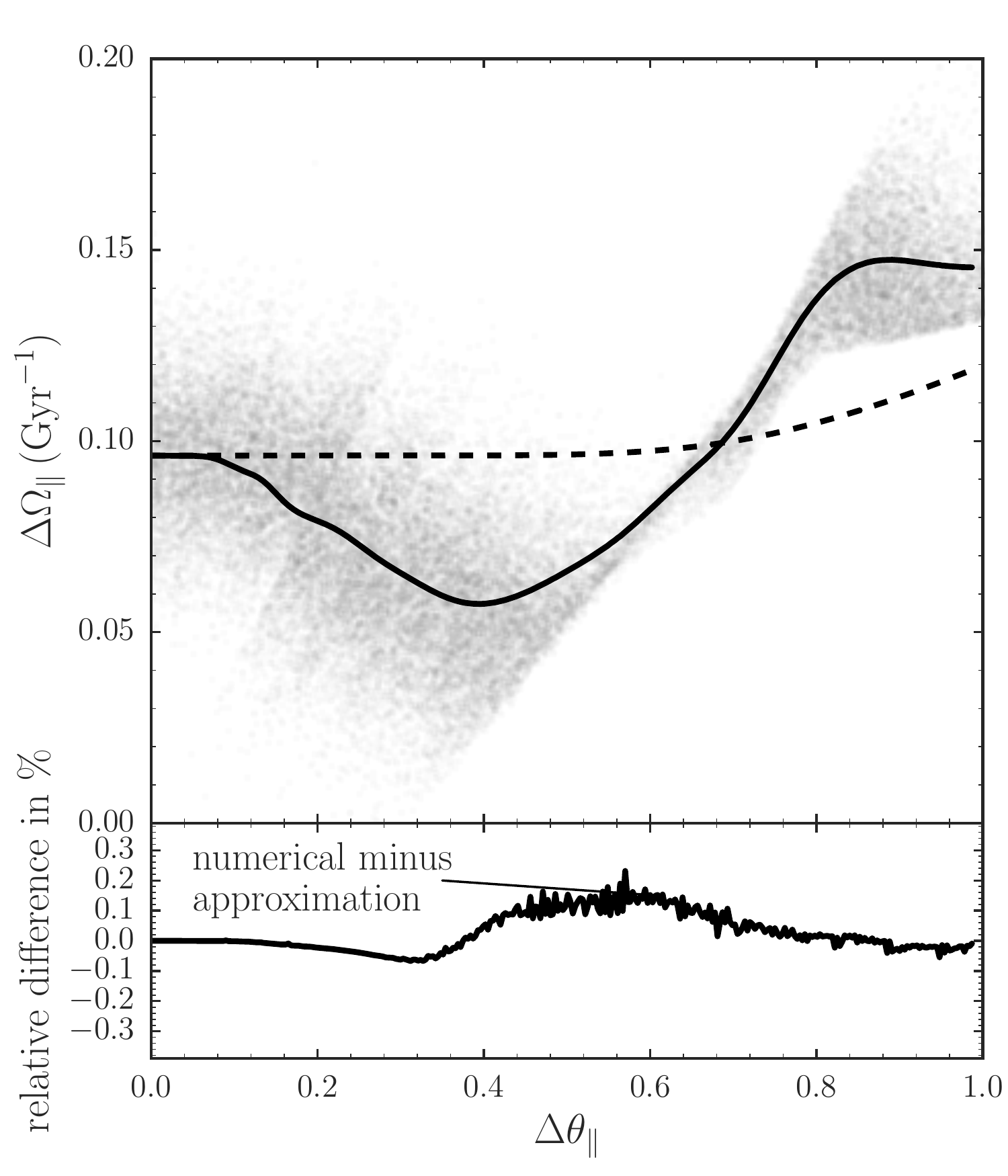}
\caption{Same as \figurename~\ref{fig:densmOparApprox}, but for
  multiple impacts. The GD-1-like mock stream has in this case
  encountered four dark--matter halos with masses $\approx10^7\msun$
  at times between $1.3\Gyr$ and $4.3\Gyr$ spaced $1\Gyr$ apart, with
  fly-by velocities of $160\kms$, $152.5\kms$, $229\kms$, and
  $161\kms$. All four encounters have impact parameters between $0.5$
  and $2.5$ scale radii of the dark--matter halos and all have a
  closest approach with a similar part of the stream, at current
  $\dapar$ between 0.6 and 0.75. Thus, the effect of the multiple
  encounters strongly overlaps. The approximate density $p(\dapar)$
  and $\mopar(\dapar)$ agrees with that obtained from mock data. The
  approximation agrees with the direct numerical evaluation to about
  $1\,\%$ for the density and better for $\mopar(\dapar)$, while being
  a factor of 100 faster.\label{fig:densmOparApproxMulti}}
\end{figure*}

\begin{figure*}
\includegraphics[width=0.48\textwidth]{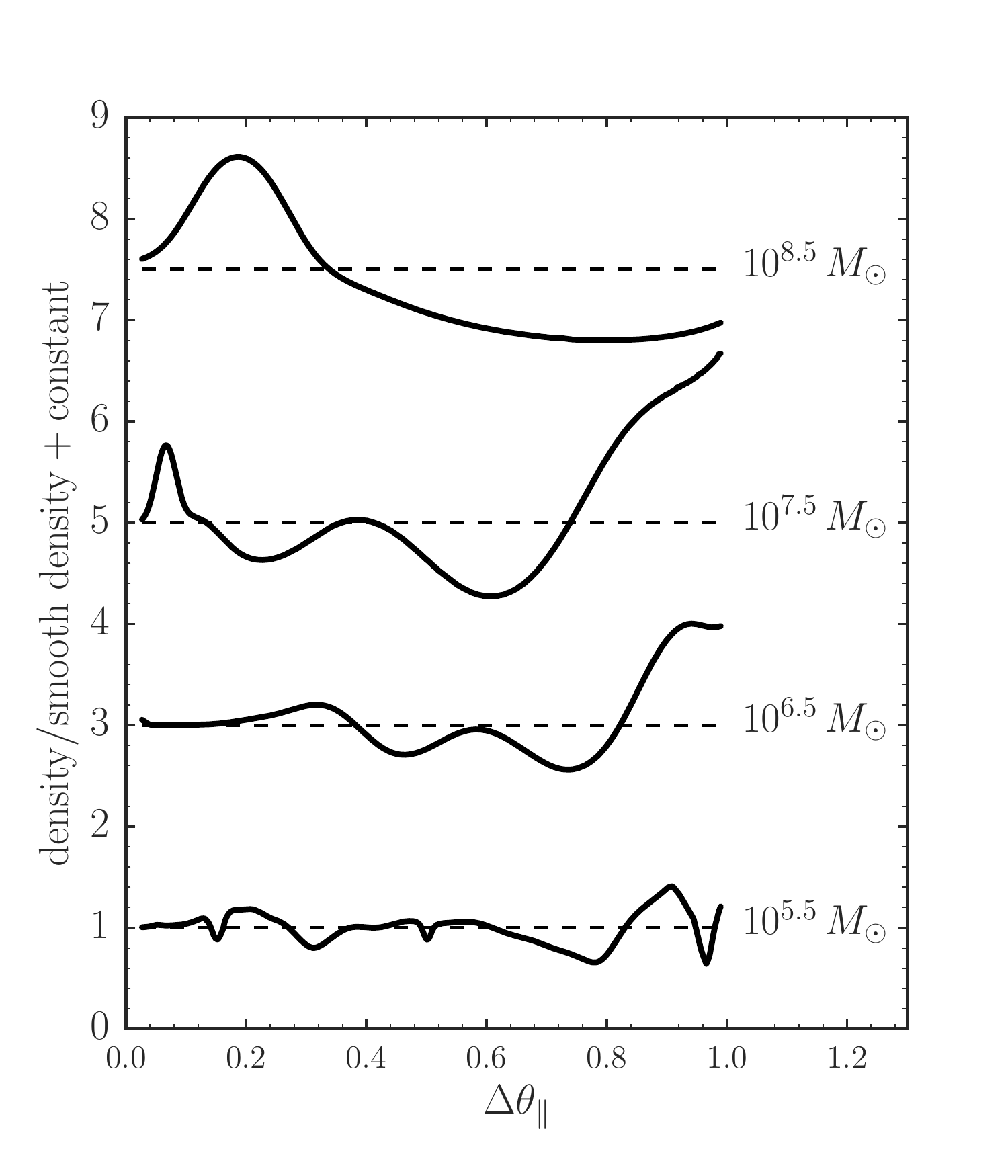}
\includegraphics[width=0.48\textwidth]{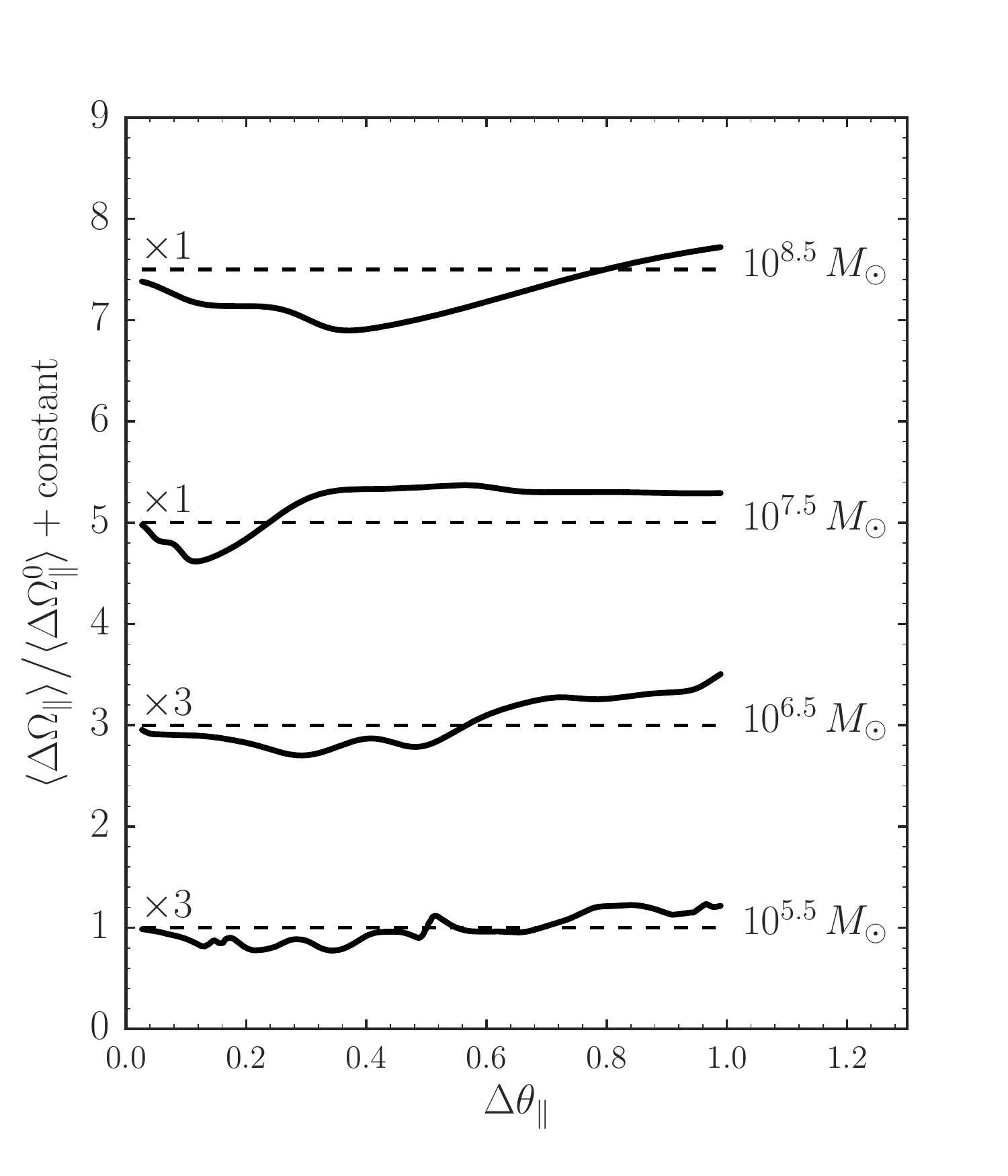}
\caption{Example perturbed stream densities (\emph{left panel}) and
  stream tracks (\emph{right panel}) computed using the formalism of
  \sectionname~\ref{sec:multiple}. Each curve in the left panel
  displays the density compared to that for the unperturbed stream
  when impacting the GD-1-like mock stream with a set of subhalos
  with the masses indicated next to the curve and otherwise sampled
  using the procedure of \sectionname~\ref{sec:sample}. The rate of
  impacts for each mass is that corresponding to its mass decade (\eg,
  $10^5$ to $10^6\msun$ for $10^{5.5}\msun$): $\approx43.3$ for $10^5$
  to $10^6\msun$, $\approx13.7$ for $10^6$ to $10^7\msun$, $\approx4.3$
  for $10^7$ to $10^8\msun$, and $\approx1.4$ for $10^8$ to
  $10^9\msun$. The mean track $\langle \dopar\rangle(\dapar)$ in
  parallel frequency for the same 4 simulations is shown in the right
  panel. The dashed lines in both panels give the locus where the
  ratio of perturbed-to-smooth density/track is equal to one. The
  deviation from one is multiplied by a factor of 3 for the two lowest
  mass sets of impacts in the right panel. The structure induced by
  different mass decades has a similar density amplitude for all
  masses (albeit slightly lower for lower masses), but different
  masses induce structure on different scales. The same holds for the
  deviations in the mean track, except that the amplitude decreases
  faster with decreasing mass.\label{fig:densexample_singlemasses}}
\end{figure*}

\begin{figure*}
\includegraphics[width=0.48\textwidth]{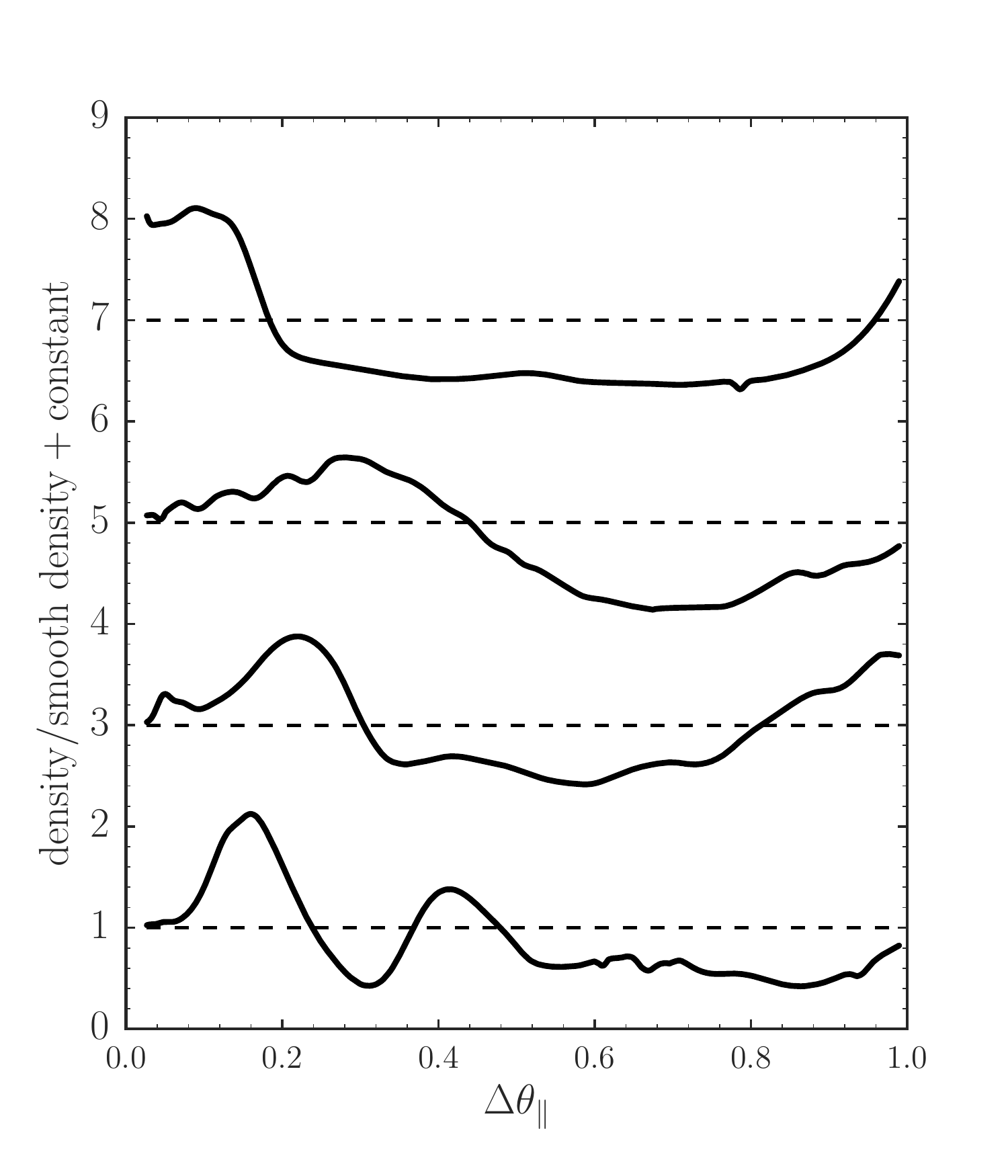}
\includegraphics[width=0.48\textwidth]{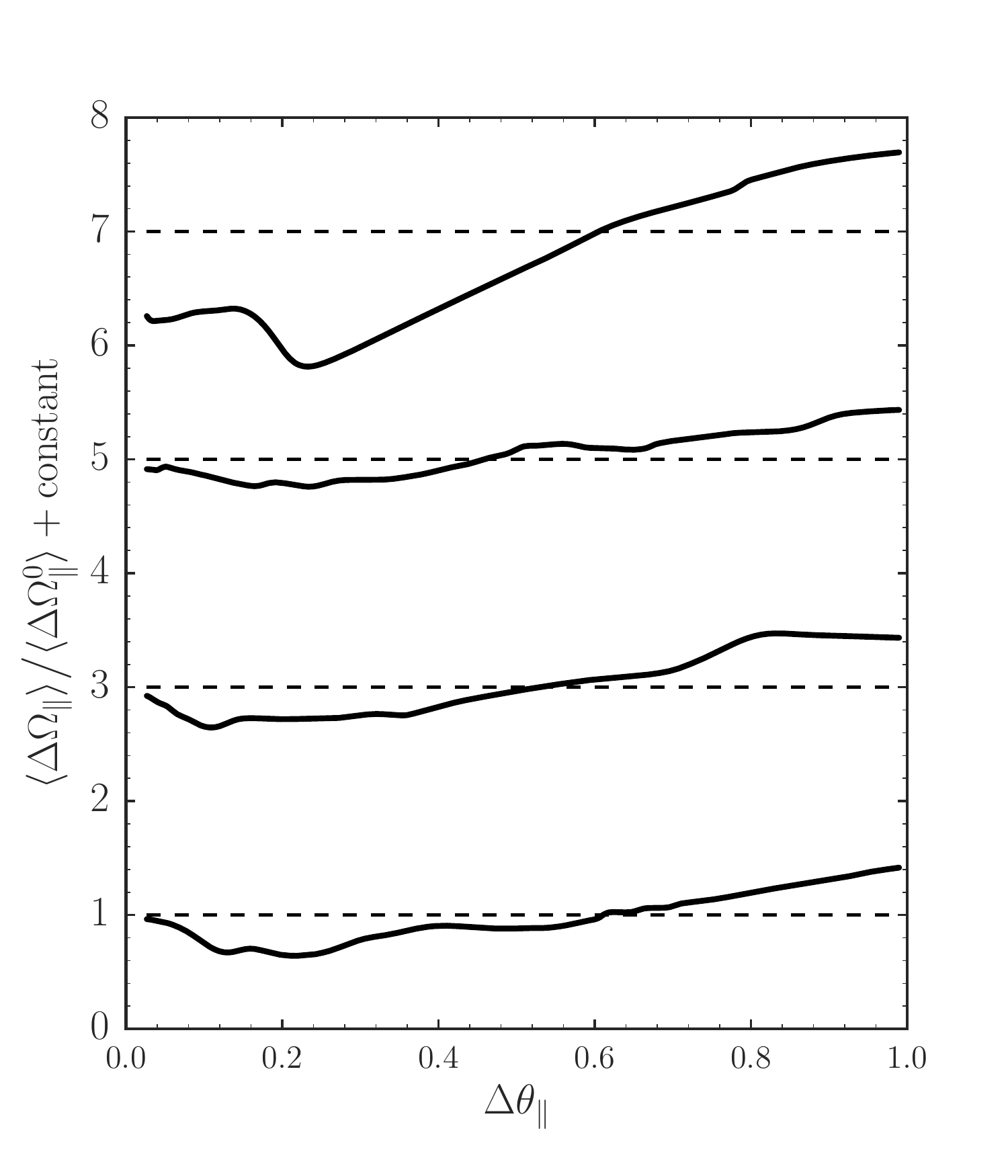}
\caption{Same as \figurename~\ref{fig:densexample_singlemasses}, but
  showing four examples with impacts sampled from the entire
  $10^5$--$10^9\msun$ mass range using the procedure of
  \sectionname~\ref{sec:sample}. The combination of impacts of
  different masses induces structure on a wide range of scales, which
  is especially clear in the density.\label{fig:densexample}}
\end{figure*}

To rapidly compute the stream structure under the influence of
multiple impacts, we proceed in a similar manner as for the
piecewise-linear approximation for the single-impact case above. We
start by combining all individual kicks $\delta \Omega_\parallel^g$
that occur at the same impact time $t_i^j$ into a single $\delta
\Omega_\parallel^{g,j}$ that is applied a time $t_i^j$ in the past. We
then order the kicks in reverse chronological sequence, that is,
starting with the most recent one, with indexing starting at one. By
propagating a line-of-parallel-angle through all previous impacts in
the piecewise-linear approximation of the kicks $\delta
\Omega_\parallel^{g,j}$, we can form a piecewise-linear representation
of this line before all impacts that we can then apply the linear
formalism from the previous section to. This procedure is illustrated
in \figurename~\ref{fig:illustrate_perturbations}. This propagation
requires one to track each piecewise-linear segment as it goes through
all previous impacts and to split up each segment into multiple
segments to account for the breakpoints of previous impacts. For this
it is necessary to update the coefficients of the piecewise-linear
representation of the line-of-parallel-angle to account for the
impacts and the splitting up of segments. In the following paragraphs,
we describe a simple method to track all of this that emerges from
writing out this straightforward, but complicated procedure. We
discuss how this algorithm is arrived at in more detail in
Appendix~\ref{sec:detail}.

We start in the same manner as for a single impact by computing
$\dopar^{\mathrm{max}}$, the maximum $\dopar$ that was released before
the final impact: $\dopar^{\mathrm{max}} = \dapar/t_i^1$. All $\dopar
> \dopar^{\mathrm{max}}$ were stripped from the progenitor after the
final impact and we can therefore again evaluate the unperturbed
$p_0(\dopar,\dapar)$ to obtain $p(\dopar,\dapar)$. We then compute the
set of breakpoints $\{\doparb\}$ using
\equationname~(\ref{eq:breakSingle}) corresponding to the final impact
$t_i^1$. With each breakpoint, we associate a set of parallel angles
$\{\daparb\}$ and times $\{t_b\}$ that are all set to $\dapar$ and
$t_i^1$, respectively
\begin{align}
  \daparb &= \dapar\,,\\
  t_b & = t_i^1\,.
\end{align}
In \figurename~\ref{fig:illustrate_perturbations}, the initial set of
$\{\doparb\}$ are the large black dots on the black line in the second
panel. 

We further associate with each breakpoint (a) $\{x_b\}$, the set of
breakpoints for the first kick, (b) $\{d^b_0\}$ and $\{d_1^b\}$, the
constant and linear coefficients for each piecewise-polynomial segment
of the first kick (denoted $c_{b0}$ and $c_{b1}$ in
\sectionname~\ref{sec:single}), (c) $\{\dd\Omega_b\}$ defined as
\begin{equation}
  \dd\Omega_b = -d^b_0-d_1^b\,(\daparb-x_b)\,,
\end{equation}
the frequency correction from the final kick for each segment, and (d)
a set of coefficients $\{c_{b0}\}$, $\{c_{b1}t\}$, and $\{c_{bx}\}$ set to
\begin{align}
  c_{b0} &= d_0^b\,,\\
  c_{b1}t & = d_1^b\,t_i^1\,,\\
  c_{bx} & = 0\,.
\end{align}
By updating these breakpoint-associated arrays through previous
impacts, we can trace the line-of-parallel-angle to its initial state.

After undoing the effect of the final impact, we have a
piecewise-linear representation of the initial line-of-parallel-angle
characterized by its breakpoints $\{\doparb\}$ and up-to-linear
polynomial coefficients $\{d_0^b\}$ and $\{d_1^b\}$ (see the middle
panel of \figurename~\ref{fig:illustrate_perturbations}). We also have
the parallel angle of each breakpoint at $t_i^1$:
$\dapar-\dopar\,t_i^1$. A segment $b$ then arrives at the time of the
second-to-last impact at the parallel angles
\begin{equation}
  \dapar-(t_i^2-t_i^1)\,\dd\Omega_b-\dopar\,[t_i^2-(t_i^2-t_i^1)\,(1+c_{b1}t)]\,.
\end{equation}
We thus update the $\{\daparb\}$ and $\{t_b\}$ as
\begin{align}\label{eq:update_dapar}
  \daparb &\rightarrow \daparb-(t_i^2-t_i^1)\,\dd\Omega_b\,,\\
  t_b & \rightarrow t_b+(t_i^2-t_i^1)\,(1+c_{b1}t)\,,\label{eq:update_t}
\end{align}
such that the segment $b$ arrives at the parallel angles
\begin{equation}
  \daparb-\dopar\,t_b\,,
\end{equation}
similar to the expression for how each original segment arrives at the
final kick. Like in \equationname~(\ref{eq:breakSingle}), we can then
determine the $\dopar$ for each current segment and for each
breakpoint $x_{b'}$ of the second-to-last impact as
\begin{equation}\label{eq:newbreak}
  \dopartb = \frac{\daparb-x_{b'}}{t_b}\,,
\end{equation}
We only keep those breakpoints ${\dopartbtilde}$ for each segment $b$
that fall within the $]\doparb,\doparbpone]$ range of the segment and
add it to the current set of breakpoints
\begin{equation}\label{eq:update_breakpoints}
\{\doparb\} \rightarrow \{\doparb\} \cup \{\dopartbtilde\}\,.
\end{equation}
The new breakpoints $\{\dopartbtilde\}$ are the small black dots in
the fourth panel of \figurename~\ref{fig:illustrate_perturbations};
the total set is both the small and large black dots (from the
previous iteration) in this panel.

Having established the set of breakpoints that describes the combined
effect of the final two impacts in a piecewise-linear manner, we
proceed to undo the effect of the second-to-last impact on
$\dopar$. For this we need to subtract the piecewise-linear
representation of $\delta\Omega_\parallel^{g,2}$ between each two
breakpoints corresponding to the second-to-last kick. To keep track of
this, we first need to update
\begin{align}\label{eq:updatecfirst}
  c_{b0} & \rightarrow c_{b0}+t_b\,d_1^b\,\doparb\,,\\
  c_{bx} & \rightarrow c_{bx}+t_b\,d_1^b\,.\label{eq:updatecfirst2}
\end{align}
These updates are necessary to account for the new breakpoints
inserted in \equationname~(\ref{eq:update_breakpoints}): this
insertion requires the constant term in the piecewise-linear
representation of the line-of-parallel-angle to be updated because of
the offset between the old and new breakpoints. All of the arrays
associated with the breakpoints are then enlarged by the addition of
the new breakpoints (\equationname~[\ref{eq:update_breakpoints}]),
with the values for the new entries set to those corresponding to the
old breakpoint associated with the new one. The latter is that
breakpoint before which the new breakpoint is inserted. The exception
to this rule are the $\{x_b\}$, $\{d_0^b\}$, and $\{d_1^b\}$, which
are set to the second-to-last impact breakpoints and coefficients;
those entries in the sequence for old breakpoints are set to the value
of the new breakpoint associated with the old breakpoint. The
objective of these is to track the effect of the latest impact. We
undo the effect of the second-to-last impact by updating
\begin{align}\label{eq:updatecsecond}
  c_{b0}&\rightarrow c_{b0}+d_0^b\,,\\
  c_{b1}t &\rightarrow c_{b1}t+d_1^b\,t_b\,,\\
  \dd\Omega_b &\rightarrow \dd\Omega_b-d_0^b+d_1^b\,(\daparb-x_b)\,.\label{eq:updatecsecond2}
\end{align}
In these updates, parts of the line-of-parallel-angle that must have
been released at times between $t_i^2$ and $t_i^1$ are automatically
left unchanged, because for these we have already arrived at their
unperturbed values for which we can evaluate $p_0(\dopar,\dapar)$.

Thus, we arrive at the same form of piecewise-linear representation of
the line-of-parallel-angle before the second-to-last impact as we had
before the last impact. We can then repeat the procedure given here
for the second-to-last impact for all previous impacts. The time
differences in \equationname s~(\ref{eq:update_dapar}) and
(\ref{eq:update_t}) becomes those between the previous impact and the
current impact, \ie, $(t_i^3-t_i^2)$ for the third-to-last
impact. Otherwise the procedure is the same. After all impacts have
been corrected for, $c_{b0}$ receives a final update
\begin{equation}\label{eq:c0finalupdate}
  c_{b0} \rightarrow c_{b0}-c_{bx}\,\doparb\,.
\end{equation}
This update is again part of the correction of the constant term of
later impacts when the breakpoints of earlier impacts are inserted
into the sequence of breakpoints, thus breaking a linear segment of a
later impact into multiple segments (this is described in detail in
Appendix~\ref{sec:detail}). At the end of this procedure, we have the
full representation of the line-of-parallel-angle before the first
impact and we can thus evaluate $p(\dopar,\dapar)$ as
$p_0(\dopar,\dapar)$.

With the representation of the line-of-parallel-angle arrived at here,
we can evaluate moments of $p(\dopar,\dapar)$ in the piecewise-linear
approximation of the kicks using the expressions in
\equationname~(\ref{eq:pdenssingle}) for the density and
\equationname~(\ref{eq:mOsingle}) for $\mopar(\dapar)$, respectively,
making the substitution $c_{b1}t_i^1\rightarrow c_{b1}t$. To determine
the lower limit $b_0$ of the sum, we solve an equation similar to
(\ref{eq:lowerlimit}), namely,
\begin{equation}
\begin{split}
  \left[\doparbpone-c_{b0}-(1+c_{b1}t)x\right]& (t_d-t_i^N) = \\
  & \daparb-(\doparbpone-x)\,t_b\,,
\end{split}
\end{equation}
but now using the time $t_i^N$ of the first impact on the left-hand
side. The lower limit $b_0$ is determined to be the segment for which
the solution $x_0$ satisfies $\doparb < x_0 \leq \doparbpone$ and we
again adjust the lower limit of this segment to be $\Delta
\Omega_{\parallel,b_0} \rightarrow
\Delta\Omega_{\parallel,b_0+1}-x_0$.

An example of the formalism from this section is displayed in
\figurename~\ref{fig:densmOparApproxMulti}. We calculate the effect of
four encounters with dark--matter halos with masses $\approx10^7\msun$
(in detail, $10^7\msun$, $10^{7.25}\msun$, $10^{6.75}\msun$,
$10^{7.5}\msun$) at times $1.3\Gyr$, $2.3\Gyr$, $3.3\Gyr$, and
$4.3\Gyr$ in the past. They have fly-by velocities of $160\kms$,
$152.5\kms$, $229\kms$, and $161\kms$, impact parameters that are
$0.5$, $2$, $1$, and $2.5$ scale radii of the dark--matter halos, and
hit at $\dapar= 0.6$, $0.4$, $0.3$ and $0.3$ at the time of
impact. Therefore, they all encounter approximately the same part of
the stream that is currently located at $\dapar \approx 0.68$. As
such, this set of impacts is a good test of the formalism here,
because the effect of the different encounters significantly
overlaps. It is clear from \figurename~\ref{fig:densmOparApproxMulti}
that the approximate calculation from this section agrees well with
both the density obtained from sampling mock data using the procedure
of \sectionname~\ref{sec:mock} and with a direct numerical integration
of the moments of $p(\dopar,\dapar)$. The densities from the
approximate and direct-numerical calculations agree to $\approx1\,\%$,
while the mean parallel frequency $\mopar(\dapar)$ agree to a fraction
of that. These precisions are much better than what could
realistically be observed for real tidal streams.

Further examples of the formalism from this section are shown in
\figurename s~\ref{fig:densexample_singlemasses} and
\ref{fig:densexample}. \figurename~\ref{fig:densexample_singlemasses}
displays the density and mean parallel frequency $\mopar(\dapar)$ for
four simulations. Each simulation consists of impacts of a single
value of the mass, but with all other impact parameters sampled as in
\sectionname~\ref{sec:sample}. The number of impacts for each
simulation was sampled from the Poisson rate corresponding to the mass
decade of each mass (\ie, the $10^{5.5}\msun$ simulation uses the rate
corresponding to the mass range $10^5$ to $10^6\msun$); see the
caption of \figurename~\ref{fig:densexample_singlemasses} for the
actual rates. The single $10^{8.5}\msun$ impact induces a perturbation
covering the entire length of the stream in both the density and
$\mopar(\dapar)$. Each individual impact with a lower-mass subhalo
affects a shorter part of the stream and the typical scale on which
structure is induced is set by the mass of the subhalo. The amplitude
of the induced density variations is similar for all mass decades,
with only slightly lower amplitudes for the lower-mass decades. The
structure induced in $\mopar(\dapar)$ behaves similar, although the
amplitude decreases more steeply with decreasing mass. Below we
quantify these observations by computing the power spectrum of these
density and $\mopar(\dapar)$ fluctuations.

\figurename~\ref{fig:densexample} gives four examples with impacts
covering the entire mass range that we consider in this paper: $10^5$
to $10^9\msun$. The density perturbations are typically of order unity
and they display structure on all scales. Our formalism properly
accounts for the interference between the effect from different
impacts and for the dispersion in the stream, \ie, the filling in or
further emptying of gaps due to these
effects. \figurename~\ref{fig:densexample} shows that structure
remains visible on all scales even when all of these effects are taken
into account. The mean parallel frequency $\mopar(\dapar)$ again
displays a similar behavior, but with smaller amplitude fluctuations
overall. Comparing the density and $\mopar(\dapar)$, it is clear that
features in the density are correlated with those in
$\mopar(\dapar)$. For example, the density dip in the top curve at
$\dapar \approx 0.78$ and the dip in the density in the
$10^{5.5}\msun$ simulation in
\figurename~\ref{fig:densexample_singlemasses} at $\dapar \approx 0.5$
have a clear counterpart in $\mopar(\dapar)$. From the four examples
shown here, it is also clear that on average $\mopar(\dapar)$ gets
slightly tilted counter-clockwise with respect to the unperturbed
stream track. This is explained by the fact that we do not consider
impacts that occur in the opposite arm or beyond the nominal length of
the stream. This only affects the structure of the stream on the
largest scales and we ignore this effect in what follows as it only
has a marginal effect on the fluctuations in the stream induced by
substructure that we are most interested in (see further discussion in
Appendix~\ref{sec:convtests}).

\begin{figure}
\includegraphics[width=0.48\textwidth]{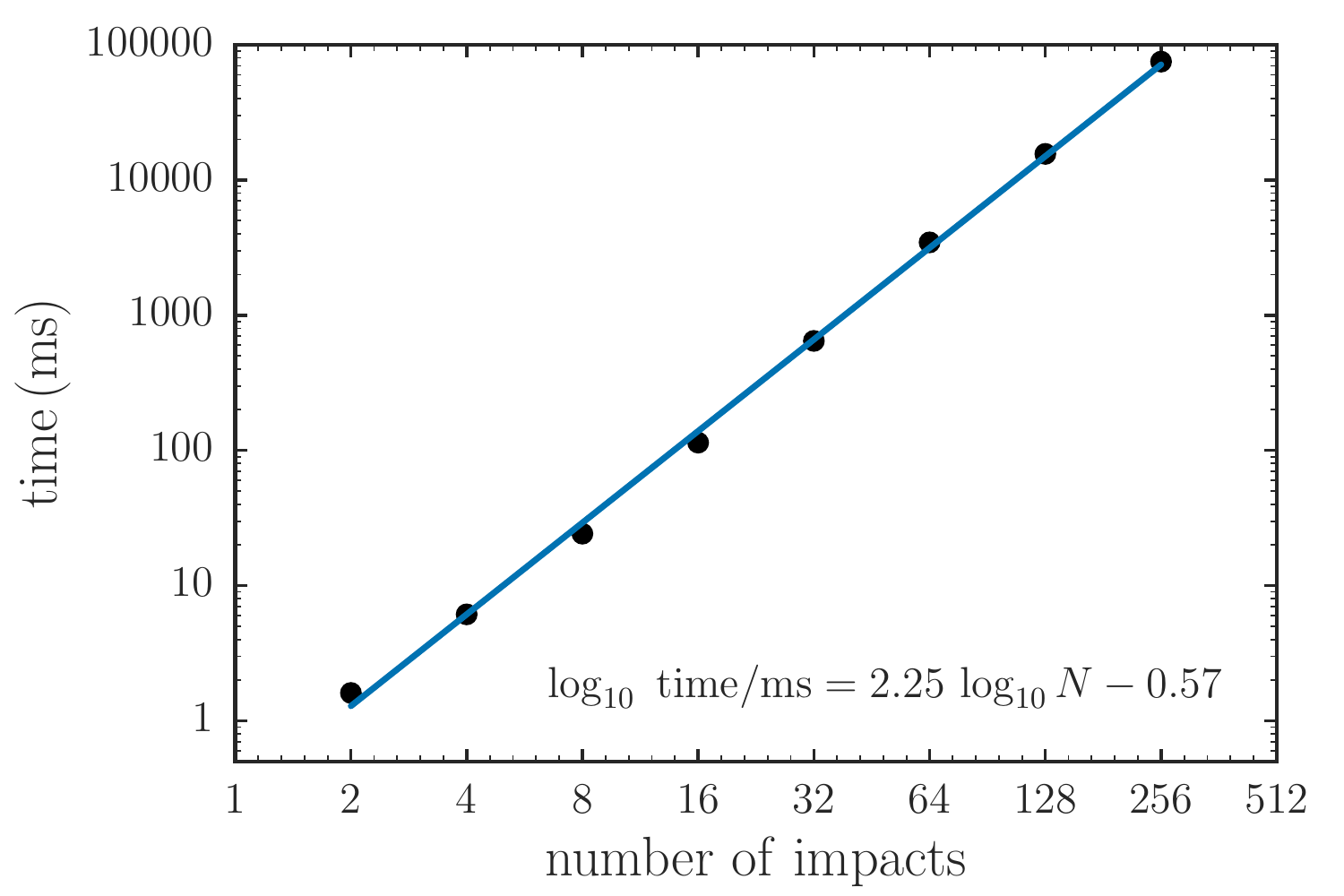}
\caption{Time to compute $p(\dopar,\dapar)$ for all $\dopar$ at a
  given $\dapar$ using the line-of-parallel-angle approximation of
  \sectionname~\ref{sec:multiple} as a function of the number of
  impacts at different times. The computational time scales
  approximately as $N^{2.25}$ for $N$ impacts. The CDM prediction for
  the number of impacts for the GD-1-like mock stream is
  approximately 63 impacts (see \sectionname~\ref{sec:sample}), for
  which the line-of-parallel-angle approximation can be computed in
  $0.3$\,s for a single $\dapar$ and a fine sampling in $\dapar$ of
  the structure of the perturbed stream can be computed in about one
  minute.\label{fig:compTime}}
\end{figure}

In \figurename~\ref{fig:densmOparApproxMulti}, the approximate
calculation is about 100 times faster than the direct numerical
integration, with the approximate calculation taking $9$\,ms for a
single $\dapar$ on a single cpu. The approximate calculation for
multiple impacts in this case is about 100 times slower than that for
a single impact (see \sectionname~\ref{sec:single}). This is because
the procedure to propagate the line-of-parallel-angle through multiple
impacts requires one to track all of the associated arrays as
described in this section and this is expensive compared to the ease
with which expressions (\ref{eq:pdenssingle}) and (\ref{eq:mOsingle})
can be evaluated. The computational time for each impact is therefore
dominated by the operations described in this section, which increase
with each previous impact due to the ever larger number of breakpoints
to track. \figurename~\ref{fig:compTime} shows the computational time
for computing the density $p(\dapar)$ at a single $\dapar$ as a
function of the number of impacts at different times (because multiple
impacts at the same time do not increase the computational cost). The
computational time increases approximately as $N^{2.25}$ for $N$
impacts, essentially because the number of breakpoints to track
increases linearly with the number of impacts at different times. As
the propagation of the line-of-parallel-angle dominates the
computational cost and this propagation returns the entire
$p(\dopar,\dapar)$, all moments can be evaluated simultaneously at
approximately zero cost compared to the propagation. Thus, in the case
of \figurename~\ref{fig:densmOparApproxMulti}, both the density and
mean parallel frequency can be evaluated in a total time of $9$\,ms
per $\dapar$, leading to a further factor of two speed-up compared to
the numerical evaluation of these moments.

While we do not consider $p(\doperp,\daperp|\dopar,\dapar)$ here,
because its contribution to the present-day stream structure is small
(as demonstrated in \figurename~\ref{fig:meanOparOperp}), it is clear
that if desired, this probability can be computed in a similar manner
as $p(\dopar,\dapar)$ in the line-of-parallel-angle approach. This is
because the kicks $\delta \veco^g$ only depend on $\dapar$, such that
all $(\doperp,\daperp)$ at a given $(\dopar,\dapar)$ have the same
history of $\delta \Omega_\perp^g$ and induced $\delta \daperp$. This
history could be straightforwardly computed by keeping track of the
changes in $\doperp$ and induced changes in $\daperp$ of each segment
for all impacts during the backward-propagation of the
line-of-parallel-angle above, as long as the piecewise-polynomial
representation of each impact's $\delta \Omega_\perp^g$ has the same
breakpoints as that of its corresponding $\delta \Omega_\parallel^g$.

\subsection{Conversion to configuration space}\label{sec:convert_obs}

\begin{figure*}
\includegraphics[width=\textwidth]{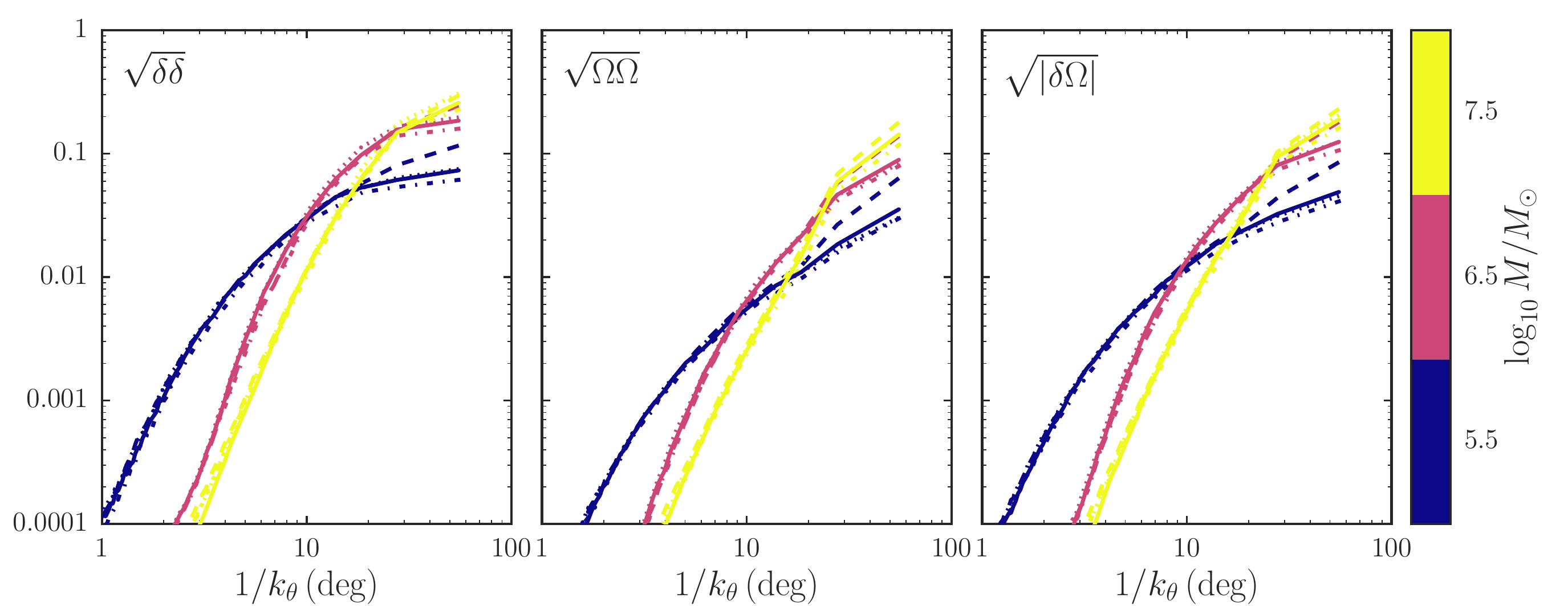}
\caption{Power spectra of the fluctuations in the density and the mean
  track $\langle \dopar\rangle(\dapar)$ relative to those in the
  unperturbed stream for impacts of different masses. Each curve has
  impacts of a single value for the mass with the rate set by the
  surrounding mass decade (\eg, that of $10^7-10^8\msun$ for
  $10^{7.5}\msun$). The left and central panels display the square
  root of the power spectra of the density and $\langle
  \dopar\rangle(\dapar)$ fluctuations, while the right panel shows the
  square root of the magnitude of the cross power spectrum between the
  density and $\langle \dopar\rangle(\dapar)$. The $x$-axis is the
  inverse of the wavenumber, such that small scales are on the
  left. All curves are the median of at least $1,000$ different
  simulations. The solid lines are computed using the fiducial
  simulation setup; with other linestyles we also show the result from
  increasing the time sampling, the factor $X$ that sets the maximum
  impact parameter, and the length along the stream at which impacts
  are considered (see text for details).\protect\\While all the
  different mass ranges contribute most of their power on the largest
  scales, the power on smaller scales is cut off in a mass-dependent
  manner. The power on a particular scale is dominated by the
  contribution of a single mass range. In particular, the power on a
  few degree scales is dominated by the effect of
  $\approx10^5-10^6\msun$ subhalos. Both the density and mean track
  display fluctuations of a similar magnitude and a similar dependence
  on scale, although the density fluctuations are typically a factor
  of a few larger. Fluctuations in the density and the track are
  strongly correlated.\label{fig:bmaxtime}}
\end{figure*}

So far we have discussed how to compute the phase--space structure of
a perturbed stream in frequency--angle space. To compare these models
to observed data, the models need to be projected into configuration
space $(\vecx,\vecv)$. Before discussing how this projection can be
performed efficiently, it is important to note the following. The
relation between $\dapar$ and more-easily observable coordinates for
the location along a stream (\eg, RA, Galactic longitude $l$, or a
custom celestial coordinate frame along the stream) is smooth without
any high-frequency power (see, for example, the middle panel of
\figurename~\ref{fig:gd1props}). Because the density, say as a
function of $l$, is given by $p(l) = p(\dapar) \left|\dd \dapar / \dd
l\right|$ and $\left|\dd \dapar / \dd l\right|$ does not vary rapidly
along a stream, the statistical properties as a function of scale
(\eg, the power spectrum or bispectrum as discussed below) of $p(l)$
will be very similar to those of $p(\dapar)$. Similarly, we will
compute the perturbed stream track by converting the stream track in
frequency--angle space to configuration space. Because the
perturbations due to subhalo impacts are small, perturbations in
observable coordinates are related to those in frequency--angle space
through Jacobians that vary smoothly over the length of the stream as
in the case of the density. Therefore, the perturbations in observable
space will be very similar to those in frequency--angle space except
for their overall amplitude.

To convert the density $p(\dapar)$ and the mean track $\langle
\dopar\rangle(\dapar)$ to configuration space---say, Galactic
coordinates $(l,b,D,V_{\mathrm{los}},\mu_l,\mu_b)$---we compute the
current track of the stream in the absence of perturbations and
convert it to configuration space using the iterative method from
\citet{Bovy14a} (see \sectionname~\ref{sec:smooth}). As discussed by
\citet{Bovy14a}, this procedure returns the track in configuration
space at a series of points $\dapari$ along the stream and the
Jacobian of the transformation $(\veco,\veca) \rightarrow
(\vecx,\vecv)$ at these points. Using these Jacobians, we can then
linearize the transformation between $(\veco,\veca)$ and
$(\vecx,\vecv)$ (or $(l,b,D,V_{\mathrm{los}},\mu_l,\mu_b)$). Improving
on \citet{Bovy14a}, we linearly interpolate these Jacobians for points
$\dapar$ between the points $\dapari$ (\citet{Bovy14a} used the
Jacobian for the nearest $\dapari$ to a given $\dapar$).

The density $p(l)$ as a function of $l$ then follows from $p(l) =
p(\dapar) \left|\dd \dapar / \dd l\right|$. The Jacobian can be
computed from the Jacobians of $(\dopar,\dapar) \rightarrow
(\vecx,\vecv)$ (in practice, this is most easily done
numerically). Below, we will consider the ratio of the perturbed
density to that of the unperturbed stream
$p(\dapar)/p_0(\dapar)$. Because both are converted to $p(l)$ using
the same Jacobian, $p(l)/p_0(l) = p(\dapar)/p_0(\dapar)$.

To convert the track $\langle \dopar\rangle(\dapar)$ to configuration
space, we simply convert the phase--space points $(\dopar=\langle
\dopar\rangle(\dapar),\doperp = 0,\dapar,\daperp=0)$ to configuration
space (see above for a discussion of why setting $\doperp = 0$ and
$\daperp=0$ is a good assumption). Because the perturbations $\langle
\dopar\rangle(\dapar)-\langle \dopar^0\rangle(\dapar)$ are small, the
linear approximation of the coordinate transformation is precise.

We do not display any examples of perturbations in configuration space
like in \figurename s~\ref{fig:densexample_singlemasses} and
\ref{fig:densexample}, because for the reasons discussed at the start
of this subsection, the perturbations in configuration space look
essentially the same as those in frequency--angle space. In
particular, the shape of, \eg, $\langle b \rangle(l)$ is the same as
that of $\langle \dopar\rangle(\dapar)$. Because the perturbations are
to a good approximation only in $\langle \dopar\rangle(\dapar)$ (see
\figurename~\ref{fig:meanOparOperp} and associated discussion),
\emph{all} projections of the perturbed stream track in configuration
space are tracing the same perturbations and they are therefore almost
completely correlated. This implies that measurements of a stream's
location in different coordinates can be stacked to obtain better
measurements of the perturbed stream.

\section{Tidal stream power spectra}\label{sec:powspec}

The line-of-parallel-angle approach for computing the density and mean
track allows the efficient computation of the structure of a tidal
stream perturbed by the $\mathcal{O}(100)$ expected number of impacts
with dark--matter subhalos that properly takes into account the
dispersion within the stream and the overlapping effects of multiple
impacts. In this section, we use this to run large suites of
simulations for the GD-1-like stream from
\sectionname~\ref{sec:smooth} and investigate its statistical
properties using the one-dimensional power spectrum of density or
track fluctuations induced by the perturbations.

\begin{figure*}
\includegraphics[width=\textwidth]{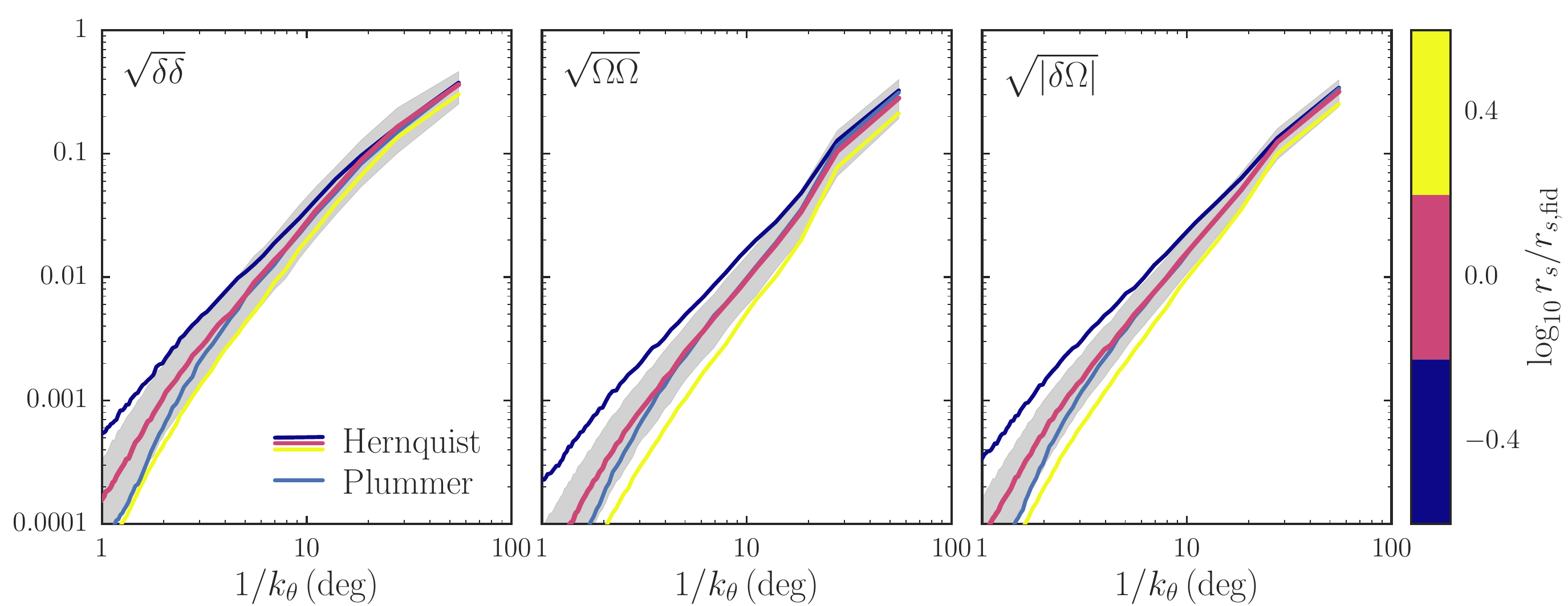}
\caption{Like \figurename~\ref{fig:bmaxtime}, but for impacts with
  masses sampled from the full $10^5\msun$ to $10^9\msun$ range. The
  curves with colors following the colorbar use dark--matter subhalos
  with a Hernquist profile following different $r_s(M)$ relations that
  roughly bracket the relation seen for halos in the Via Lactea II
  simulation; these have $r_s$ a factor of 2.5 larger or smaller than
  the fiducial relation. The lighter blue curve models the
  dark--matter subhalos as Plummer spheres, with $r_s(M) =
  1.62\kpc\,\left(M/10^8\msun\right)^{0.5}$, which produces subhalos
  with similar concentrations as the fiducial Hernquist model. The
  gray band shows the interquartile range within which individual
  simulations lie for the fiducial setup. The concentration of the
  subhalo modifies the power spectrum on the smallest scales, with
  larger fluctuations for more compact subhalos. The power spectra are
  almost indistinguishable at scales larger than a few degrees when
  the subhalos are modeled with Hernquist or Plummer
  profiles.\label{fig:rs}}
\end{figure*}

We compute the power spectrum of the density fluctuations for a single
simulation by taking the perturbed density and dividing it by the
density of the unperturbed stream (\eg,
\figurename~\ref{fig:densexample}) and computing the one-dimensional
power spectrum using a Hann window \citep[\eg,][]{Press07a}. In
practice, we use the \texttt{csd} routine in \texttt{scipy}
\citep{Jones01a} and do not divide by the sampling frequency. In all
of the figures, we plot the square root of the power spectrum versus
the inverse of the frequency, such that the $y$ and $x$ axes
represent the typical value of fluctuations at a given spatial
scale. In all figures, we perform at least 1,000 simulations of the
specified setup and display the median power spectra of all of these
simulations. We do this, because individual power spectra scatter are
noisy and scatter around the median. In many figures we also show the
interquartile range of the $>1,000$ simulations as a gray band, such
that the scatter among different realizations can be assessed.

Similarly, we compute the power spectrum of fluctuations in the mean
track $\langle \dopar\rangle(\dapar)$ by dividing the perturbed
location of the track by the unperturbed location and follow the same
procedure as for the density. To investigate the amount of correlation
between fluctuations in the density and in the track, we also compute
the cross power spectrum and look at the square root of its absolute
value (\ie, we do not consider the phase information here, although
that will also contain useful information about the impacts).

In this section, we consider the density and track in frequency--angle
space. As discussed above, the stream track is simplest in this space,
because all of the relevant structure is in the direction parallel to
the stream, which is simplest in this coordinate system. In a later
section we will compute power spectra of the density and track as they
would be observed, but these are very similar to the power spectra in
frequency--angle space for the reasons discussed above in
\sectionname~\ref{sec:convert_obs}.

Our sampling setup has a number of parameters that need to be
fixed. We investigate these in detail in Appendix~\ref{sec:convtests}
and briefly describe the results from these tests here. We only allow
impacts to happen at a set of equally-spaced discrete times along the
past orbit of the stream, because of the computational savings that
come from having multiple impacts at the same time (multiple impacts
at the same time do not increase the computational cost of the
line-of-parallel-angle algorithm). The sampling of the discrete set of
times needs to be high enough such that the structure of the stream is
not affected by the discrete time sampling. In
Appendix~\ref{sec:convtests} we find that the statistical properties
of the perturbed GD-1-like tidal stream converge quickly when the time
sampling is increased above a few times and we use a standard value of
64 times, which corresponds to a time interval of $\approx
140\Myr$. This is somewhat smaller than the radial period of the
stream, which is $400\Myr$, which makes intuitive sense. In general,
we therefore conclude that allowing impacts to happen at times sampled
slightly more frequently than the radial period of the orbit is
sufficient for investigating the statistical properties of perturbed
streams (as long as the time interval is not a simple fraction of the
period).

We further have to choose a value for the parameter $X$, which
describes the maximum impact parameter that we consider in the
equation $b_{\mathrm{max}} = X\,r_s(M)$. Appendix~\ref{sec:convtests}
demonstrates that the largest-scale structure of the stream converges
very slowly with increasing $X$, but the small-scale structure
converges for $X > 1$. We fix $X=5$ as the standard value in our
simulations and it should be kept in mind that the largest-scale modes
at $1/k_\theta \gtrsim 25^\circ$ in the power spectrum (and bispectrum
below) are not fully converged and likely underestimated by a few tens
of percent. The largest-scale structure of a stream is significantly
affected by distant encounters.

\begin{figure*}
\includegraphics[width=\textwidth]{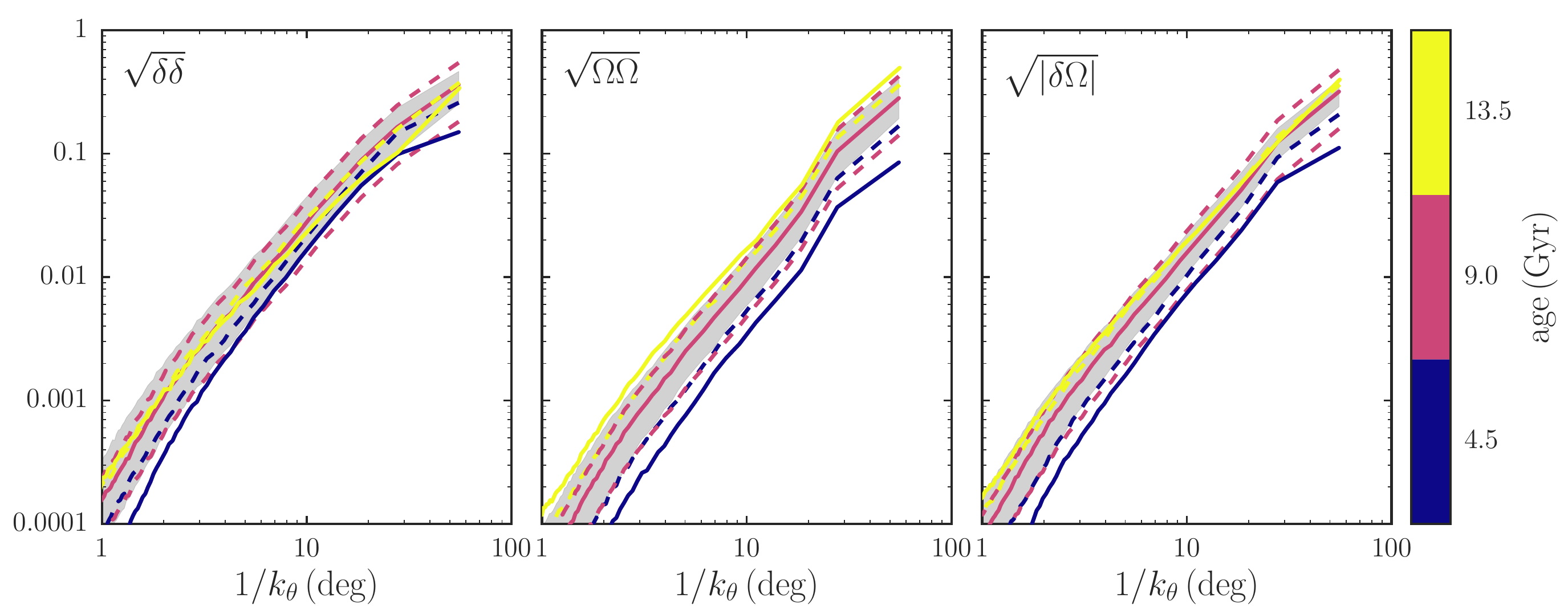}
\caption{Like \figurename~\ref{fig:rs}, but varying the age of the
  stream. The solid lines vary the age of the stream from the fiducial
  value of $9\Gyr$ to $4.5\Gyr$ and $13.5\Gyr$, while keeping the
  product of the age and the stream's velocity-dispersion parameter
  $\sigma_v$ constant. This keeps the current length of the stream the
  same. Because the impact rate depends linearly on the age at fixed
  length (cf.~\equationname~[\ref{eq:nenc}]), the younger/older stream
  has a rate of impacts that is 50\,\% lower / 50\,\% higher. The blue
  and yellow dashed curves vary the age while keeping the number of
  impacts the same. The dashed lines with the same color as the
  fiducial age are rescalings of the fiducial power spectrum with
  factors of $0.5$ and $1.5$. Changing the age of the stream while
  keeping the rate of impacts the same changes the power spectrum of
  the track fluctuations (middle panel) by approximately these two
  factors. However, adjusting the age and using the appropriate number
  of impacts for the adjusted age gives much larger changes. The
  density power spectra or the density--track cross power spectra do
  not change linearly when the number of impacts increases due to the
  older age of the stream.\label{fig:age}}
\end{figure*}

\begin{figure*}
\includegraphics[width=\textwidth]{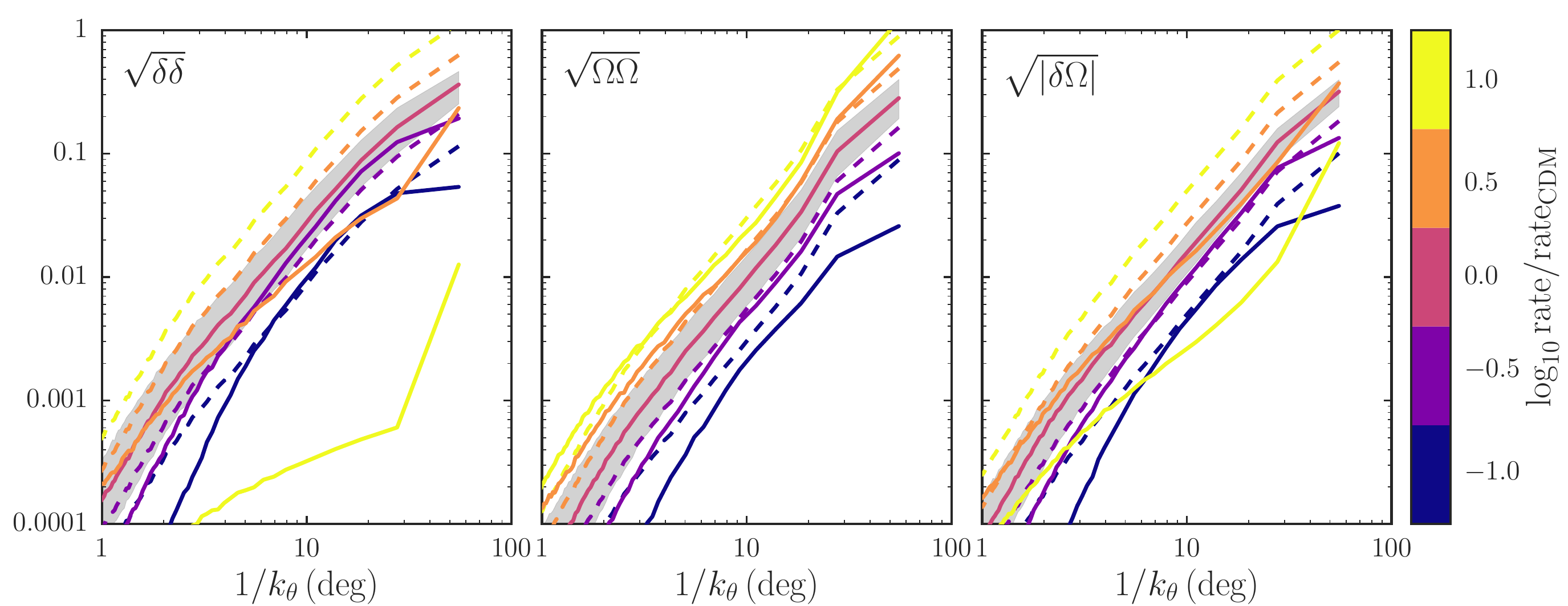}\\
\includegraphics[width=\textwidth]{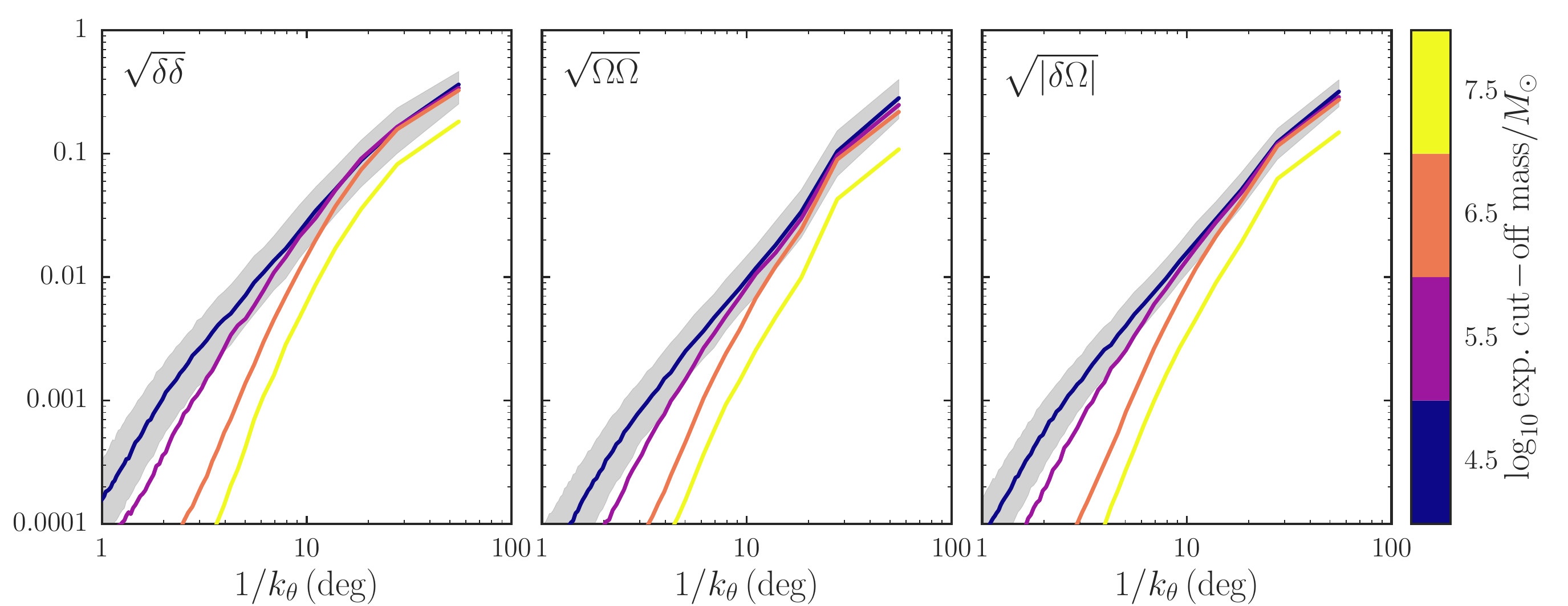}\\
\caption{Like \figurename~\ref{fig:rs}, but varying the rate of
  impacts (top row) and using a subhalo mass spectrum that is
  exponentially cut-off below a given mass. In the top row, the rate
  of impacts is increased or decreased by factors of 3 and 10; the
  rate of impacts with masses between $10^5\msun$ and $10^9\msun$
  therefore varies between 6.27 and 627. The dashed curves display the
  fiducial power spectrum rescaled by the square root of the changes
  in rate; this is the expected dependence if all impacts produce
  independent fluctuations in the stream. The power spectrum of mean
  track fluctuations is remarkably close to this expectation. This
  makes the mean-track power spectrum a powerful and unambiguous probe
  of the dark--matter subhalo mass spectrum (however, it is also the
  most difficult to measure; see below). The density hews close to the
  expectation from independent fluctuations when the rate is lowered,
  but the power spectrum is severely depressed when the rate is
  increased. For the case of a ten times higher rate, this is because
  the stream essentially gets destroyed by the subhalos, leaving a
  density peak near the progenitor without a stream. The bottom row
  demonstrates that a low-mass cut-off in the subhalo mass spectrum
  leads to a cut-off in the stream power spectra, at the expected
  locations from the single-mass simulations in
  \figurename~\ref{fig:bmaxtime}. In the bottom row, the fiducial
  case, which has no cut-off, is colored according to a cut-off at
  $10^{4.5}\msun$, which is below the minimum mass that we
  consider.\label{fig:multcutoff}}
\end{figure*}

\begin{figure*}
\includegraphics[width=\textwidth]{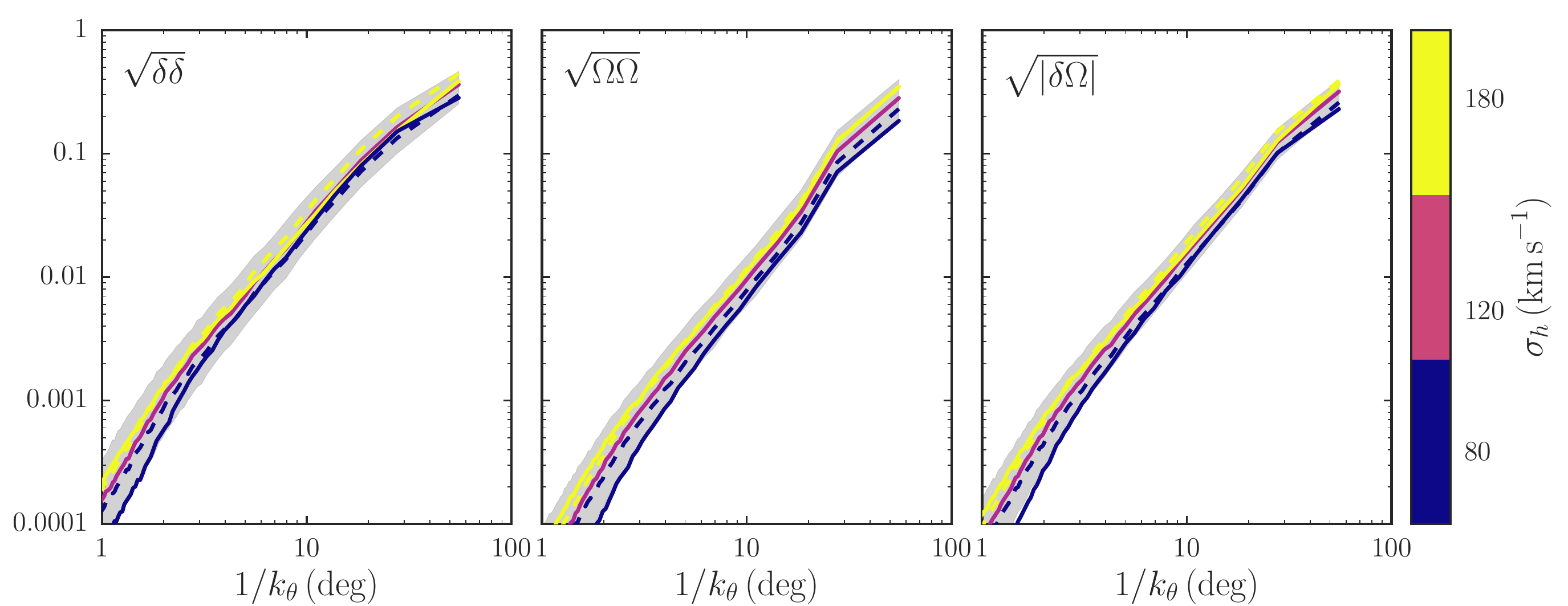}
\caption{Like \figurename~\ref{fig:rs}, but varying the velocity
  dispersion $\sigma_h$ of the population of dark--matter
  subhalos. The solid lines vary $\sigma_h$ by $50\,\%$ from the
  fiducial value of $120\kms$ to $180\kms$ (yellow curve) and to
  $80\kms$ (purple curve). Because the impact rate depends linearly on
  $\sigma_h$ (cf.~\equationname~[\ref{eq:nenc}]), the hotter/colder
  dark--matter population has rate of impacts that is 50\,\% higher /
  33\,\% lower. The dashed curves are rescalings of the fiducial power
  spectrum with factors of $\sqrt{3/2}$ and $\sqrt{2/3}$, appropriate
  if the impacts would act independently. The track power spectrum
  lies close to this expectation, while the density power spectrum and
  the density--track cross power spectrum vary less with
  $\sigma_h$.\label{fig:sigmah}}
\end{figure*}

Finally, we investigate how far along the stream we have to consider
impacts. Stellar streams do not have sharp edges and subhalo
perturbations could push low-surface brightness stream material
towards the higher-density part of the stream. In our fiducial setup
we simply consider impacts up to the length of the stream as defined
at the end of \sectionname~\ref{sec:smooth} evaluated at the time of
the impact. The convergence tests in Appendix~\ref{sec:convtests}
demonstrate that impacts further along the stream have a negligible
impact on the present-day structure of the stream (up to its nominal
length). However, if one were only computing the perturbed structure
of a stellar stream up to, say, half the length of the stream
(because, \eg, high-quality observations only extend over that range),
it would still be necessary to consider impacts over the full length
of the stream, as their effect at later times could impact the
observed half of the stream.

To understand the level at which power is induced on different scales
by different decades in subhalo mass, we perform simulations of
impacts with impact rates for a single mass decade while setting the
mass of all subhalos to the (logarithmically) central value in the
range. That is, we perform simulations of $M=10^{7.5}\msun$ subhalo
impacts with a rate of 4.3 that is the expected number for the range
$10^7\msun$ to $10^8\msun$ and compute the power spectra and cross
power spectra of the density and track fluctuations. We do the same
for $10^{6.5}\msun$ and $10^{5.5}\msun$
impacts. \figurename~\ref{fig:densexample_singlemasses} displays some
examples of the perturbed stream densities and tracks produced by
these kinds of simulations.

The power spectra for the different mass decades are displayed in
\figurename~\ref{fig:bmaxtime}. The solid lines represent the results
from the fiducial simulation setup, with impacts considered at $64$
different times, impact parameters up to $5\times r_s(M)$, and out to
the length of the stream (as defined at the end of
\sectionname~\ref{sec:smooth}). The other line styles summarize the
results from the convergence tests in Appendix~\ref{sec:convtests}
discussed above: The dot-dashed lines consider impacts at 256
different times; these are in almost all cases underneath the solid
lines, demonstrating that the power spectra have converged at 64
times. The dotted lines have impacts up to $125\,\%$ times the length
of the stream; these are also close to the solid lines. Finally, the
dashed curves take into account impacts out to $10\times
r_s(M)$. These dashed curves deviate from the solid curves on the
largest scales, demonstrating that the large-scale structure of the
stream is significantly affected by distant encounters ($10\times
r_s(M)\approx6\kpc$ for $M=10^{7.5}\msun$). We do not show the result
from $>10^8\msun$ impacts, because those are subdominant on all scales
for this stream.

It is clear from \figurename~\ref{fig:bmaxtime} that the different
mass ranges dominate the structure on particular scales for CDM-like
impact rates. The highest-mass impacts dominate the structure on the
largest scales and lower-mass impacts dominate on smaller
scales. Thus, by carefully measuring the power spectrum, the
contribution of different mass decades to the impact rate could be
determined. Because this rate is a direct reflection of the subhalo
mass spectrum, this allows the subhalo mass spectrum to be measured
from tidal stream power spectra.

Because of the dispersion in the stream and the dynamics of the
induced gaps, the effect of different impacts at the present time
overlaps and we need to consider impacts of all masses
simultaneously. The power spectra resulting from simulations over the
entire mass range of $10^5\msun$ to $10^9\msun$ are displayed in
\figurename~\ref{fig:rs}. The reddish line represents our fiducial
sampling setup. The power spectra considering impacts from all masses
closely follow the upper envelope of those from the individual mass
decades in \figurename~\ref{fig:bmaxtime}. This indicates that to a
good approximation the total power is simply the combination of the
power from each individual mass decade. Therefore, the identification
of power on a certain scale with a certain mass decade is robust.

\figurename~\ref{fig:rs} further considers the impact of the radial
profile of the dark--matter subhalos. The fiducial model uses a simple
$r_s(M)$ relation that is close to the average relation in the Via
Lactea II simulation (see discussion above). \figurename~\ref{fig:rs}
also shows the power spectra when increasing or decreasing $r_s(M)$ by
a factor of $2.5$, which approximately brackets the
mass--concentration relation of Via Lactea II subhalos. Increasing the
concentration of the dark--matter subhalos increases the power on
small scales, but only has a minor effect on the structure on the
largest scales. \figurename~\ref{fig:rs} also shows power spectra for
the case where the dark--matter subhalos are modeled as Plummer
spheres rather than Hernquist spheres that similarly follow the
average mass--concentration relation in Via Lactea II. This leads to
very similar power spectra as in the fiducial Hernquist
case. Therefore, the concentration is the primary dark--matter-profile
parameter that matters.

The time since the disruption of the progenitor started---the ``age''
of the stream---is a quantity that is observationally difficult to
establish, as it will generally be different from the age of the stars
in the stream. A reasonable upper limit on the age of the stream is
the age of the youngest stars near the ends of the stream, because no
star formation occurs in low-mass streams. The age of the stream is an
important ingredient in any modeling of perturbations to the stream
structure due to subhalo impacts, as the encounter rate depends
linearly on the age of the stream if the length of the stream is
known. This follows from \equationname~(\ref{eq:nenc}), because
$t_d\,\Delta \Omega^m$ is proportional to the length of the stream,
leaving a single factor of $t_d$ that gives the linear age
dependence. At a fixed rate of impacts, the age of the stream affects
the time that individual perturbations have to grow and decay (due to
internal dispersion and the effect of other impacts).

\figurename~\ref{fig:age} displays what happens to the various
tidal-stream power spectra that we consider when the age of the stream
is changed. We vary the age of the stream by $50\,\%$ up or down and
adjust the stream's velocity-dispersion parameter $\sigma_v$ by the
inverse of this factor, such that the stream remains approximately the
same length.  The solid lines use the rate that follows directly from
\equationname~(\ref{eq:nenc}) and which is thus $50\,\%$ larger or
smaller than the fiducial case; we refer to these as the
``fully-varied realizations''. This produces large changes in the
power spectra. The dashed lines vary the age while keeping the number of
encounters the same (thus nominally falling below or above the CDM
rate); we denote these as the ``constant-rate age realizations''. This
produces power spectra that are less different from the standard
case. To give a sense of the magnitude of the changes, we display
rescaled versions of the fiducial power spectra by factors of $1.5$
and of $0.5$. The fully-varied age realizations lie close to the $1.5$
and $0.5$ lines in the track power spectrum and the cross power
spectrum, except on the largest scales for the older realizations. The
younger stream's track power spectrum also falls below the $0.5$
scaling. The constant-rate realizations typically lie closer to the
fiducial setup than the $0.5/1.5$ scalings.

The fully-varied older stream has a significantly suppressed density
power spectrum. This is because the large rate of impacts severely
batters the stream, making it appear smoother again. We will
investigate this effect further below, where we vary the rate of
impacts at constant age. The fully-varied younger stream has a track
power spectrum that is more reduced from the fiducial case than its
density power spectrum is. This is because a subhalo impact
immediately gives a large change in $\langle \dopar\rangle(\dapar)$
that gets smoothed out at later times. The density perturbation due to
a subhalo impact requires time to grow (\eg, \citealt{Sanders15a}) and
only starts to get smoothed by the stream's dispersion and the effect
of other impacts at much later times. It is clear from comparing
\figurename~\ref{fig:age} to the other figures in this section that
the age of the stream is an important parameter for the tidal stream
power spectra and in particular is more important than the internal
structure of the subhalos (\figurename~\ref{fig:rs}) or their velocity
dispersion (\figurename~\ref{fig:sigmah} below).

\begin{figure*}
\includegraphics[width=\textwidth]{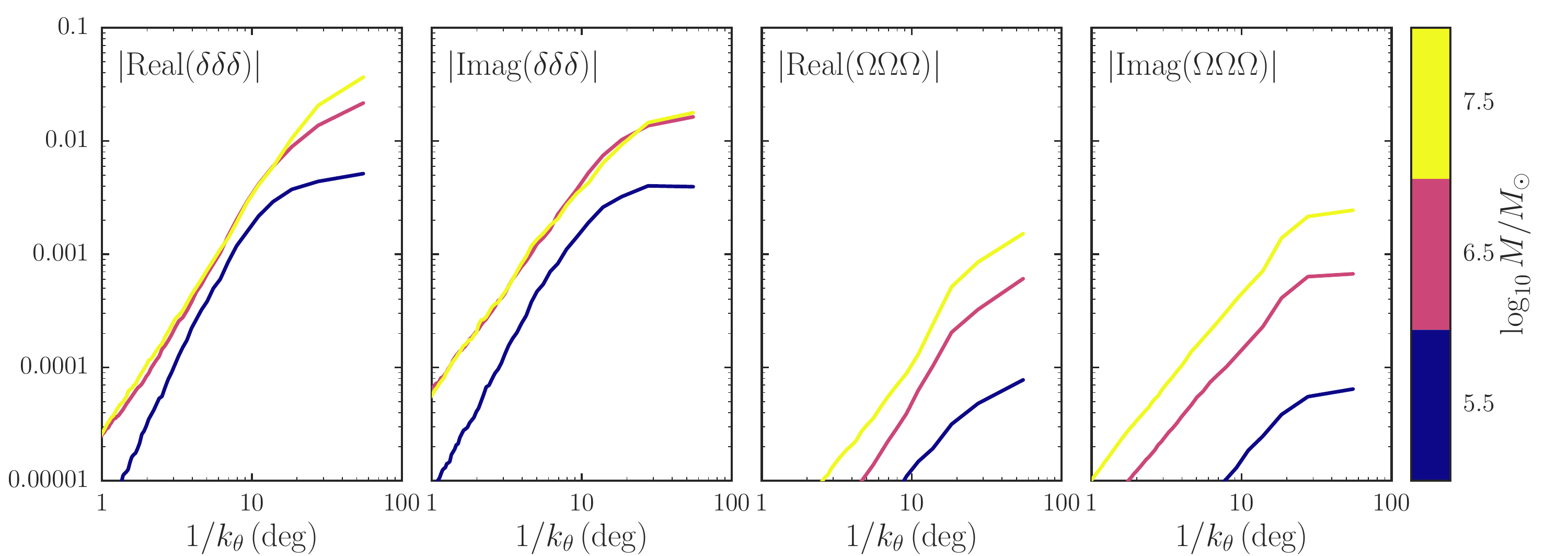}
\caption{Bispectra of the density and mean-track $\langle
  \dopar\rangle(\dapar)$ fluctuations for simulations with a single
  value for the mass (those described in
  \figurename~\ref{fig:bmaxtime}). The first and second panel from the
  left display the magnitude of the real and imaginary parts of the
  bispectrum for the density; the third and fourth panels display the
  same for the mean track. We only show a one-dimensional slice
  through the two-dimensional bispectrum at a scale of
  $\approx11^\circ$. The density has large real and imaginary parts of
  the bispectrum, because of the asymmetry between troughs and peaks
  and the higher impact rate at the ends of the stream,
  respectively. The bispectrum of mean-track fluctuations is smaller,
  but the imaginary part is large because of the asymmetry of the
  parallel-frequency kick (\ie, the fact that they always push stream
  material away from the impact point). The bispectrum is dominated by
  the highest mass impacts on all scales.\label{fig:bispmass}}
\end{figure*}

\begin{figure*}
\includegraphics[width=\textwidth]{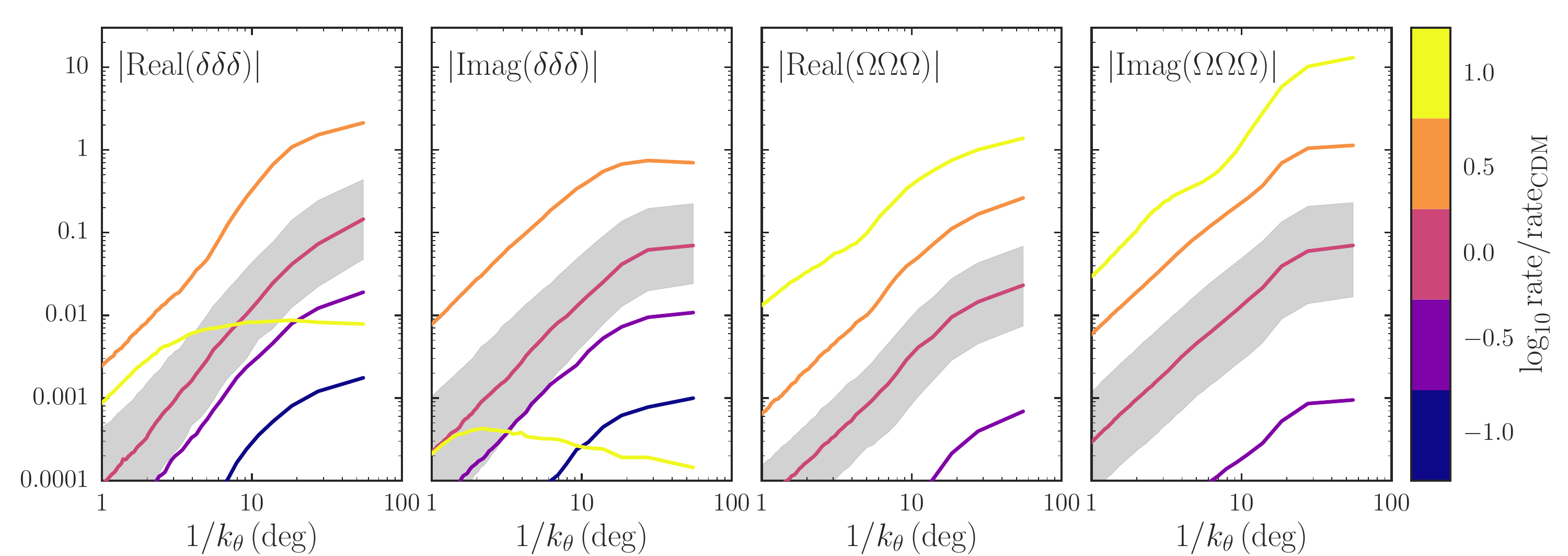}
\caption{Bispectra like in \figurename~\ref{fig:bispmass}, but for
  simulations covering the entire mass range $10^5\msun$ to
  $10^9\msun$. We display the bispectra for the simulations that vary
  the rate of impacts (those in the top row of
  \figurename~\ref{fig:multcutoff}). The gray band again shows the
  interquartile range within which individual simulations lie for the
  fiducial setup. The bispectrum is highly sensitive to the rate of
  impacts, except that for high rates, the density bispectrum is
  suppressed because the stream gets destroyed
  (cf. \figurename~\ref{fig:multcutoff}).\label{fig:bisprate}}
\end{figure*}

The effect of different impact rates is explored in
\figurename~\ref{fig:multcutoff}. The top row multiplies the CDM-like
rate of impacts over the entire $10^5\msun$ to $10^9\msun$ range by
factors of 10, 3, 1/3, and 1/10. The dashed lines display the
fiducial, CDM-like power spectra rescaled by the square root of these
factors. This is the expected scaling if the impacts give rise to
independent fluctuations. The lower-rate density power spectra hew
close to this expectation, except on the smallest and largest
scales. However, the higher-rate density power spectra are
significantly suppressed compared to the fiducial case. The reason for
this behavior is that the large number of impacts essentially destroys
the stream. Because each impact pushes stream material away from the
impact point, the total effect of a large number of impacts is to push
a large amount of the stream all the way to the progenitor and the
rest far away from the progenitor. As we do not track impacts that
happen in the opposite arm of the stream, our modeling does not handle
this case very well, but it is clear that the outcome does not produce
a realistic stream\footnote{The material pushed close to the
  progenitor or to the opposite arm would likely be pushed back again
  by the large number of impacts to the opposite arm that we ignore,
  such that it should remain close to the progenitor in a game of
  cosmic table tennis. The large number of long observed streams
  implies that this behavior likely does not occur in the Milky
  Way.}. Thus, a low density power spectrum due to a \emph{low} number
of dark--matter subhalos would not be confused with a low power
spectrum due to a \emph{high} number of subhalos, because in the
latter case the overall morphology of the stream would be very
different from that in the former case.

While the density power spectrum reacts non-linearly to a higher or
lower rate of subhalo impacts, the track power spectrum changes close
to linear over almost the entire factor of 100 difference in the
impact rate. The one exception is the case where the rate is lowered
by a factor of ten, which is more strongly suppressed than the linear
expectation. Because the rate of impacts in this case is only 6.27,
this is in large part due to the great Poisson variation in the number
of impacts in different realizations. That the case with a ten times
higher rate has an approximately $\sqrt{10}$ times greater power than
the fiducial CDM-like case, even though the stream is essentially
destroyed is a remarkable illustration of the fact that the track
power spectrum is a well-behaved tracer of the rate of impacts. The
density--track cross power spectrum behaves similarly to the density
power spectrum in all cases, although it suffers from less suppression
for the higher rates than the density power spectrum. 

Comparing \figurename~\ref{fig:age} to the top panel of
\figurename~\ref{fig:multcutoff}, we see that varying the age of the
stream is approximately equivalent to changing the overall rate of
impacts in how it affects the various power spectra. Thus, knowing a
stream's age is important when determining the rate of impacts from
the observed power spectrum.

The bottom row of \figurename~\ref{fig:multcutoff} displays what
happens when the subhalo mass spectrum has an exponential cut-off
below a certain mass. As expected from the behavior of different mass
subhalos in \figurename~\ref{fig:bmaxtime}, the cut-off in the mass
spectrum leads to a cut-off in the stream power spectra below the
scale at which the cut-off mass starts dominating the power in the
stream (\eg, $\approx 10^\circ$ for $10^{5.5}\msun$). This happens
similarly in the density and track power spectra and in the cross
power spectrum. For a cut-off at $10^{7.5}\msun$, the power on the
largest scales is already suppressed. Therefore, a drop in the power
in the density or track of a tidal stream below a certain scale is
indicative of a cut-off in the subhalo mass spectrum.

\figurename~\ref{fig:sigmah} explores what happens to the tidal stream
power spectra when we vary the velocity dispersion $\sigma_h$ of the
dark--matter subhalos. While this parameter can be estimated based on
the kinematics of halo stars \citep[\eg,][]{Sirko04a} or by
equilibrium modeling of the dark--matter halo
\citep[\eg,][]{Piffl15a}, the velocity distribution of subhalos is not
necessarily the same as either of those and is therefore somewhat
uncertain \citep[\eg,][]{Diemand04a}. Varying $\sigma_h$ changes the
overall rate of encounters, which is proportional to $\sigma_h$ (see
\equationname~[\ref{eq:nenc}]), as well as the velocity distribution
of fly-bys. Relative fly-by speeds will typically be larger if
$\sigma_h$ is larger, which should lead to smaller
kicks. \figurename~\ref{fig:sigmah} demonstrates that the track power
spectra again behave simply in that the power scales close to
$\sigma_h$ (dashed lines). This indicates that any differences in the
velocity distribution of fly-bys are subdominant compared to the
overall change in the rate. The density power spectra and
density--track cross power spectra display smaller changes, with the
density being largely insensitive to $\sigma_h$ on scales larger than
a few degrees.

Similar kinds of simulations that vary the spectral index of the
subhalo mass spectrum or the subhalo concentration--mass relation
(beyond what was explored in \figurename~\ref{fig:rs}), or that change
the phase or Galactocentric radius at which the stream is observed, or
other variations are straightforward and fast to perform using the
line-of-parallel-angle approach from
\sectionname~\ref{sec:lineofpar}. The exploration in this section
focused on some of the most important variations and we postpone
further exploration to future work.

\section{Tidal stream bispectra}\label{sec:bispec}

So far we have only considered the power spectrum of the density and
track variations induced by dark--matter subhalo impacts. But the
density and track changes induced by individual impacts are manifestly
non-Gaussian and it should be expected that the fluctuations in the
density and track induced by the combination of many impacts might
have non-trivial higher-order moments.

In particular, the density variation induced by a single impact some
time after the impact is a deep, wide trough with narrow ridges whose
height depends on the potential
\citep[\eg,][]{Carlberg12a,Erkal15a}. This profile has a strong
up/down asymmetry missed by the power spectrum. The track variation
$\langle \dopar\rangle(\dapar)$ from a single impact is downward up to
a minimum deviation followed by a sharp upturn through the impact
point to a maximum upward change that declines further from the impact
point (for the leading arm and opposite for the trailing arm;
\citealt{Sanders15a}). This has a strong upstream/downstream asymmetry
that is also missed by the power spectrum. Both of these signals
should show up strongly in the bispectrum, because the bispectrum's
real part is sensitive to up/down asymmetries and the imaginary part
is responsive to left/right asymmetries in one-dimensional signals
\citep{Masuda81a,Elgar87a}. That the oldest, end part of the stream
will have suffered more impacts than the youngest part close to the
progenitor will also cause a strong left/right asymmetry (which may
provide a method for determining whether a stream is leading or
trailing for progenitor-less streams).

\begin{figure}\begin{center}
\includegraphics[width=0.4\textwidth]{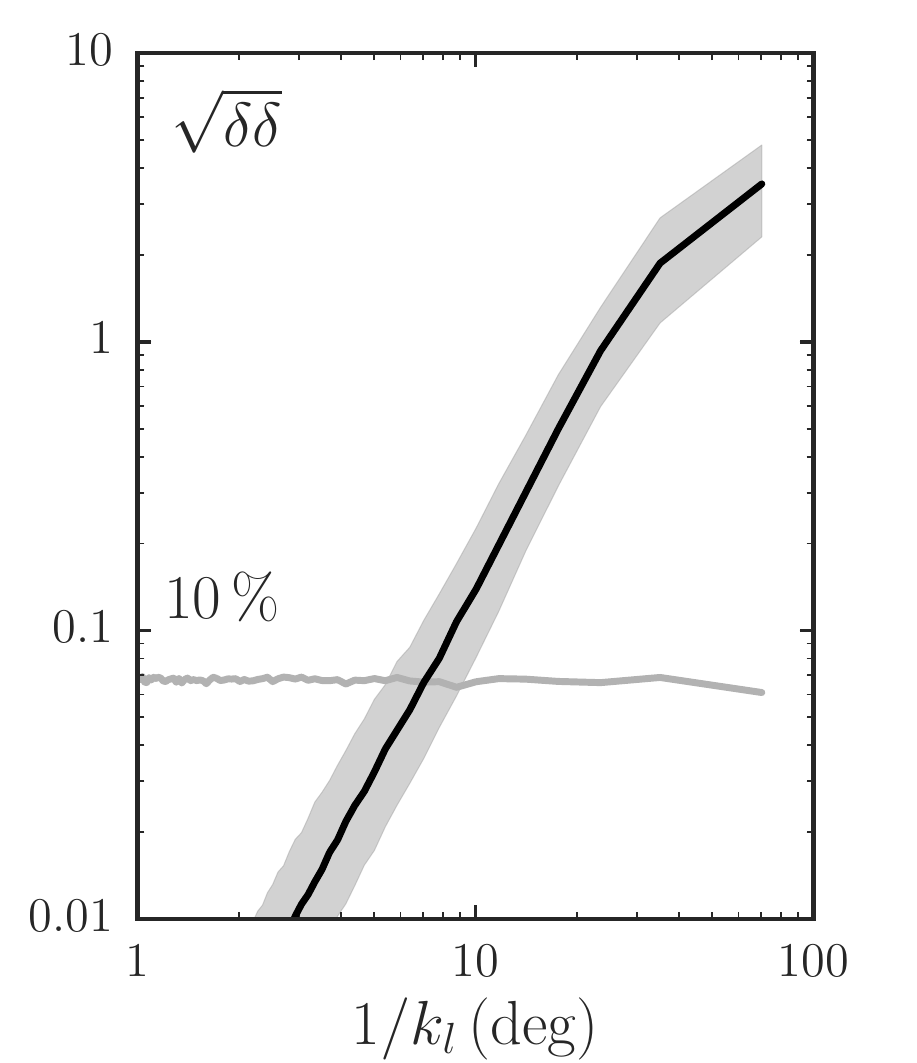}
\end{center}
\caption{Power spectrum for the fiducial simulation setup for the
  GD-1-like stream (cf. \figurename~\ref{fig:rs}), but measuring the
  density in $\Delta l = 0.3^\circ$ bins as a function of Galactic
  longitude rather than $\dapar$. The gray band shows the
  interquartile range within which individual simulations lie. The
  gray horizontal line shows the noise level for $10\,\%$ measurements
  of the density in each bin, a realistic goal for future
  experiments.\label{fig:densObs}}
\end{figure}

\begin{figure}\begin{center}
\includegraphics[width=0.45\textwidth]{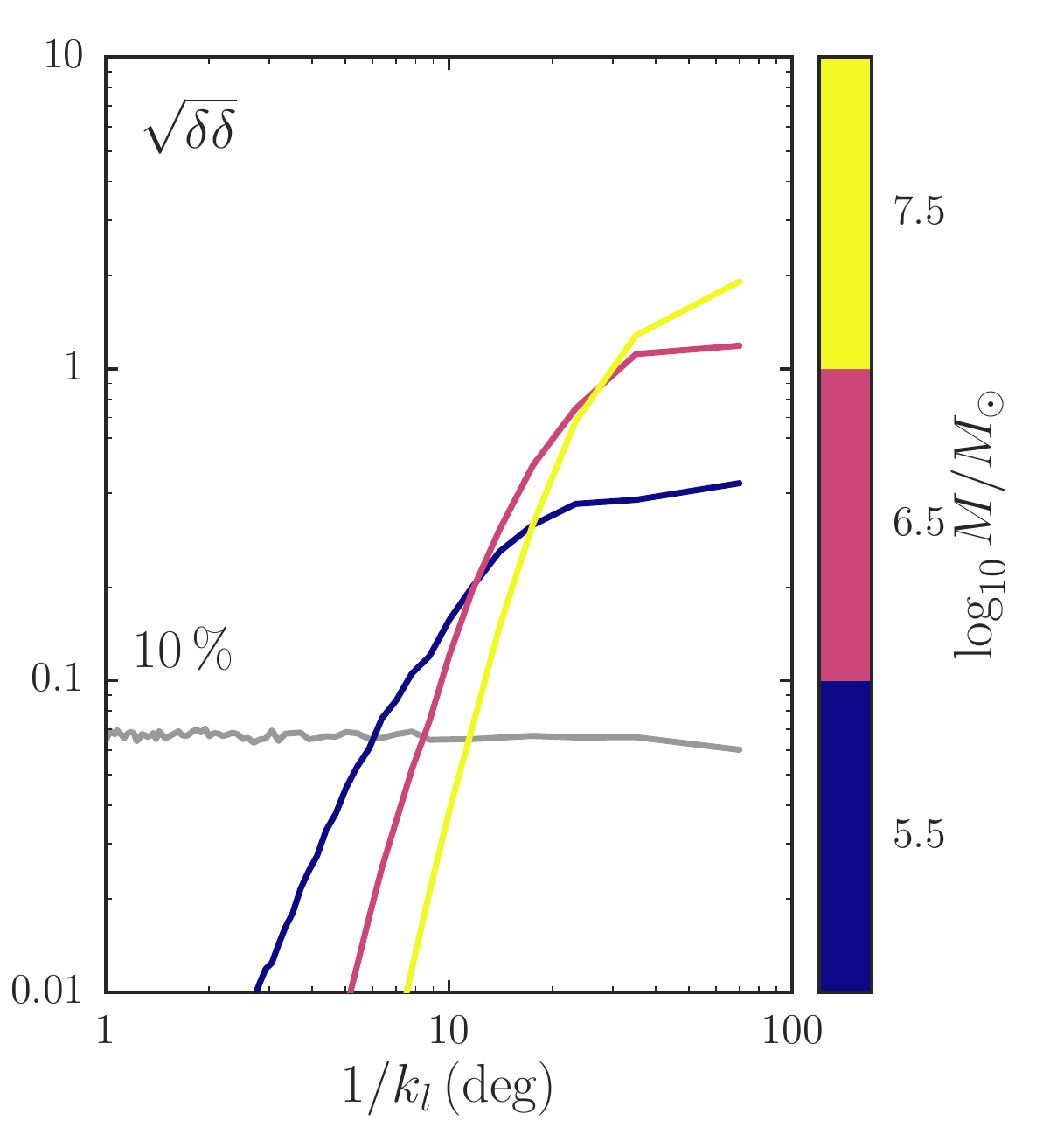}
\end{center}
\caption{Like \figurename~\ref{fig:densObs}, but for the simulations
  with a single value for the mass (those described in
  \figurename~\ref{fig:bmaxtime}). This demonstrates that $10\,\%$
  density measurements can determine the number of $>10^5\msun$
  dark--matter subhalos.\label{fig:densObs_massranges}}
\end{figure}

The bispectrum $B(k_1,k_2)$ of a signal $f(x)$ is defined as
\begin{equation}\label{eq:bispec}
  B(k_1,k_2) = F^*(k_1+k_2)\,F(k_1)\,F(k_2)\,,
\end{equation}
where $F(k)$ is the Fourier transform of $f(x)$ and $F^*(k)$ is
its complex conjugate. We compute this using a Rao-Gabr window
function with size 7 \citep{Rao84a}. The bispectrum is a
two-dimensional function, although it has many symmetries
\citep{Rao84a}. We will only display one-dimensional slices through
the bispectrum at an intermediate scale $k_1$; other slices that are
not shown display similar behavior.

\begin{figure*}
\includegraphics[width=\textwidth]{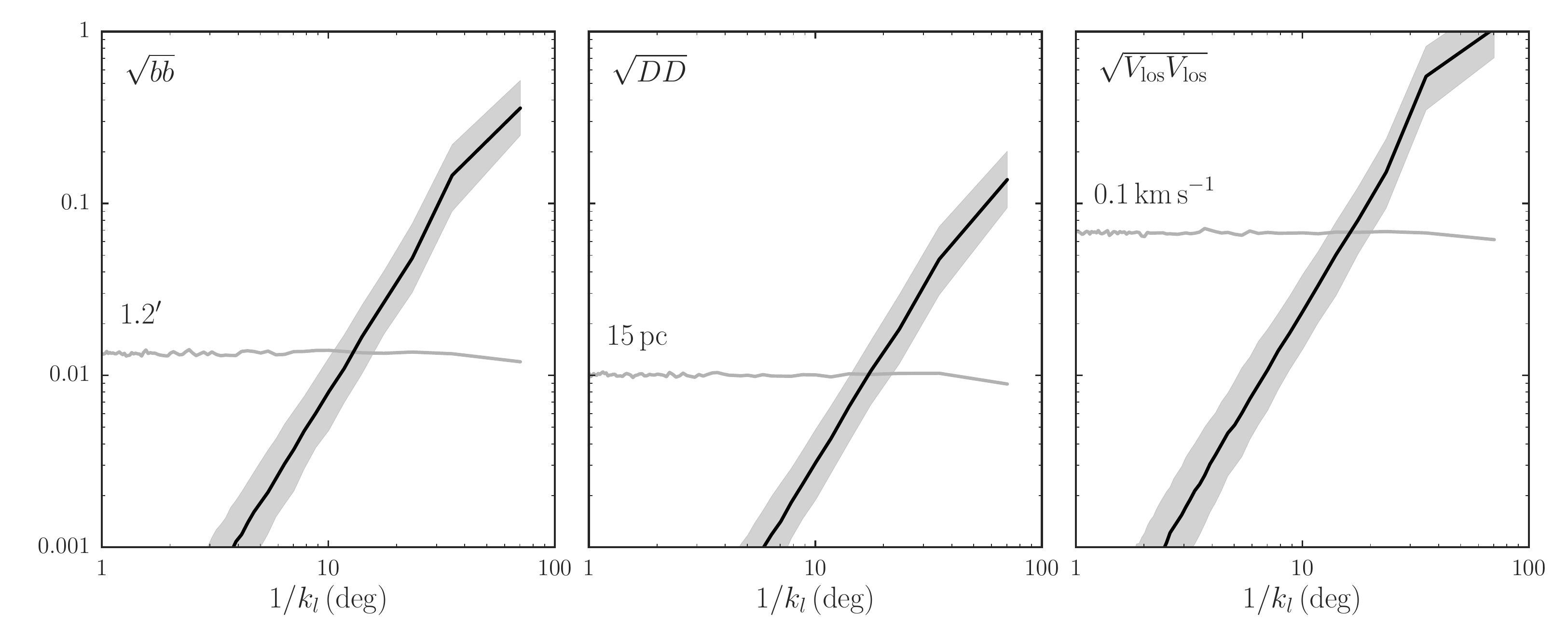}
\caption{Like \figurename~\ref{fig:densObs}, but for fluctuations of
  the mean track in observed Galactic latitude, distance, and
  line-of-sight velocity, as a function of $l$. The horizontal gray
  lines display the noise level for measurements of the mean location
  of the track with the given uncertainties. These uncertainties are
  realistic if $10\,\%$ density measurements are possible. Because all
  track measurements determine the same underlying one-dimensional
  $\langle \dopar\rangle(\dapar)$ fluctuations, different coordinates
  are highly correlated and could be combined to provide
  higher-precision measurements.\label{fig:trackObs}}
\end{figure*}

\begin{figure*}
\includegraphics[width=\textwidth]{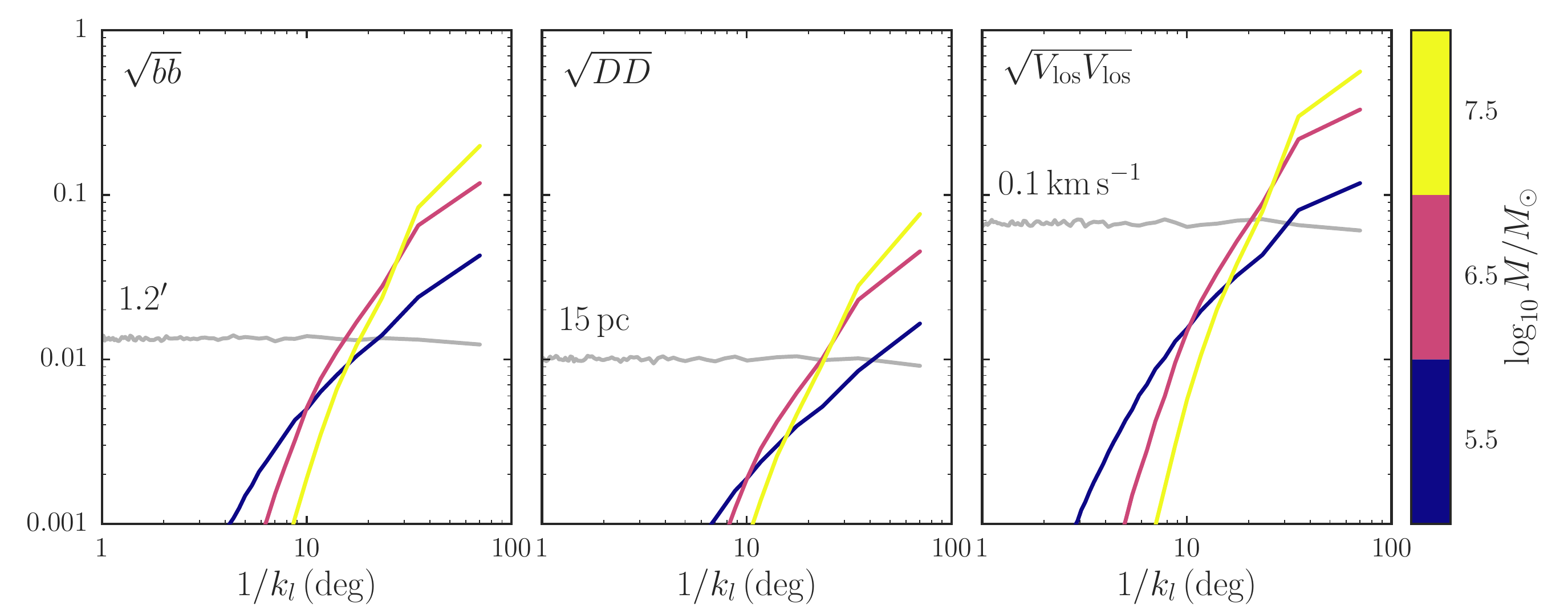}
\caption{Like \figurename~\ref{fig:trackObs}, but for the simulations
  with a single value for the mass (those described in
  \figurename~\ref{fig:bmaxtime}). Because track-location fluctuations
  are smaller than density fluctuations, track fluctuations are
  limited to determining the number of $>10^6\msun$ dark--matter
  subhalos.\label{fig:trackObs_massranges}}
\end{figure*}

The density and mean-track bispectra for the same single-valued-mass
simulations as in \figurename~\ref{fig:bmaxtime} are displayed in
\figurename~\ref{fig:bispmass}. These represent a one-dimensional
slice at $1/k_1 = 11^\circ$, a scale at which different mass ranges
contribute similarly to the power spectrum (see
\figurename~\ref{fig:bmaxtime}). The bispectra act largely as
expected, with a large real part of the density bispectrum because of
the trough/ridge asymmetry and a large imaginary part of the track
bispectrum because of the upstream/downstream asymmetry. However, the
imaginary and real part of the density and track bispectrum,
respectively, are only a factor of a few smaller. For the density, the
$10^6\msun$ to $10^7\msun$ range and the $10^7\msun$ to $10^8\msun$
range give a similar contribution to the bispectrum, while for the
track the $10^7\msun$ to $10^8\msun$ range provides the largest
contribution on all scales. This is also the case for other values of
$k_1$ and we conclude that most of the signal in the bispectrum comes
from the highest-mass impacts.

In \figurename~\ref{fig:bisprate} we show the bispectrum for
simulations of the entire mass range $10^5\msun$ to $10^9\msun$ for
the fiducial CDM-like setup (red line) and for higher and lower impact
rates (same as in the top row of
\figurename~\ref{fig:multcutoff}). The bispectrum acts in a similar
way as the power spectrum when the rate of impacts is
varied. Reductions in the rate for the density and any change for the
mean track produce well-behaved changes in the bispectrum, but upward
changes in the rate suppress the density bispectrum similar to the
power spectrum. This is again caused by the destruction of the stream
(see discussion of the power spectrum above). Overall, the bispectrum
is strongly sensitive to changes in the rate and is therefore an
excellent probe of the higher-mass end of the subhalo mass spectrum.

\begin{figure*}
\includegraphics[width=\textwidth]{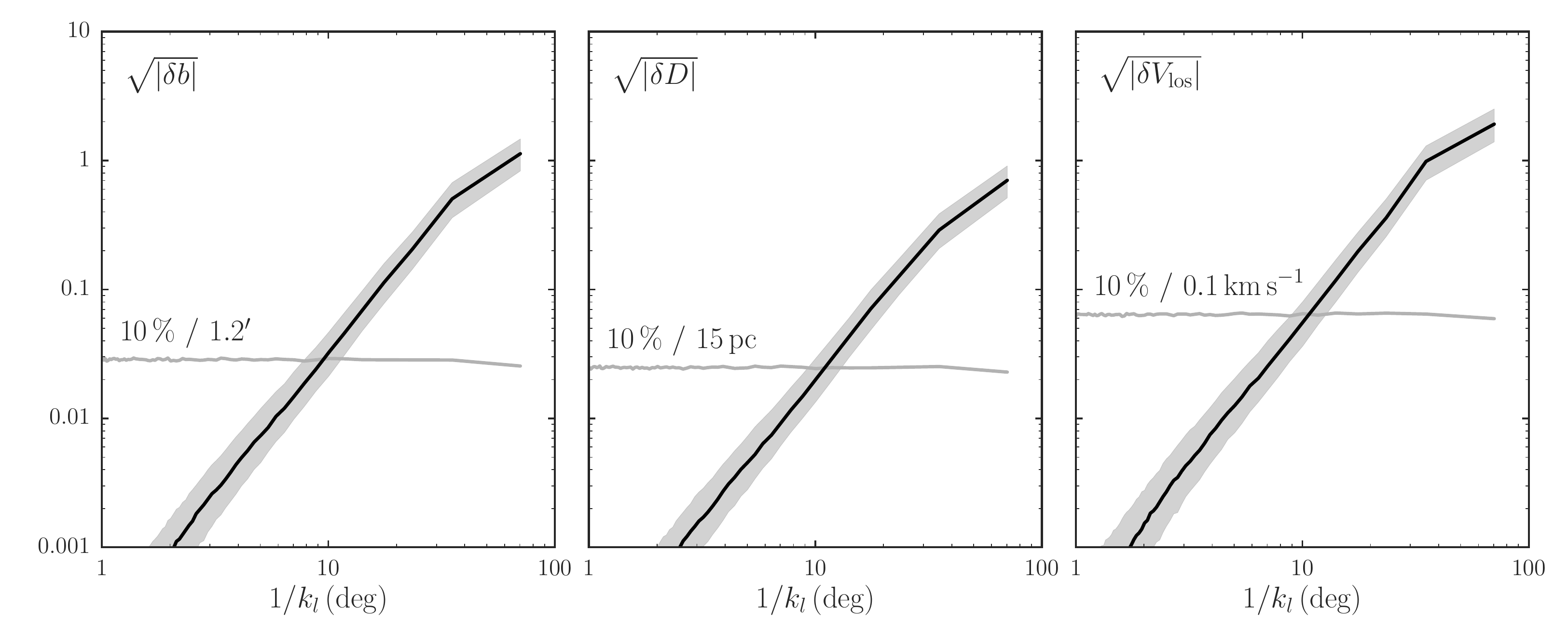}
\caption{Like \figurename~\ref{fig:densObs}, but for the cross power
  spectra of density and track fluctuations. We employ the same noise
  in both as in \figurename s~\ref{fig:densObs} and \ref{fig:trackObs}
  to determine the noise level in the cross correlation. Different
  cross power spectra are again strongly correlated and could be
  combined to form more precise measurements.\label{fig:crossObs}}
\end{figure*}

\begin{figure*}
\includegraphics[width=\textwidth]{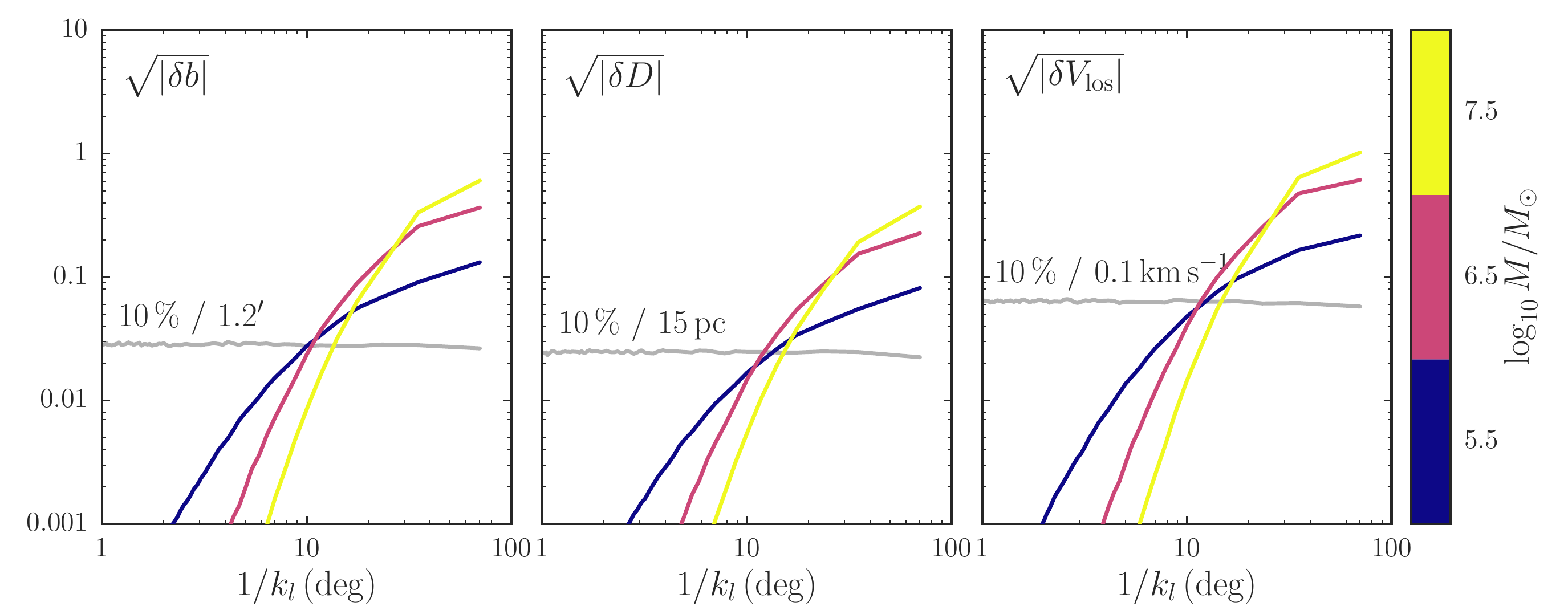}
\caption{Same as \figurename~\ref{fig:densObs_massranges}, but for the
  cross power spectra. At low noise levels, the cross correlation
  between the density and track fluctuations due to $>10^5\msun$
  dark--matter subhalos should be detectable. This would provide an
  important confirmation of measurements based on the density
  fluctuations alone.\label{fig:crossObs_massranges}}
\end{figure*}

\begin{figure*}
\includegraphics[width=\textwidth]{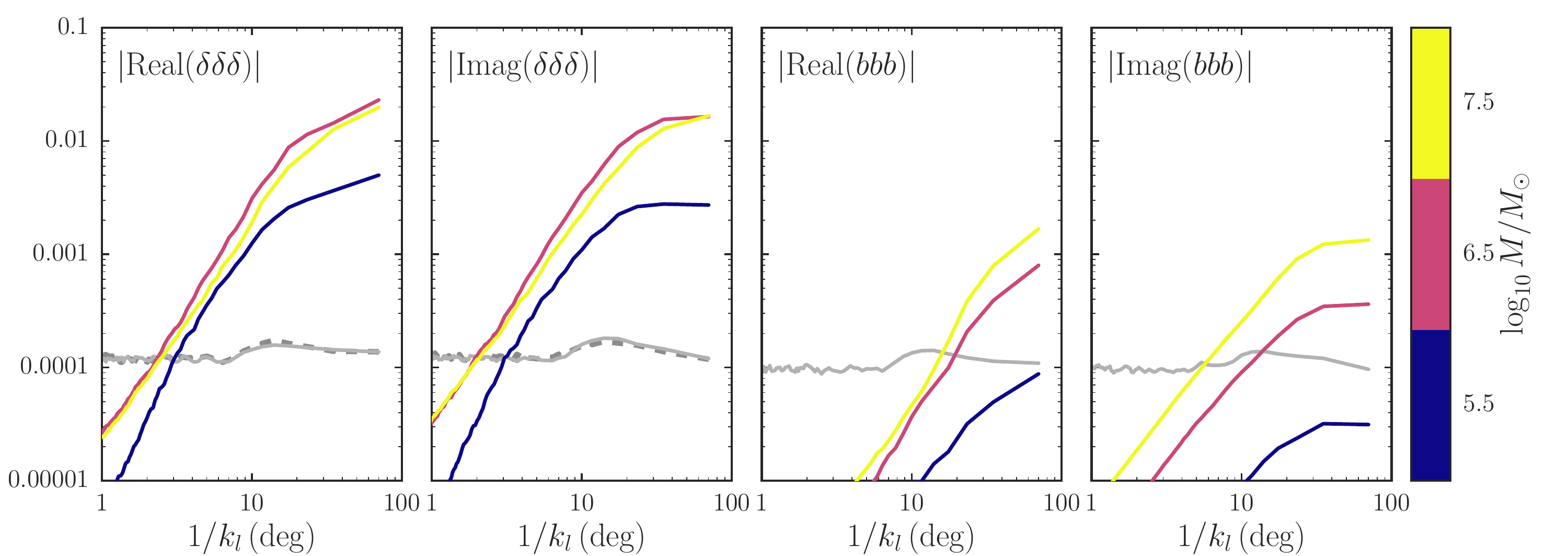}
\caption{Bispectra of the density and mean track location in
  observable coordinates. As in \figurename
  s~\ref{fig:densObs_massranges} and \ref{fig:trackObs_massranges},
  the density and track are computed in $\Delta l = 0.3^\circ$ bins in
  Galactic longitude $l$. The gray horizontal line displays the
  $2\sigma$ (5\,\%) upper noise level for Gaussian uncertainties of
  $10\,\%$ in the density and $1.2'$ in the track location $b$ (the
  median noise is very close to zero for Gaussian uncertainties). The
  darker, dashed gray curve in the left two panels gives the $2\sigma$
  upper noise level when the noise has a Poisson distribution; for
  $10\,\%$ uncertainties this is close to Gaussian. The bispectrum
  should be easily measurable in the near
  future.\label{fig:bispmass_obs}}
\end{figure*}

\begin{figure*}
\includegraphics[width=\textwidth]{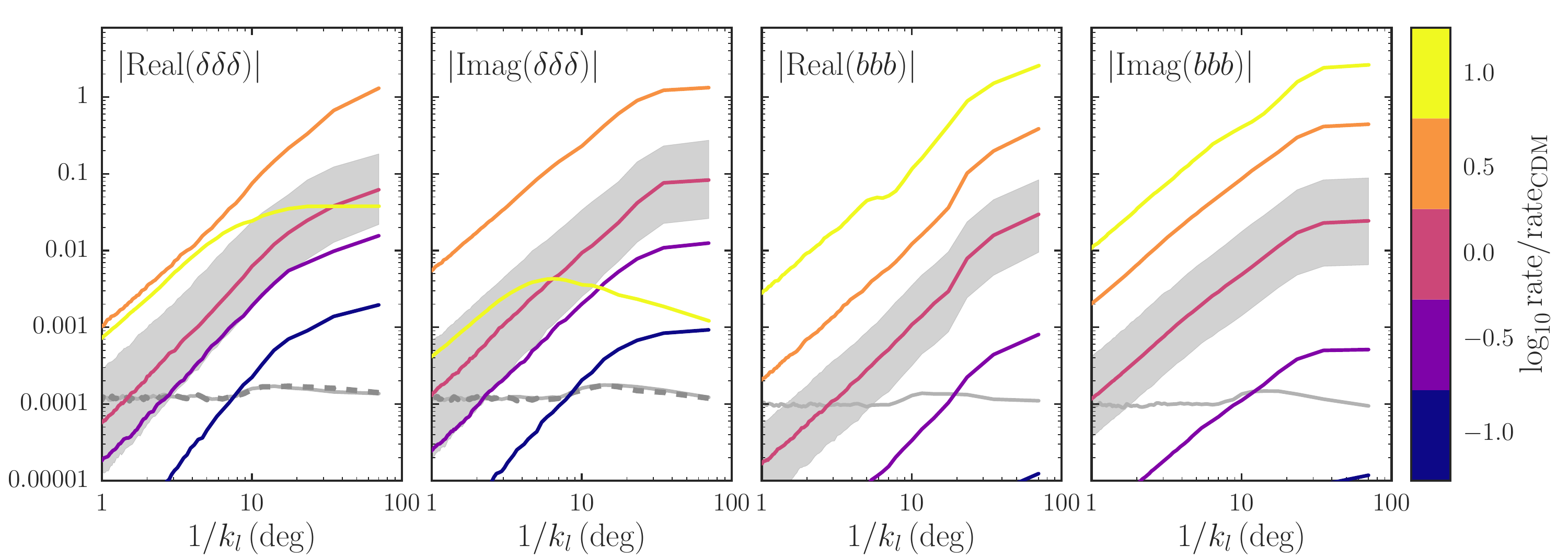}
\caption{Like \figurename~\ref{fig:bispmass_obs}, but for simulations
  where the rate of impacts is varied (those in the top row of
  \figurename~\ref{fig:multcutoff}). The gray band again shows the
  interquartile range within which individual simulations lie for the
  fiducial setup. From this and the previous figure, it is clear that
  the bispectrum is extremely sensitive to the number of
  $\gtrsim10^6\msun$ subhalos.\label{fig:bisprate_obs}}
\end{figure*}

We have also computed the bispectrum for simulations in which the age
of the stream is varied (cf. \figurename~\ref{fig:age}). Similar to
the power sectrum, changes in the bispectrum due to varying the age
are almost equivalent to changes because of rate variations, with
approximately the same relation between age and rate as for the power
spectrum. Thus, the bispectrum cannot be used to break the degeneracy
between age and rate that is inherent in the power spectrum. It is
possible that higher-order moments, such as the trispectrum, or the
phase contain information on the age of the stream, but we do not
investigate this here. The age can also be determined from the
observed length and width of a stream---independent of the
perturbations due to subhalos---through dynamical modeling of the
formation of the stream \citep[\eg,][]{Bovy14a,Erkal16a}. Thus, the
degeneracy between rate and age will not be a major limitation with
future data on streams.

It is straightforward to compute the bispectra of the other
simulations for which we discussed the power spectrum in
\sectionname~\ref{sec:powspec}. We do not show these here, but the
bispectra generally behave similarly as the power spectra, \ie, when
the power spectrum is increased, so is the bispectrum and vice
versa. If the noise in observational determinations of the density and
the mean track is well understood, the bispectrum will be a crucial
ingredient in any analysis of the structure of tidal streams in terms
of subhalo perturbations.

\section{Tidal stream power spectra and bispectra in configuration space}\label{sec:obsspace}

As discussed in \sectionname~\ref{sec:convert_obs}, we can easily
convert the density $p(\dapar)$ and the mean track $\langle
\dopar\rangle(\dapar)$ to configuration space using a linearized
$(\veco,\veca)\rightarrow(\vecx,\vecv)$ transformation. Therefore, we
can determine the tidal stream power spectra and bispectra from the
preceding sections in configuration space where they are more easily
compared to observations. In particular, we convert the density and
track to the Galactic coordinate system
$(l,b,D,V_{\mathrm{los}},\mu_l,\mu_b)$ and use $l$ as the coordinate
along the stream (see \figurename~\ref{fig:gd1props}). As argued in
\sectionname~\ref{sec:convert_obs}, the smoothness of the
$(\veco,\veca)\rightarrow(\vecx,\vecv)$ transformation and the small
size of the track perturbations implies that the density and track
perturbations induced by subhalo impacts are very similar in
frequency--angle and configuration space. The scale dependence of the
power spectra and bispectra in particular closely tracks that of the
power spectra and bispectra in frequency--angle space. Therefore, we
only show and discuss the power spectra and bispectra of some of the
simulations from \sectionname\sectionname~\ref{sec:powspec} and
\ref{sec:bispec}. The main advantage of computing the stream
properties in configuration space is that it is easier to assess the
impact of observational noise and we discuss the contributions from
noise in detail.

\begin{figure*}
\includegraphics[width=0.99\textwidth]{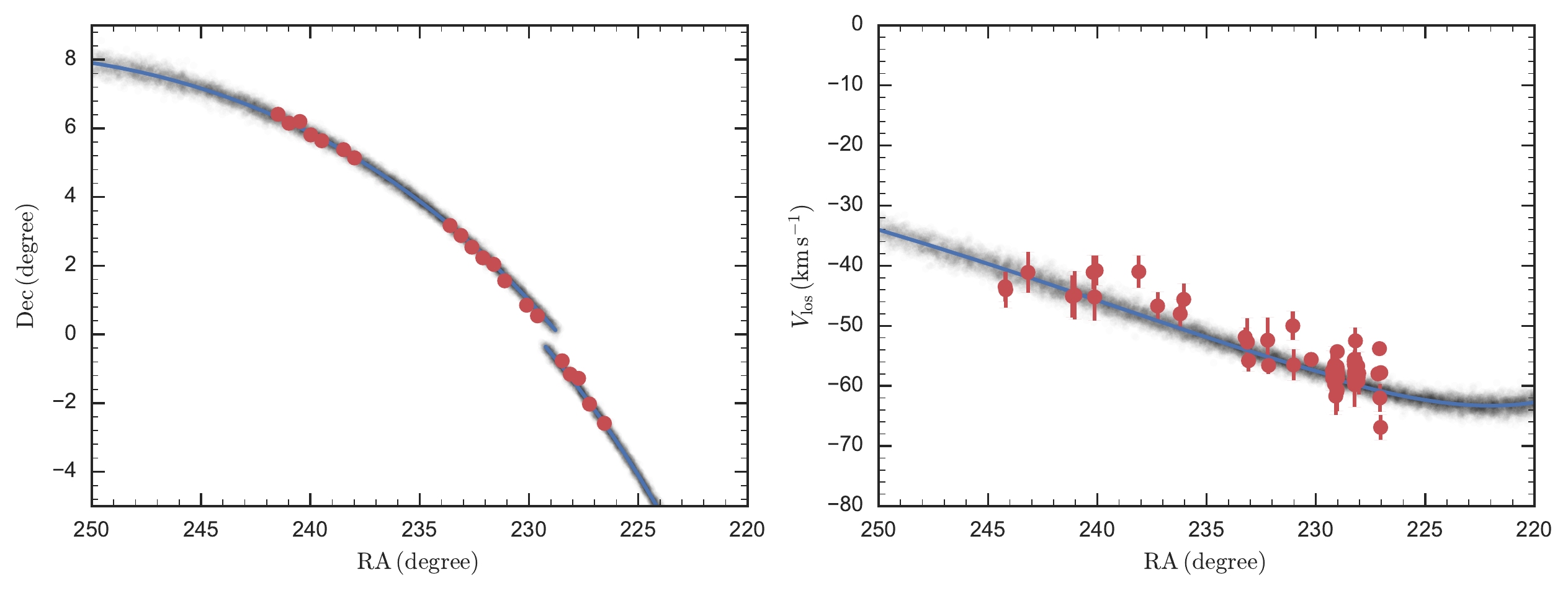}
\caption{Model for the Pal 5 stream generated using the method of
  \citet{Bovy14a} using \texttt{galpy}'s \texttt{MWPotential2014} for
  a stream age of $5\,\Gyr$, a velocity-dispersion parameter $\sigma_v
  = 0.5\kms$, and using the phase-space coordinates of the Pal 5
  globular cluster from \citet{Fritz15a}'s Table 2 (with a distance to
  the cluster of $23.2\kpc$). The blue line displays the mean stream
  track in angular coordinates (left panel) and in line-of-sight
  velocity (right panel), the grayscale shows a sampling of mock
  stream data from the model, and the red points are the stream
  positions from \citet{Fritz15a} on the left and the velocity
  measurements from \citet{Kuzma15a} on the right. The angular width
  of the stream is 15' FWHM, in good agreement with the measurement of
  18' of \citet{Carlberg12b}. The stream model in the
  \texttt{MWPotential2014} potential provides an excellent match to
  the positional and velocity data of the Pal 5
  stream.\label{fig:pal5fit}}
\end{figure*}

\figurename~\ref{fig:densObs} displays the power spectrum of density
fluctuations as a function of Galactic longitude for the fiducial
CDM-like setup for the GD-1-like stream. Thus, this power spectrum is
similar to that in \figurename~\ref{fig:rs}, except that it is
computed from the density fluctuation that is a function of $l$ rather
than $\dapar$. It is clear that the power spectrum is very
similar. That the angular scales are almost the same whether the
density is a function of $\dapar$ or $l$ is due to the fact that the
stream is observed between peri- and apocenter; the length of the
stream in $\dapar$ and configuration space is in general different
depending on which phase of the stream orbit that the stream is
observed at. We have also included the noise contribution when the
density of the stream is measured to $10\,\%$ in $0.3^\circ$ bins over
the entire length of the stream. The noise power spectrum is computed
in the same way as the tidal stream power spectrum (both for the
density and the track below), by generating 1,000 realizations of
Gaussian noise and computing their median. As expected, this noise
power spectrum is approximately flat. This noise level is feasible if
a clean stream map below the main-sequence turn-off can be constructed
with future surveys, \eg, through proper-motion selection.

To determine what dark--matter subhalo masses we are sensitive to, we
also compute the density power spectrum for the single-valued-mass
simulations of \figurename~\ref{fig:bmaxtime}. This is shown in
\figurename~\ref{fig:densObs_massranges}. These power spectra are
again very similar to those in \figurename~\ref{fig:bmaxtime} and we
see that we can still associate angular scales with specific
dark--matter subhalo mass ranges. With $10\,\%$ density measurements
we can easily see the fluctuations induced by subhalo with masses down
to $10^5\msun$ and potentially even below this mass if the trend in
this figure continues. Even with density measurements that are a
factor of a few worse, we should still be able to see impacts down to
$\approx 10^{5.5}\msun$.

Going beyond the density, we further compute the mean location of the
stream in angular position on the sky $\langle b\rangle(l)$, distance
$\langle D \rangle(l)$, and line-of-sight velocity $\langle
V_{\mathrm{los}}\rangle(l)$ (the proper motion perturbations are tiny
and likely unobservable far into the future). The power spectrum of
these three projections of the track for the fiducial setup is shown
in \figurename~\ref{fig:trackObs} (we subtract the unperturbed track
rather than dividing by it for these projections). We also include
optimistic estimates of the noise in the measured track position in
$\Delta l = 0.3^\circ$ bins. If $10\,\%$ density measurements are
possible, then these track measurements are possible as well given the
width of the stream (in $b$ and $V_{\mathrm{los}}$) and photometric
precision (for the distance). Thus, we should be able to measure the
$b$-track power spectrum down to $10^\circ$ for this stream and to
slightly larger scales for $D$ and $V_{\mathrm{los}}$.

The breakdown of these track power spectra in terms of subhalo mass
decades is displayed in
\figurename~\ref{fig:trackObs_massranges}. These demonstrate that we
may be able to measure the contribution to the track power spectra of
subhalo masses down to $10^6\msun$. Because each projection of the
track ultimately derives from the one-dimensional $\langle
\dopar\rangle(\dapar)$, the different projections are fully correlated
and could in principle be combined to produce a higher signal-to-noise
rate measurement of the track power spectrum. This could push the
sensitivity down to $10^{5.5}\msun$ dark--matter subhalos.

Like in frequency--angle space, the density and track fluctuations are
strongly correlated, which we can use as another powerful measure of
the subhalo mass spectrum. The density-track cross power spectra for
the three track projections considered in the previous paragraphs are
displayed in \figurename~\ref{fig:crossObs}. The breakdown into
different subhalo mass decades is shown in
\figurename~\ref{fig:crossObs_massranges}. For the same observational
uncertainties as for the density and track measurements above, the
cross power spectra are observable at scales $\gtrsim 8^\circ$, which
reaches a sensitivity of $\approx10^{5.5}\msun$. The cross power
spectra corresponding to the different track projections are again
strongly correlated and could be combined to produce higher
signal-to-noise-ratio observations. These may increase the sensitivity
down to $\approx10^{5}\msun$. The cross power spectra combine the
sensitivity of the density fluctuations to subhalo impacts with the
observational robustness of track fluctuations. They will play a major
role in confirming and sharpening the subhalos signal detected in the
density fluctuations.

Using the density and mean track in configuration space we can further
compute bispectra of the density and mean-track fluctuations. These
are shown in \figurename s~\ref{fig:bispmass_obs} and
\ref{fig:bisprate_obs} for the same simulations for which we displayed
the bispectra in frequency--angle space in \figurename
s~\ref{fig:bispmass} and \ref{fig:bisprate}. We only show the
bispectra of the mean angular location $\langle b\rangle(l)$; those of
the mean distance and line-of-sight velocity are similar. It is clear
that the bispectra in configuration space are almost the same as those
in frequency--angle space. For Gaussian uncertainties, the median
noise bispectrum is zero, however, because we display the absolute
value, the median noise is simply very small. In \figurename
s~\ref{fig:bispmass_obs} and \ref{fig:bisprate_obs}, we show the noise
as the $5\,\%$ upper limit to the noise level as a gray line for the
same assumed noise level in the density and track as for the observed
power spectra above. That is, only $5\,\%$ of simulations of the noise
are above the gray line. For the density bispectrum, the dashed gray
curve displays the same noise level for noise with a Poisson
distribution, assuming the $10\,\%$ uncertainties in the relative
density come from a Poisson distribution with a mean of $100$
counts. The noise level in the bispectrum is nearly identical for
Poisson or Gaussian noise. The bispectrum extends a few orders of
magnitude above the noise, which is scale independent. The bispectrum
is observable, even if the rate is ten times less than the CDM rate,
in which case the power spectrum becomes more difficult to observe for
our assumed uncertainties.

\section{Density power spectrum of Pal 5}\label{sec:pal5}

Some measurements of the density of tidal streams already exist
\citep[\eg,][]{Odenkirchen03a,Carlberg12b,Carlberg13b}. In this
Section, we illustrate the formalism of this paper by performing a
measurement of the density power spectrum of the Pal 5 stream and the
first rigorous constraint on the number of dark--matter subhalos with
masses between $10^{6.5}\msun$ and $10^9\msun$ using the density data
from \citet{Ibata16a}.

We build a model for the Pal 5 stream in a smooth potential using the
formalism of \citet{Bovy14a}. As discussed in
\sectionname~\ref{sec:smooth}, such a model is specified by 8 free
parameters in a given gravitational potential: the current
phase--space position of the progenitor, the velocity-dispersion
parameter $\sigma_v$, and the time $t_d$ since disruption
started. Because the progenitor of the Pal 5 stream is the Pal 5
globular cluster, we can use the measured phase--space position of
this cluster as the current phase--space position. We employ the
position and velocity from \citet{Fritz15a}, who measured the proper
motion in addition to existing measurements of the celestial position,
distance, and line-of-sight velocity of Pal 5. The distance to Pal 5
has some uncertainty and could plausibly lie between $19$ and
$24\kpc$; we use a distance of $23.2\kpc$ as this gives a good match
to the stream location. As our model for the Milky Way's smooth
gravitational potential we use \texttt{MWPotential2014} from
\citet{BovyGalpy}. This potential has been fit to a variety of
kinematic data on the bulge, disk, and halo of the Milky Way, as
discussed by \citet{BovyGalpy}. The parameters $\sigma_v$ and $t_d$
need to be determined from the width and length of the Pal 5
stream. We do not perform a rigorous fit, but simply try a few
common-sense values. We find that $\sigma_v = 0.5\kms$ gives a stream
width of $15'$ (FWHM), in good agreement with the measurement of
\citet{Carlberg12b}, who find $18'$. We use an age of $t_d = 5\Gyr$ of
the stream, which gives a decent match to the data below, although it
is not well constrained as the length of the Pal 5 stream has not been
measured because the stream hits the edge of the SDSS survey.

We compare our model to the measurements of the location of the Pal 5
stream from \citet{Fritz15a} and to line-of-sight velocities for
stream members from \citet{Kuzma15a} in
\figurename~\ref{fig:pal5fit}. The model for the Pal 5 stream using
the observed phase--space position of the Pal 5 globular cluster in
the \texttt{MWPotential2014} potential provides an excellent match to
the data. This gives additional credence to the
\texttt{MWPotential2014} from \citet{BovyGalpy}, because many
otherwise reasonable models for the Milky Way's potential fail to give
a good match to both the celestial position of the stream and its
line-of-sight velocity (see \citealt{Fritz15a}). We employ this Pal 5
model here as a model for the unperturbed stream and apply the
formalism from this paper to predict the stream density in the
presence of subhalo encounters using the power spectrum.

\begin{figure}
\includegraphics[width=0.49\textwidth]{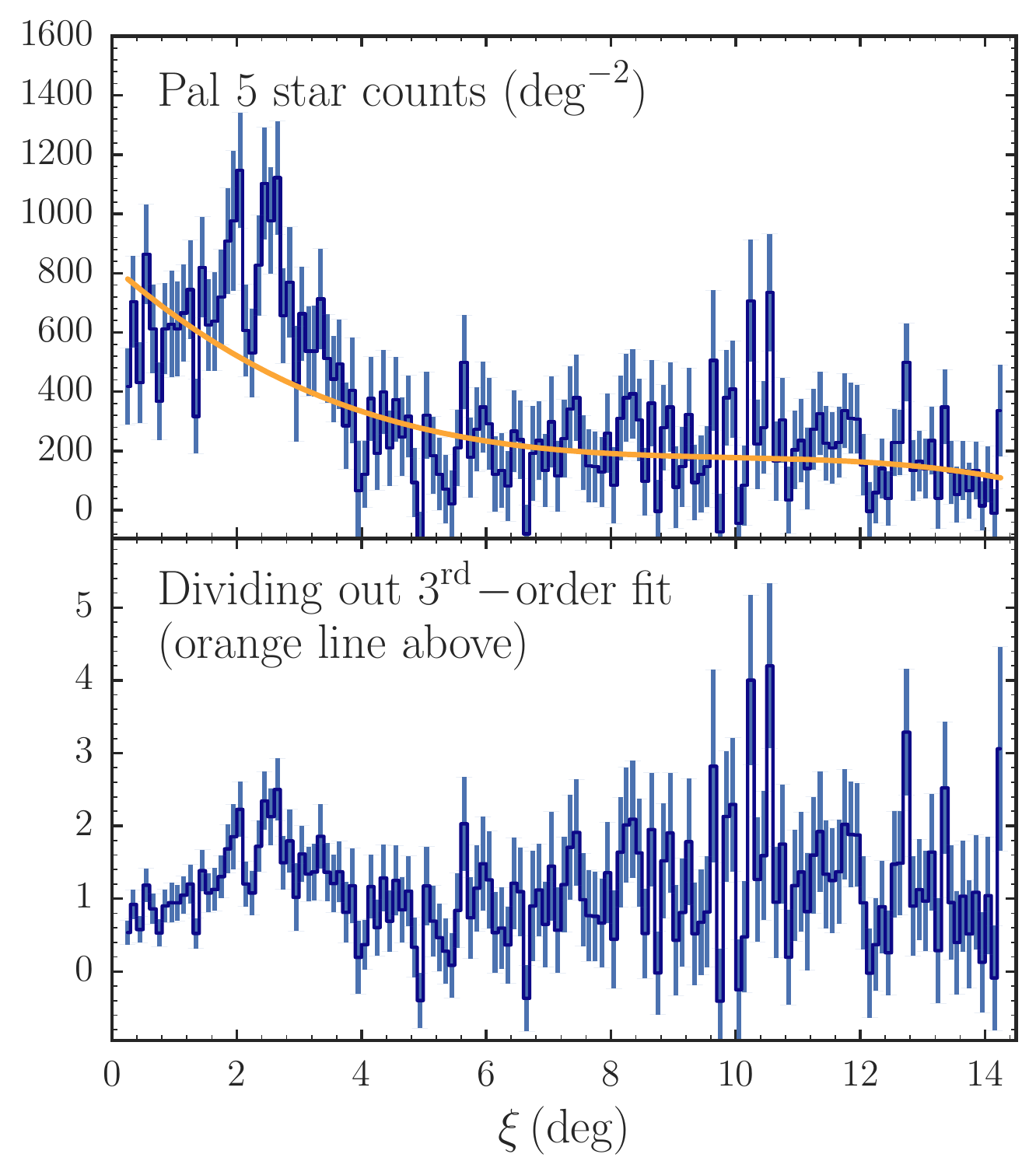}
\caption{Star counts for the trailing arm of the Pal 5 stream from
  \citet{Ibata16a}. The top panel displays the background-subtracted
  counts of stars with $20 < g_0 < 23$ and $0.2^\circ < \xi <
  14.3^\circ$, along with the polynomial fit that we employ to
  normalize the star counts in the bottom panel. The bottom panel
  shows the relative density variations along the stream after
  dividing out the third-order polynomial fit (orange curve in the top
  panel).\label{fig:pal5starcounts}}
\end{figure}

We use the data from \citet{Ibata16a} to determine the density power
spectrum along the trailing arm of Pal 5. \citet{Ibata16a} use deep
CFHT $g$ and $r$-band data to determine star counts in the vicinity of
the Pal 5 stream down to $g_0 = 24$, the deepest map of any known
stream. In particular, we use the star counts for the Pal 5 stream in
the range $20 < g_0 < 23$ from their Figure 7 using their $(\xi,\eta)$
coordinate system. These data are obtained with a simple
color--magnitude filter centered on the main-sequence of the Pal 5
globular cluster and it has not been background-subtracted. We use the
CFHT photometric data to estimate a constant background level of
$\approx400$ stars deg$^{-2}$ and subtract this from the density
data. This determination of the Pal 5 density along the trailing arm
is shown in the top panel of \figurename~\ref{fig:pal5starcounts}. The
background may not be entirely uniform and current data lack a wide
enough area away from the stream to determine the importance of the
background. Because background variations should be largely
uncorrelated with the stream star counts (except, \eg, if variable
extinction is important), background variations will simply add power
to the power spectrum and lead us to overestimate the number of
subhalos from the current data.

The density data in \figurename~\ref{fig:pal5starcounts} have a clear
large-scale trend, with a peak at $\xi < 4^\circ$. Such a peak is
absent in our model, which has a constant stripping rate over time. It
is likely that this peak at least partially occurs because of an
increase in the stripping rate over the last few orbits, because Pal 5
is close to being fully disrupted \citep{Dehnen04a}. To account for
this, we fit a third-order polynomial to the density data and divide
this out. The thus normalized density is displayed in the bottom panel
of \figurename~\ref{fig:pal5starcounts}. It is this density that we
compare to simulations of the effect of subhalos for different rates
of impacts and different mass ranges of subhalos. The median
uncertainty on the normalized density is $\approx60\,\%$. The
normalization has a big effect on the power on the largest scale,
reducing it to $\approx0.6$ from $\approx2$. For this reason, we
cannot fully use the power on the largest scale to constrain the
impact rate below. Power on smaller scales is less affected by this
procedure.

\begin{figure}
\includegraphics[width=0.49\textwidth]{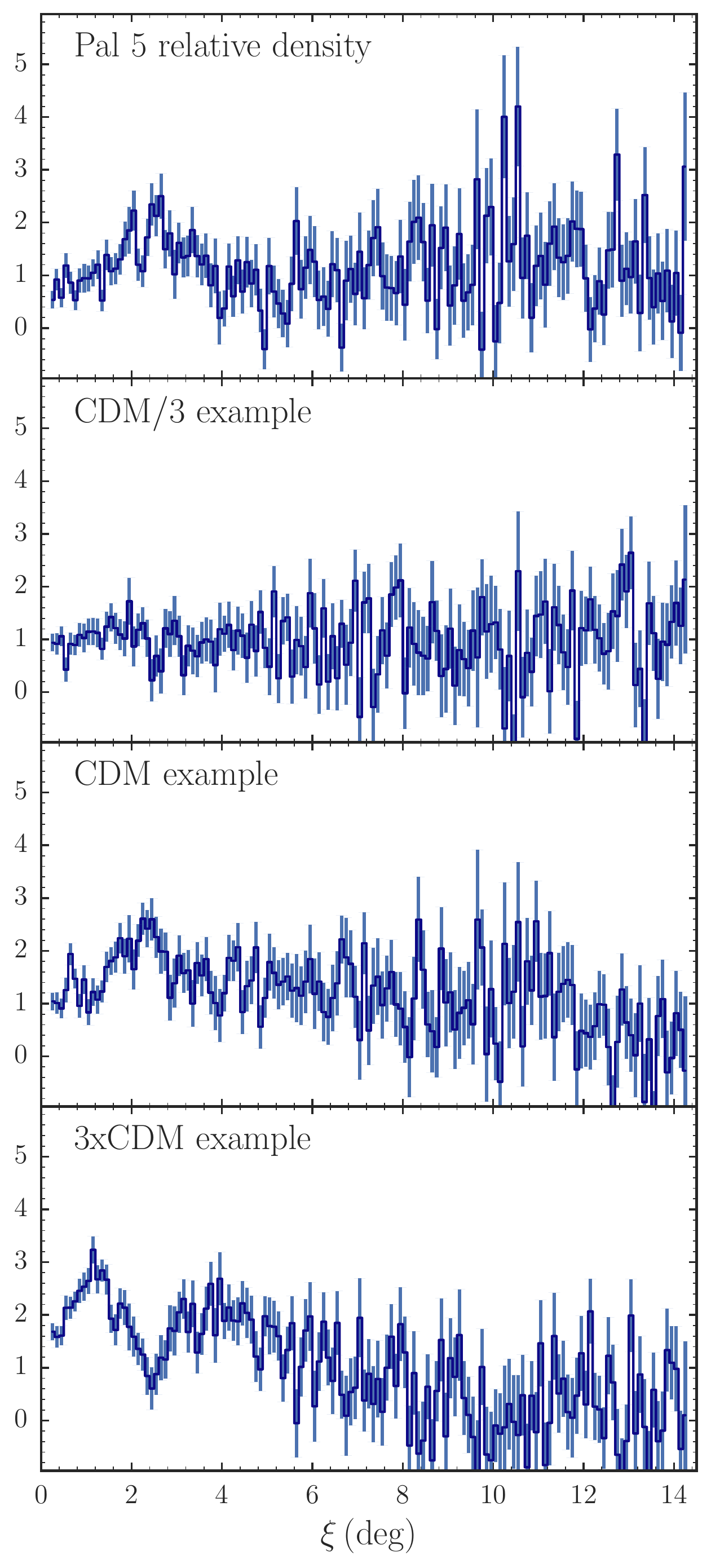}
\caption{Example simulations of the relative density along the
  trailing arm of the Pal 5 stream. The top panel displays the
  observed relative density (bottom panel of
  \figurename~\ref{fig:pal5starcounts}). The three lower panels show
  example simulations where the model Pal 5 stream from
  \figurename~\ref{fig:pal5fit} has been perturbed by different rates
  of impacts between $10^6\msun$ and $10^9\msun$. Rates much smaller
  than the CDM rate do not typically cause density fluctuations above
  the noise level, while rates much larger than the CDM rate produce
  obvious large-scale fluctuations. The CDM example was hand-picked
  out of a set of $\approx200$ simulations to be similar to the
  observed relative density, but it is not atypical; it provides a
  remarkable match to the observed
  density.\label{fig:pal5starcountexamples}}
\end{figure}

We explore the expected density structure of the Pal 5 stream using
simulations of the density using the formalism from this paper in
\figurename~\ref{fig:pal5starcountexamples}. The top panel repeats the
normalized observed density and the remaining panels display example
simulations of impacts with masses between $10^6\msun$ and $10^9\msun$
for different impact rates, with the fiducial CDM rate of 42.9 impacts
corresponding to 42.54 subhalos in this mass range within $25\kpc$
from the center (see \sectionname~\ref{sec:sample}). Because we can
currently only measure the largest-scale density fluctuations, we
include impacts up to $X=10$ to include the effect of distant
fly-bys. Gaussian noise has been added according to the observational
uncertainties. It is clear that the density data, even with the
current large uncertainties, holds information on the number of
encounters with dark--matter subhalos that the Pal 5 stream has
experienced over its lifetime. If the rate is three times less than
the fiducial CDM rate---as is predicted from simulations that model
the disruption of subhalos by the Milky Way's disk
\citep{DOnghia10a}---essentially no intrinsic density perturbations
are induced above the noise. If the rate is three times higher than
the fiducial CDM rate, large-scale density features become
apparent. Simulations for the CDM rate often give a good match to the
observed data, as exemplified by the example shown.

\begin{figure}
\includegraphics[width=0.49\textwidth]{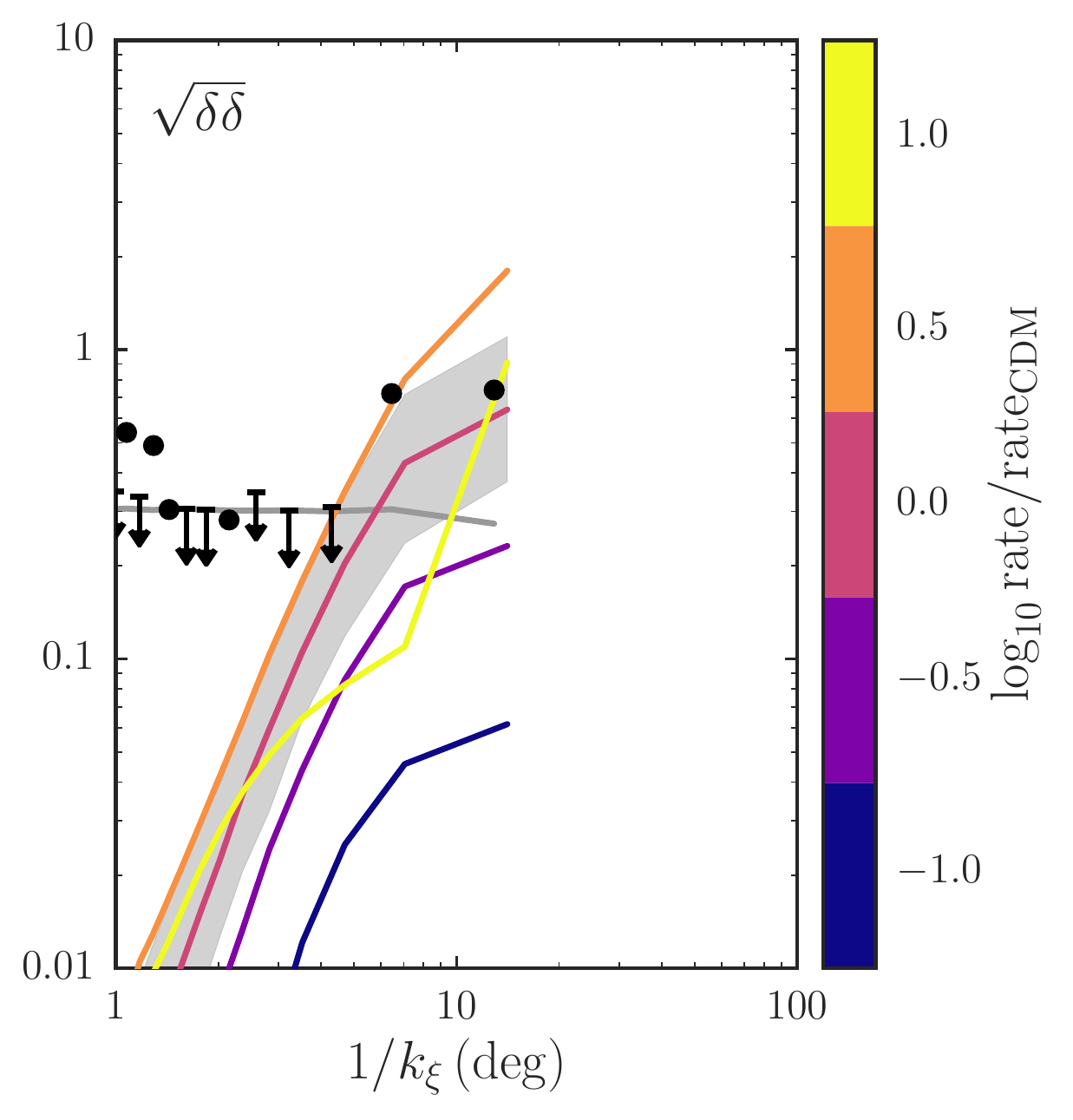}
\caption{Power spectrum of Pal 5. The black points display the power
  spectrum of the relative density along the trailing arm of Pal 5
  (bottom panel of \figurename~\ref{fig:pal5starcounts}) after
  subtracting the approximate noise level ($\approx0.3$; points below
  or just above this level are indicated as upper limits); the gray
  line shows the noise power spectrum based on the median
  uncertainty. The colored curves display the median density power
  spectra of simulations with different subhalo-encounter rates. The
  gray band shows the interquartile range for simulations with the
  fiducial CDM rate. Pal 5's observed power on the largest scales is
  consistent with the power induced by a CDM-like population of
  dark--matter subhalos; it is inconsistent with much higher
  rates.\label{fig:pal5denspowspec}}
\end{figure}

The density power spectrum of the Pal 5 stream is displayed in
\figurename~\ref{fig:pal5denspowspec}. This figure shows an estimate
of the intrinsic power spectrum, subtracting off the median power
spectrum from simulations of the noise. On scales $\lesssim5^\circ$,
the power spectrum is consistent with the noise, but the Pal 5 data
display excess power on scales larger than five degrees. This figure
also displays median density power spectra for simulations with
different rates of impacts. As expected from
\figurename~\ref{fig:pal5starcountexamples}, rates below the CDM rate
only produce power below the noise level, while rates a few times
larger than the CDM rate produce more power than is observed. Similar
to the GD-1-like example in \sectionname~\ref{sec:powspec} above, very
high impact rates again lead to a smoother stream, because most of the
stream becomes very low in surface brightness. As discussed above, the
power on the largest scales is significantly affected by the
density-normalization procedure and should not be trusted too much.

To determine a rigorous constraint on the number of subhalos near Pal
5's orbit, we use Approximate Bayesian Computation (ABC) to construct
an approximation to the posterior probability distribution (PDF) of
the rate of impacts \citep{Marin12a}. ABC approximates the PDF of the
rate based on the Pal 5 density data without evaluating a likelihood
function, but rather using simulations of the density, which we can
easily produce using our formalism. ABC works by drawing a rate from
the prior distribution of the rate, which we choose to be uniform in
$\log_{10}$ of the rate between CDM/10 and 10xCDM, and simulating the
observed relative density for this rate. ABC then constructs the PDF
by keeping only those simulations that are within a certain tolerance
of the real data or a set of summaries of the real data. Ideally, data
summaries are sufficient statistics for the inference in question;
summary statistics that are not sufficient will create wider PDFs,
because they do not make full use of the data. As the data summaries,
we use the power $\sqrt{\delta\delta}$ on the three largest observed
scales and the bispectrum on the second largest scale, because these
are the only scales on which the power and the bispectrum can be
measured from the current data (\eg,
\figurename~\ref{fig:pal5denspowspec}; the power on the third largest
scale is $\approx0.05$, but shown as an upper limit at the noise level
in this figure). We keep those simulations that (a) match the power on
the largest scale to within 1.5 (loose, to account for the effect of
the density normalization, see below), (b) the power on the
second-largest scale to within 0.15, (c) the power on the
third-largest scale to within 0.2, and (d) match the bispectrum at
$6.5^\circ$ (specifically, $B(1/6.5^\circ,1/6.5^\circ)$) to within
0.03 (for both the real and imaginary part). The tolerances on the
power on the second- and third-largest scales and on the bispectrum
are approximately as small as makes sense given the noise in the data
and the observed power; the power spectrum and the bispectrum on
smaller scales are too noisy to lead to a useful constraint. The
resulting PDFs do not depend on the exact values of the
tolerances. Because the most time-consuming part of the simulations is
computing the relative densities, we perform 100 simulations of the
noise for each rate (a sort of Rao-Blackwellization). That is, we
produce 100 simulations for each rate by adding 100 realizations of
the noise to each relative-density realization.

\begin{figure}
\includegraphics[width=0.47\textwidth]{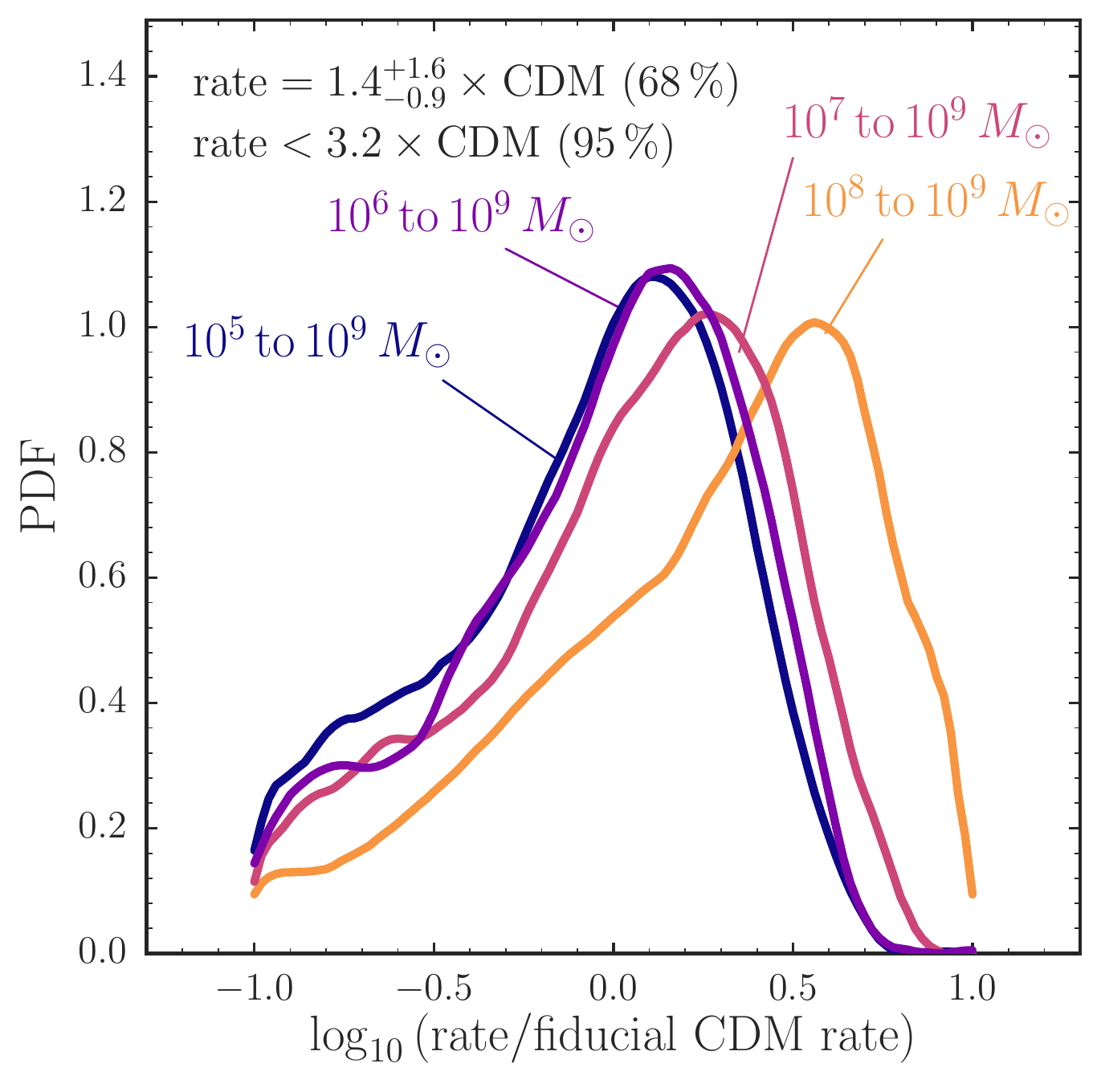}
\caption{Posterior probability distribution functions for the rate of
  dark--matter subhalo encounters based on Pal 5 data. The PDFs are
  obtained by using simulations of Pal 5 and ABC to match the power on
  the three largest scales in \figurename~\ref{fig:pal5denspowspec}
  and to match the bispectrum at $6.5^\circ$. Different curves include
  impacts in different mass ranges; they converge below
  $10^7\msun$. The current Pal 5 data prefer a rate of impacts equal
  to or smaller than that predicted by CDM. The data are inconsistent
  with high impact rates. The data are sensitive to subhalos with
  masses down to $\approx10^{6.5}\msun$.\label{fig:pal5ratepdf}}
\end{figure}

The PDF produced by this procedure including impacts in different mass
ranges is displayed in \figurename~\ref{fig:pal5ratepdf}. Simulations
of impacts of a single mass with the CDM rate expected for the
surrounding mass decade (\eg, that of $10^6\msun$ to $10^7\msun$ for
$M = 10^{6.5}$; cf. \figurename~\ref{fig:bmaxtime}) indicate that we
should be sensitive down to $\approx10^{6.5}\msun$. This is borne out
by the PDFs for simulations with masses between $10^x\msun$ and
$10^9\msun$, which converge for $x$ between 6 and 7. The PDF peaks at
a rate that is $1.4^{+1.6}_{-0.9}\times$CDM or that has
$10^{+11}_{-6}$ dark--matter subhalos with masses between
$10^{6.5}\msun$ and $10^9\msun$ within $20\kpc$ from the Galactic
center (Pal 5's approximate apocenter). The $95\,\%$ upper limit is
$3.2\times$CDM or 23 subhalos within the central $20\kpc$; the
$99\,\%$ limit is $4.2\times$CDM. This conversion between rate
relative to CDM and number of subhalos uses the expected CDM rate from
the Via Lactea II simulation \citep{Diemand08a}, but the number is
what we directly measure. The CDM rate computed from the Aquarius
simulations \citep{Springel08a} for a $M = 10^{12}\msun$ Milky Way
halo is about twice that from Via Lactea II (see \citealt{Erkal16a});
interpreted using the Aquarius simulations, our number measurement
would correspond to $0.7^{+0.8}_{-0.5}\times$CDM. Both rates are
clearly consistent with CDM. Expressed as a mass fraction
$f_{\mathrm{sub}}$ in subhalos for a total halo mass within 20 kpc of
$\approx10^{11}\msun$ \citep{BovyRix}, our result is
$f_{\mathrm{sub}}(r<20\kpc)\approx0.2\,\%$ for $M = 10^{6.5}\msun$ to
$10^9\msun$.

The PDF based on the bispectrum alone is almost the same as that shown
in \figurename~\ref{fig:pal5ratepdf} and the bispectrum is thus fully
consistent with the power spectrum. This demonstrates the power of the
bispectrum. While the data are most compatible with a rate around the
fiducial CDM rate, we cannot rule out a much lower rate, such as that
expected from the depletion of dark--matter subhalos by the massive
Milky Way disk. We have repeated the analysis for impacts with masses
between $10^6\msun$ and $10^9\msun$ using different order polynomials
to normalize Pal 5's density. For linear or quadratic normalizations,
the PDF shifts by $\approx0.5\sigma$ and is essentially the same as
that for masses between $10^7\msun$ and $10^9\msun$ in
\figurename~\ref{fig:pal5ratepdf}. Therefore, the normalization has
only a marginal effect for the current data. We have repeated these
tests for the mock data analyzed in Appendix~\ref{sec:mockpal5}. The
mock data display similar behavior to the real data, lending further
support to our inferential procedure.

As discussed in \sectionname\sectionname~\ref{sec:powspec} and
\ref{sec:bispec}, the rate of impacts inferred from observations of
the power spectrum and bispectrum is likely to be degenerate with the
assumed age of the stream. To investigate how this affects the
analysis of the Pal 5 data, we have repeated the inference in this
section for Pal-5 stream ages of $3\Gyr$ and $10\Gyr$. For these ages
the peak of the rate PDF shifts to $\approx4\times$CDM and
$\approx0.25\times$CDM, respectively. As expected, there is no
preference for any of the ages. We do not consider $3\Gyr$ to be a
plausible age for the Pal 5 stream, because \citet{Kuepper15a} found
an age of $3.4\Gyr$ for the segment of the stream that they extracted
from the SDSS footprint and the data from \citet{Ibata16a} trace the
stream a few degrees further, requiring an age of at least $4\Gyr$. We
thus consider the measurement of the rate quoted above to be a good
upper limit. Nevertheless, marginalizing over all three considered
ages by assuming (for simplicity) that any of these ages are a priori
equally probable produces a rate PDF that is approximately flat up to
$\approx3\times$CDM and has a $95\,\%$ upper limit of $5\times$CDM.

We can confidently rule out very large rates of impacts corresponding
to substructure that is $\gtrsim5$ times more abundant than expected
from CDM simulations. However, there are effects that we have ignored
in our analysis. We have assumed a constant background level in
determining the Pal 5 density and have disregarded potential density
fluctuations due to variable extinction. We have further assumed a
constant stripping rate in our model, rather than concentrating
stripping episodes near pericentric passages. Simulations using the
formalism of \citet{Bovy14a} and \citet{Sanders14a}, but concentrating
$73\,\%$ of the stripping at pericentric passages induce power on the
largest scales that is $<0.1$, far below the noise and data level (see
\figurename~\ref{fig:pal5denspowspec}). This demonstrates that
pericentric stripping plays an unimportant role in the density
structure of the Pal 5 stream (see also \citealt{Dehnen04a}). We have
assumed a rather low velocity dispersion for the population of
dark--matter subhalos of $\sigma_h = 120\kms$. And we have ignored the
contribution from giant molecular clouds (GMCs), the largest of which
may act much like dark--matter subhalos. However, there are very few
GMCs with $M \gtrsim10^{6.5}\msun$ in the inner Milky Way and
practically none within Pal 5's orbital volume \citep{Rosolowsky05a};
a close enough encounter with even a single massive GMC is unlikely to
have occurred. All of these effects except for the age would
(effectively) increase the power in the model and thus reduce the
inferred rate. Therefore, the measurement here provides a robust upper
limit on the rate of impacts. Appendix \ref{sec:mockpal5} tests the
procedure from this section further using mock $N$-body realizations
of perturbed Pal-5-like streams. These tests demonstrate that our
procedure recovers the correct rate for a a Pal-5-like stream
perturbed by a realistic subhalo population.

\section{Discussion}\label{sec:discussion}

In this paper we have introduced a novel calculus for computing the
effect of encounters with dark--matter subhalos on tidal streams and
we have discussed the perturbations to the density and track location
that they induce. In this section, we discuss various aspects of the
approximations that we have made, new insights on how to determine the
presence of subhalo impacts, and the prospects for observationally
measuring the power spectra and bispectra that we have computed.

{\bf Episodic stripping and epicyclic motions:} All of the simulations
in this paper use a constant stripping rate to create the unperturbed
stream model. Realistic streams are probably better modeled including
a component of stripping at pericentric passages and may have
correlations between the actions and angles upon exiting the
progenitor that lead to intricate epicyclic motions
\citep[\eg,][]{Kuepper10a,Kuepper12a}. These effects can be easily
incorporated into the fast line-of-parallel-angle approach presented
here. The line-of-parallel-angle approach provides a fast way to
relate a present-day $(\veco,\veca)$ to the $\veco$ at stripping
($\dapar = 0$) and does not depend on a particular form of the initial
distribution of frequency--angle offsets from the progenitor or of the
stripping times. Any initial distribution of $\veco$ and $\aperp$ can
be evaluated with any stripping time distribution $p(t_s)$; we chose
to work with a constant stripping rate because it is the easiest
case. Because of mixing within the stream, episodic stripping will
only significantly affect the youngest part of the stream, that is,
the part closest to the progenitor, while the subhalo impacts mostly
affect the oldest part of the stream (see also
\citealt{Ngan14a}). Density and track variations due to subhalo
impacts and episodic stripping will also be largely uncorrelated and
thus add up independently in the power spectrum. Thus, we expect that
episodic stripping is of minor importance in constraining the subhalo
mass spectrum and that density--track cross power spectra and the
bispectrum can distinguish between these two effects.

{\bf The impulse approximation:} All of our calculations use the
impulse approximation to determine the velocity kicks due to a
dark--matter subhalo fly-by. However, the impulse approximation may
not hold for all impact geometries or if the dark--matter subhalos are
in the process of being disrupted \citep{Bovy16a}. In the linear
approximation where the effect of all impacts can be computed based on
the unperturbed stream track, velocity kicks could be computed
accounting for the full trajectory of the subhalo and then converted
to frequency--angle coordinates. This would increase the computational
cost, but only marginally because the effect of the fly-by could still
be computed along a one-dimensional approximation to the stream as the
stream will still be perturbed coherently at a given $\dapar$. Further
investigation of the impulse approximation and when it breaks down is
warranted for getting precise measurements of the subhalo mass
function. However, the details of the interaction are largely
unimportant, because the $\delta \opar^g$ kicks dominate the late-time
effects of subhalo encounters and these kicks are mainly determined by
the overall amplitude and spatial scale of the impacts; they care
little about the details of the velocity kicks.

{\bf Tidal streams as corrugated sheets:} To a first approximation a
tidal stream is a one-dimensional object along the $\dapar$ direction
in six-dimensional phase--space. To a second approximation, a stream
is a two-dimensional sheet in $(\dapar,\dopar)$. For cold streams
originating from globular-cluster progenitors, the extent of a stream
in the remaining four phase--space dimensions is negligible and will
likely not be observable in the coming decades. This follows from the
structure of phase--space near the orbit of a tidal stream: the
largest eigenvalue of the Hessian matrix $\partial \veco/\partial
\vecj$ is typically a factor of $\approx30$ larger than the second
largest eigenvalue \citep{Sanders13a}. The same Hessian controls the
effect of subhalo encounters: \emph{any} small perturbation to a
stream primarily affects the parallel direction $\dopar$ by the same
factor (see, for example, \figurename~\ref{fig:meanOparOperp} and the
parallel-only models in Appendix~\ref{sec:nbody}). Thus, subhalo kicks
turn a tidal stream from a smooth sheet into a corrugated sheet, but
the overall dimensionality remains two. At a given position along a
stream, the stream is therefore one dimensional. This implies that
\emph{all} projections of the stream are almost exactly
correlated. That is, perturbations due to subhalo kicks in the sky
location, distance, and velocity of a stream are entirely
correlated. The amplitude of the deviations in different coordinates
is different (and may have a smooth trend along the stream if the
stream does not follow an orbit), but the shape is the same. This
means that we can combine different projections to better measure
track deviations and that we can use cross-correlations between
different projections to clean the signal from background
contamination, extinction variations, or any part of the signal that
is not intrinsic to the stream. Data are already nearly good enough
that cross-correlations between the density and track could be
informative. It is imperative that we start measuring the track of
tidal streams in detail now.

{\bf Using the density or track to measure the subhalo mass function:}
We have computed the effect of subhalo impacts on both the density and
the track of a tidal stream. It is clear from the power spectrum that
the density is a more sensitive tracer: fluctuations in the density
are larger than those in the track and are easier to measure.  This is
because density perturbations build up over time, while track
perturbations start out large at the time of impact and then fade
because material streams away from the impact point and because of
internal mixing. However, this sensitivity comes at a price. Density
perturbations do not typically respond linearly to increases or
decreases in the rate of impacts (\eg, \figurename
s~\ref{fig:multcutoff} and \ref{fig:sigmah}). The track deviations do
largely respond linearly, as if they all add up independently of each
other. Track deviations are therefore a much more well-behaved tracer
of the rate of impacts. However, \figurename s~\ref{fig:trackObs} and
\ref{fig:trackObs_massranges} demonstrate that the track power
spectrum will be difficult to observe, especially for $M
\lesssim10^{6.5}\msun$. The cross power spectrum of the density and
track is therefore the best compromise between sensitivity and
well-behavedness. Density fluctuations due to subhalos confirmed
through the cross power spectrum with a track projection will make for
a compelling case for cold dark--matter structure.

{\bf Scale dependence of fluctuations:} One of the main advantages of
the line-of-parallel-angle approach for computing the effect of
subhalo impacts over direct simulations or using tracer particles is
its ability to make noiseless predictions for the structure of streams
(that is, Poisson noise in the particle distribution in $N$-body
simulations is absent). This allows us to trace the induced structure
on very small scales. Previous work has claimed that structure on
scales smaller than a few times the width of a tidal stream is
suppressed \citep[\eg,][]{Carlberg13a}. We find no evidence for this
here. For the GD-1-like stream considered here, we find structure on
scales a few times the width of the stream. The only cut-off that is
induced is due to a cut-off in the subhalo mass function (see
\figurename~\ref{fig:multcutoff}).

{\bf Heating of tidal streams:} Much of the early work on the effect
of subhalo encounters on tidal streams focused on the increase in the
velocity dispersion of the stream
\citep[\eg,][]{Johnston02a,Ibata02a,Carlberg09a}. Large effects were
found because these studies significantly overestimated the number of
subhalos present in the inner Milky Way. We have focused on the mean
stream track rather than its dispersion. It is clear that the velocity
dispersion of a perturbed stream could be computed using a similar
approach as that in \sectionname~\ref{sec:lineofpar}. It is unclear
whether this will be a useful exercise, especially if one is
interested in subhalos with masses $<10^7\msun$. These induce very
little increase in the dispersion that would be largely swamped by the
increase because of higher-mass encounters. Increases in dispersion
are also much more difficult to measure observationally.

{\bf The importance of small scales:} We have analyzed the density of
the Pal 5 stream in \sectionname~\ref{sec:pal5} above. Because of the
large uncertainties, we were only able to use the largest
scales. These are dominated by large subhalo masses $M
\gtrsim10^7\msun$, which are most susceptible to modeling errors such
as the breakdown of the impulse approximation and the maximum impact
parameter, and these scales are also strongly affected by
uncertainties in the smooth stream model. All of these problems become
much less important on smaller scales, which also allow the more
interesting mass range $M \lesssim10^7\msun$ to be accessed. Thus, a
high priority for this field is to push density and track measurements
to few degree scales. Made-to-order modeling for potential
observational targets like GD-1 is necessary to establish the optimal
strategy, but a focus on some parts of a stream with sparse sampling
of the full stream to constrain the unperturbed-stream model and to
access larger-scale modes is likely to be a good way forward.

{\bf Statistical versus direct measurements of impacts:} We have
focused in this paper on predicting the statistical properties of the
fluctuation pattern of the density and track of perturbed tidal
streams. However, with good phase--space measurements individual
impacts may be fully characterized in terms of mass, fly-by velocity,
impact parameter, etc., especially at the high-mass end
\citep[\eg,][]{Erkal15b}. It is clear that the fast
line-of-parallel-angle algorithm developed here can also be used in
fitting the parameters of a single impact or multiple impacts, because
it computes the full density and track structure in configuration
space that can be compared to observational data. This may prove to be
useful in two ways: (a) fitting single impacts to find evidence of
individual low-mass dark--matter subhalos, improving on the method of
\citet{Erkal15b} by accounting for the eccentricity and internal
dispersion of tidal streams; and (b) fast fitting of a small number of
impacts that induce the largest changes in a tidal stream and
subtracting their effect to reveal the statistical fluctuations due to
lower-mass subhalos. Point (b) will make it easier to detect
$M<10^6\msun$ subhalos, because it would allow their statistical
effect to be seen on the larger scales where they induce more power
(cf. \figurename~\ref{fig:densObs_massranges}).

{\bf Observational requirements to observe $10^5\msun$ subhalos:} The
density power spectra and density--track cross power spectra in
observable coordinates in \figurename s~\ref{fig:densObs_massranges}
and \ref{fig:crossObs_massranges} make it clear that we can determine
the abundance of $10^5\msun$ dark--matter subhalos in the Milky Way's
halo, if we can measure the scale dependence of the power down to few
degree scales for a few long, cold tidal streams such as GD-1. While
this prediction is somewhat on the optimistic side (the GD-1-like
stream is at the cold, old side of what is expected for a stream like
GD-1) and it is by no means clear that GD-1 will be the best target
for such studies, this prospect makes a compelling case for pursuing
better measurements of the density and track of cold tidal streams.

The observational uncertainties in \figurename
s~\ref{fig:densObs_massranges} and \ref{fig:crossObs_massranges}
correspond to $10\,\%$ density measurements in $0.3^\circ$ bins along
the stream. For GD-1, which has a stellar mass of $\approx20,000\msun$
\citep{Koposov10a}, this would require a background-free map of about
half of all of the stars down to the bottom of the
main-sequence. While this may seem unrealistic, a wide-field proper
motion survey with WFIRST or a wide-field AO imager on a large
telescope may get close to this goal in ten years from now. Until then
and because background contamination is the biggest issue limiting the
current measurements (see the analysis of Pal 5 above), a
line-of-sight velocity survey of GD-1 or a similar stream may be the
best way forward. A line-of-sight velocity precision of
$\lesssim10\kms$ would significantly reduce the background from halo
stars, especially when combined with abundance information from the
spectra. GD-1 has approximately 3,000 stars down to $r =22$
\citep{Koposov10a}. A background-free map of those would allow $25\%$
density measurements, which would still allow the contribution of
$\approx10^{5.5}\msun$ subhalos to be seen in the density (see
\figurename~\ref{fig:densObs_massranges}). Observing all stream
members down to $r=24$ would essentially bring the noise level down to
that in \figurename~\ref{fig:densObs_massranges}. This will not be
easy and GD-1 might not be the best target, but investing in this will
help settle the question of whether dark--matter clumps on scales
smaller than those of galaxies.

It is clear from \figurename s~\ref{fig:trackObs} and
\ref{fig:crossObs} that measuring deviations in the track of a stream
using the line-of-sight velocity will be a frustrating experience. The
predicted deviations are $<1\kms$ from the largest impacts and
$<0.1\kms$ from lower-mass subhalos. We have not shown the predicted
power spectra for the proper motions, because they are far beyond what
can be measured. While line-of-sight velocity measurements can be
helpful on the largest scales corresponding to $\gtrsim10^7\msun$
subhalos, their main use will be in removing background
contamination. This is simply a reflection of the fact discussed above
that track deviations at a given distance along the stream are
one-dimensional: every projection measures the same deviation and this
is much easier to observe in sky position and distance than velocity.

\section{Conclusion}\label{sec:conclusion}

In this paper we have developed a novel method for computing the
phase--space structure of tidal streams perturbed by dark--matter
subhalo impacts. We have used it to study the fluctuations in the
density and location (the ``track'') of a stream due to subhalo
impacts with different overall rates and masses. We have also
performed a first rigorous measurement of the abundance of
dark--matter subhalos with $M \gtrsim10^{6.5}\msun$ within about
$20\kpc$ from the center of the Milky Way using density data for the
Pal 5 stream. Our main findings and conclusions are the
following:\\ $\bullet$ The line-of-parallel angle approach to compute
the phase--space structure of a tidal stream developed in
\sectionname~\ref{sec:lineofpar} is a fast method to compute the
effect of subhalo impacts, allowing the perturbed structure of a tidal
stream for a given, realistic set of impacts to be computed in a
matter of minutes. We have made a number of assumptions in this
approach and its application in this paper, but most of these are not
of fundamental importance to the speed of the method. The basic
assumption that allows the fast computation is that of the
\emph{linearity} of the impacts, \ie, that we can compute the velocity
kicks for all kicks based on the unperturbed stream track, rather than
the perturbed stream track. We have extensively tested this assumption
in Appendix~\ref{sec:nbody} and find it to work well, especially for
the relatively low number of impacts expected for a CDM-like
population of subhalos in the inner Milky Way. Other assumptions, such
as the validity of the impulse approximation, our ignoring of kicks in
the perpendicular frequency and angle directions, and the specific
assumptions made about the initial frequency distribution of the tidal
debris and the rate at which material is stripped are not essential to
the method and could be easily relaxed at little additional
computational cost.\\$\bullet$ Tidal streams, when perturbed by a
population of subhalo impacts, are unlikely to display clearly
identifiable gaps in their density profiles (see, \eg,
\figurename~\ref{fig:densexample}), even when the number of impacts is
relatively small and especially for those gaps that are due to
low-mass subhalos ($M \lesssim10^7\msun$). But the subhalo impacts
induce a rich structure of fluctuations on different scales that can
be observed through the power spectrum of the density and track. The
structure in the density and that in the track are strongly
correlated---indeed, it is the structure in the track that gives rise
to the density fluctuations at later times. Because cold--dark--matter
subhalos follow a somewhat tight concentration--mass relation,
different mass subhalos give rise to structure on different scales,
with smaller scales dominated by very low-mass subhalos ($M
\lesssim10^7\msun$). Observations of this power spectrum in the
density and track, of its scale dependence, and of the
cross-correlation between the density and track and of different
projections of the track are a clear path forward to determining the
subhalo mass spectrum between $10^5\msun$ and $10^9\msun$ from
observations of tidal streams in the Milky Way. Going beyond the power
spectrum, the time dependence of the impact rate and of the evolution
of density fluctuations induces significant higher-order moments such
as can be observed through the bispectrum that may be observable in
future data and that would constitute a powerful confirmation of the
subhalo-impact origin of density and track fluctuations.\\$\bullet$
Using this new framework, we have performed the first rigorous
inference of the dark--matter subhalo population using data on the
density of the Pal 5 stream. We find a rate of impacts that is
$1.4^{+1.6}_{-0.9}\times$CDM, equivalent to $10^{+11}_{-6}$
dark--matter subhalos with masses between $10^{6.5}\msun$ and
$10^9\msun$ within $20\kpc$ from the Galactic center or a subhalo mass
fraction of $f_{\mathrm{sub}}(r<20\kpc)\approx0.2\,\%$ over the same
mass range. While the uncertainty on the rate is large and this
measurement comes with caveats because we can only use the power on
the largest scales given the current observational uncertainties and
because of the uncertainty on the Pal 5 stream's age, this constraint
is already at an interesting level and the upper limit in particular
is robust. Furthermore, simulations of the structure in the Pal 5
stream induced by a CDM-like population of subhalos (see
\figurename~\ref{fig:pal5starcountexamples}) provide a remarkable
match to the observed density profile. This first analysis makes it
clear that modest improvements in the data quality will soon lead to
the best available constraints on the low-mass subhalo population of a
Milky-Way-sized halo, especially if we can push the analysis to
smaller scales.

All code used in this paper is made publicly available, except for the
GADGET-3 code used to run the $N$-body simulations, as we are not at
liberty to release this. The methods from
\sectionname~\ref{sec:perturb} are contained in
\texttt{galpy}\footnote{Available at
  \url{http://github.com/jobovy/galpy}~.} for the case of a single
impact (\texttt{galpy.df.streamgapdf}) and as the \texttt{galpy}
extension \texttt{galpy.df.streampepperdf}\footnote{Available
  at\\ \centerline{\url{https://gist.github.com/jobovy/1be0be25b525e5f50ea3}~.}}
for the case of multiple impacts. All of the analysis in this paper
can be reproduced using the code found
at\\ \centerline{\url{https://github.com/jobovy/streamgap-pepper}}.

{\bf Acknowledgments} We thank the anonymous referee for a
constructive report. This research received financial support from the
Natural Sciences and Engineering Research Council of Canada. The
research leading to these results has received funding from the
European Research Council under the European Union's Seventh Framework
Programme (FP/2007-2013)/ERC Grant Agreement no. 308024.

\bibliography{ms} \bibliographystyle{mn2e}

\appendix

\section{Detailed tests of the frequency--angle framework with $N$-body simulations}\label{sec:nbody}

\begin{figure*}
\includegraphics[width=\textwidth]{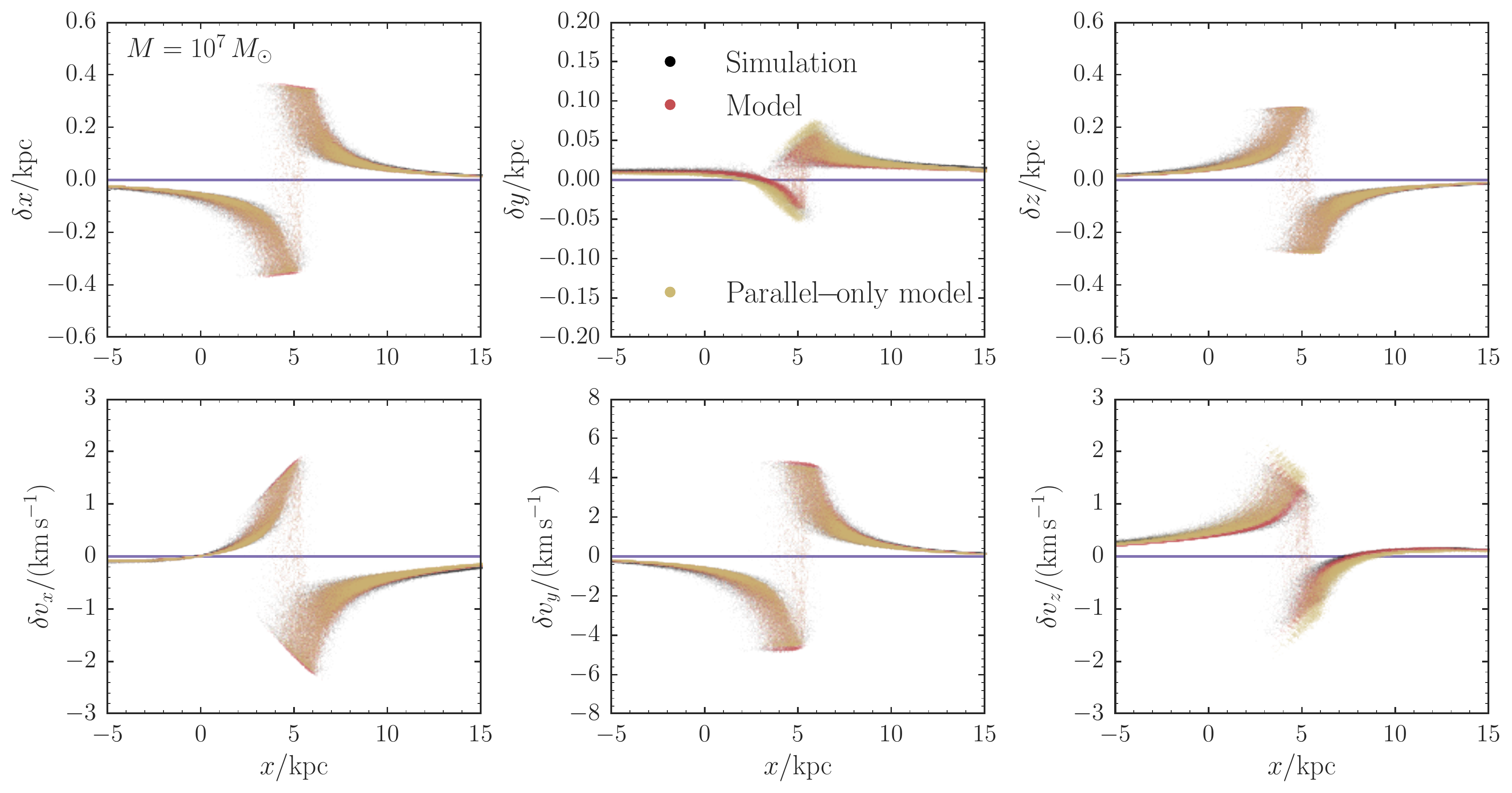}
\caption{Comparison between an $N$-body simulation of an impact with a
  $10^7\msun$ subhalo (black points) and mock data generated from the
  modeling in frequency--angle space. The red points are for the
  generative model in which kicks $(\delta \veco^g,\delta\veca^g)$ in
  all coordinates are included, the yellow points are for the case
  where only $\delta \opar^g$ kicks in the parallel frequency
  direction are applied. The phase-space differences are shown as a
  function of unperturbed $x$ position (that in the absence of the
  impact) near the current location of the impact point, which is
  known in both the simulation and the mock data. As expected from the
  phase--space structure near the stream, most of the effect of the
  subhalo is in the parallel direction and the yellow, red, and black
  points almost completely overlap (both in their mean trend and their
  scatter at a given $x$). Our frequency--angle-with-impulsive-kicks
  modeling produces a configuration-space phase-space structure that
  agrees with the $N$-body simulation.\label{fig:nbody_1e7}}
\end{figure*}

\begin{figure*}
\includegraphics[width=\textwidth]{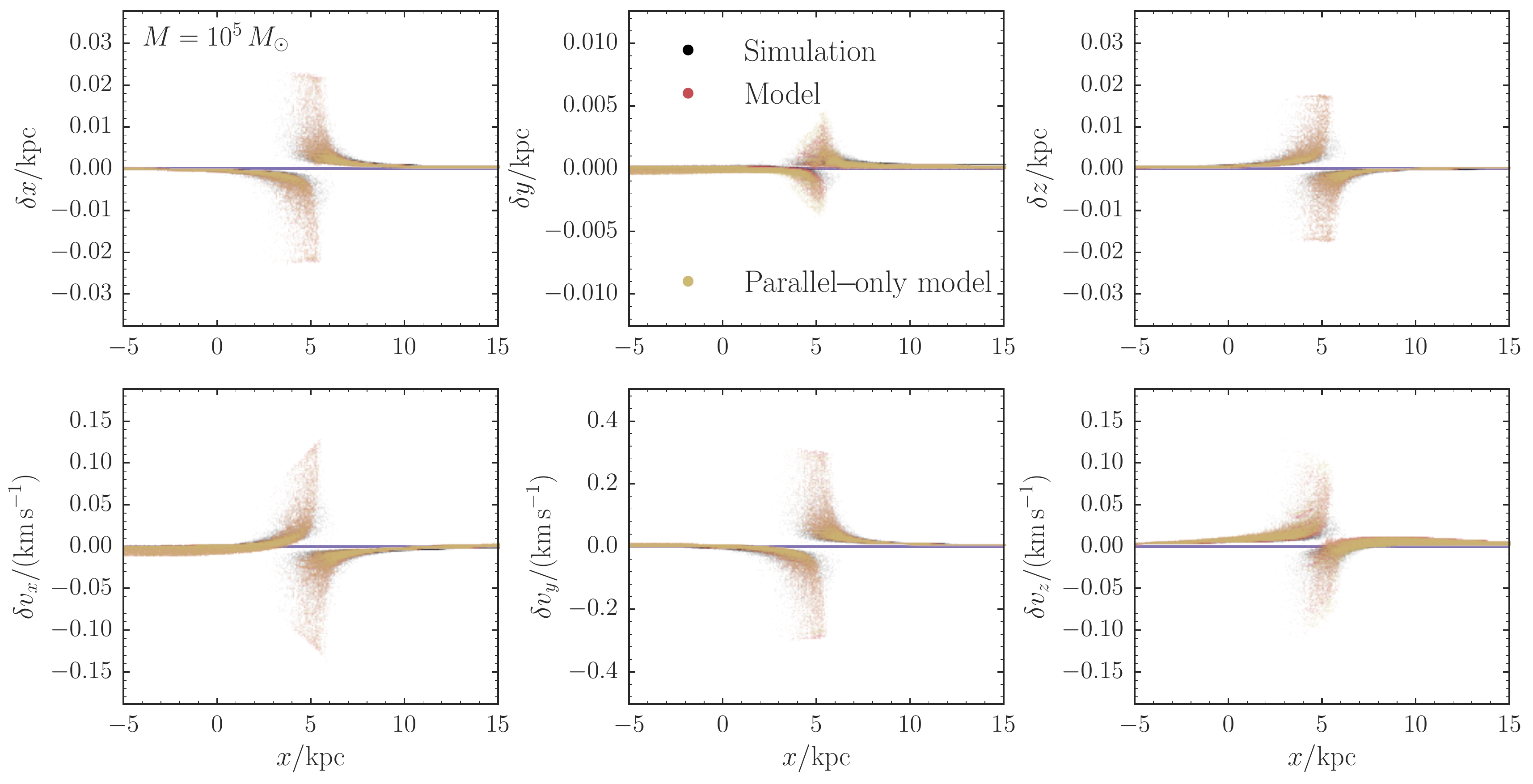}
\caption{Same as in \figurename~\ref{fig:nbody_1e7}, but for an impact
  with a $10^5\msun$ subhalo. The impact induces configuration-space
  deviations on the order of $100\pc$ and
  $50\,\mathrm{m\,s}^{-1}$. The agreement between the $N$-body model
  and the data generated from the frequency--angle modeling is
  excellent.\label{fig:nbody_1e5}}
\end{figure*}

\begin{figure*}
\includegraphics[width=\textwidth]{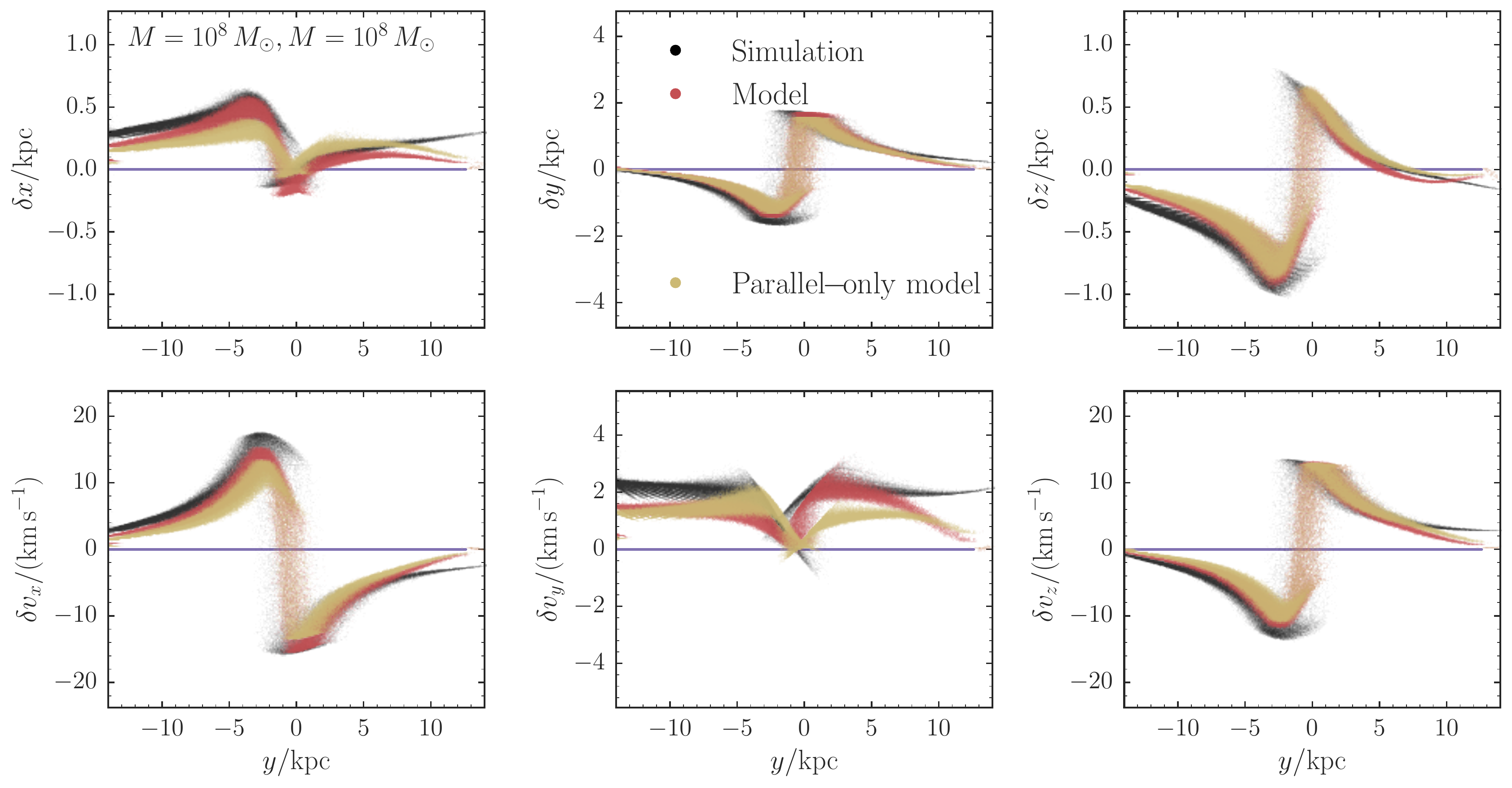}
\caption{Like \figurename~\ref{fig:nbody_1e7}, but for a simulation
  where the stream has been perturbed by two $10^8\msun$ subhalos at
  nearly the same location along the stream separated by $1\Gyr$ in
  time and observed $1\Gyr$ after the second impact (see text for the
  exact setup). The phase-space differences are shown as a function of
  unperturbed $y$ position near the current location of the impact
  points. The agreement between the $N$-body simulation and the
  frequency--angle modeling is good, with only minor differences (note
  that the $y$-range in the middle, lower panel is much smaller than
  those in the left and right lower panels).\label{fig:nbody_1e81e8}}
\end{figure*}

\begin{figure*}
\includegraphics[width=\textwidth]{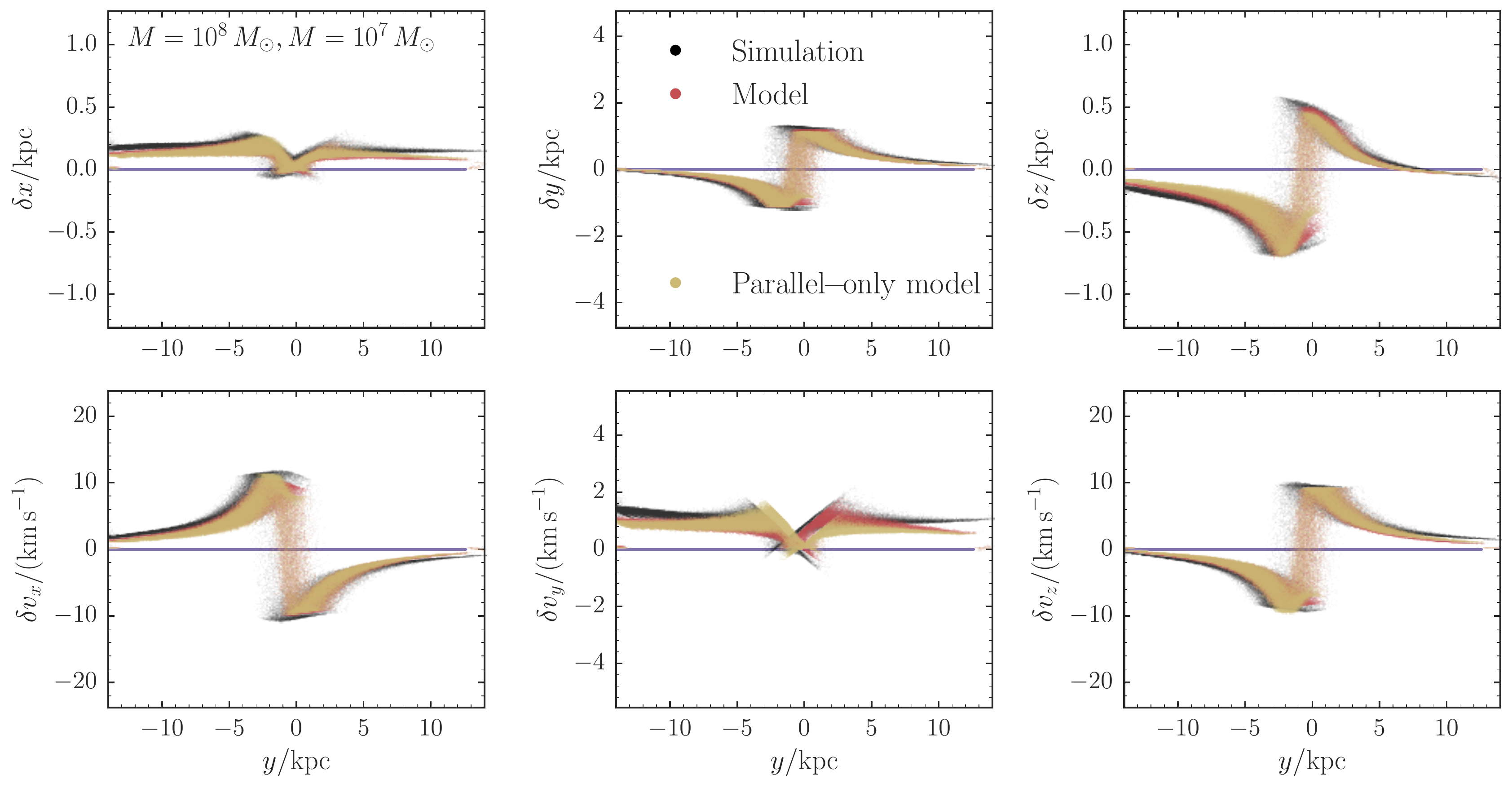}
\caption{As \figurename~\ref{fig:nbody_1e81e8}, but for the case where
  the second impact is that of a $10^7\msun$
  subhalo.\label{fig:nbody_1e81e7}}
\end{figure*}

In this Appendix, we investigate the applicability of our
stream-gap-modeling framework in frequency--angle space by comparing
to full $N$-body simulations of the interaction between dark--matter
subhalos and stellar streams. These tests are similar to those
described by \citet{Sanders15a}. There, a detailed comparison between
the frequency--angle framework for a single impact with a $10^8\msun$
subhalo and a cold stellar stream was presented in configuration
space. To as closely follow the $N$-body simulation of the stream as
possible, a custom unperturbed stream model was constructed that
combines stripping events at pericenter with a contribution from a
continuous stripping process; the combination matches the simulation
of the unperturbed stream. This allowed a fair comparison between the
frequency--angle framework for applying subhalo kicks ($\delta
\veco^g,\delta\veca^g)$. Good agreement between the frequency--angle
framework and the $N$-body simulation of the interaction was found.

Here, we are interested in testing various additional aspects of our
framework. First and foremost is whether the framework remains
accurate down to the lowest-mass subhalos that we consider in this
paper ($M=10^5\msun)$. Additionally, we want to test the accuracy of
our approximation that all kicks $(\delta \veco^g,\delta \veca^g)$ are
computed on the basis of the unperturbed stream track (\ie, that the
effect of the kicks is linear at the time of impact). Lastly, we want
to explicitly test the approximation used in the fast
line-of-parallel-angle approach that the main effect of subhalo
impacts is in the parallel frequency direction, such that we only need
to consider $\delta \opar^g$.

We test that our framework works down to $M=10^5\msun$ by repeating
the simulations in \citet{Sanders15a} for impacts with $M =
10^7\msun$, $10^6\msun$, and $10^5\msun$. These simulations were run
with the $N$-body part of \textsc{gadget-3} which is similar to
\textsc{gadget-2} \citep{Springel05a} and are identical to the setup
in \cite{Sanders15a}. The impact in \citet{Sanders15a} was with a
Plummer sphere with $M=10^8\msun$ and $r_s = 625\pc$. For the lower
masses, we change the scale radii to $250\pc$, $80\pc$, and $40\pc$,
respectively. Otherwise all of the impact parameters (location along
the stream, impact parameter [0], fly-by velocity, impact time
[$880\Myr$ ago]) are the same. We generate mock data from the
frequency--angle model (see \sectionname~\ref{sec:mock} and
\citealt{Sanders15a}) in configuration space for both an unperturbed
stream and a stream perturbed by a subhalo using the same initial
conditions. This way, we can compare the present-day position of each
individual mock data point between the perturbed and unperturbed
model. These differences in position for all phase--space coordinates
in the vicinity of the gap are displayed in \figurename
s~\ref{fig:nbody_1e7} and \ref{fig:nbody_1e5} for the $10^7\msun$ and
$10^5\msun$ impacts, respectively (red points). The same differences
in present-day position for the $N$-body simulation with and without
perturbation (starting from the same initial condition) are shown as
the black points. The yellow points show the mock data that are
generated by only applying the kick in parallel frequency $\delta
\opar^g$, that is, without applying kicks in the perpendicular
direction and without any angle kicks. All three simulations agree,
demonstrating that our frequency--angle-with-impulsive-kicks framework
works well down to $M=10^5\msun$ (the $M=10^6\msun$ comparison is
similar, but not shown here). This is especially impressive
considering that the effect of the kicks is accurately modeled at the
few pc and $50\,\mathrm{m\,s}^{-1}$ level for the lowest-mass
interaction. That the model that only considers $\delta \opar^g$ kicks
agrees with the full model shows that this approximation that we make
throughout most of this paper is valid.

\begin{figure*}
\includegraphics[width=\textwidth]{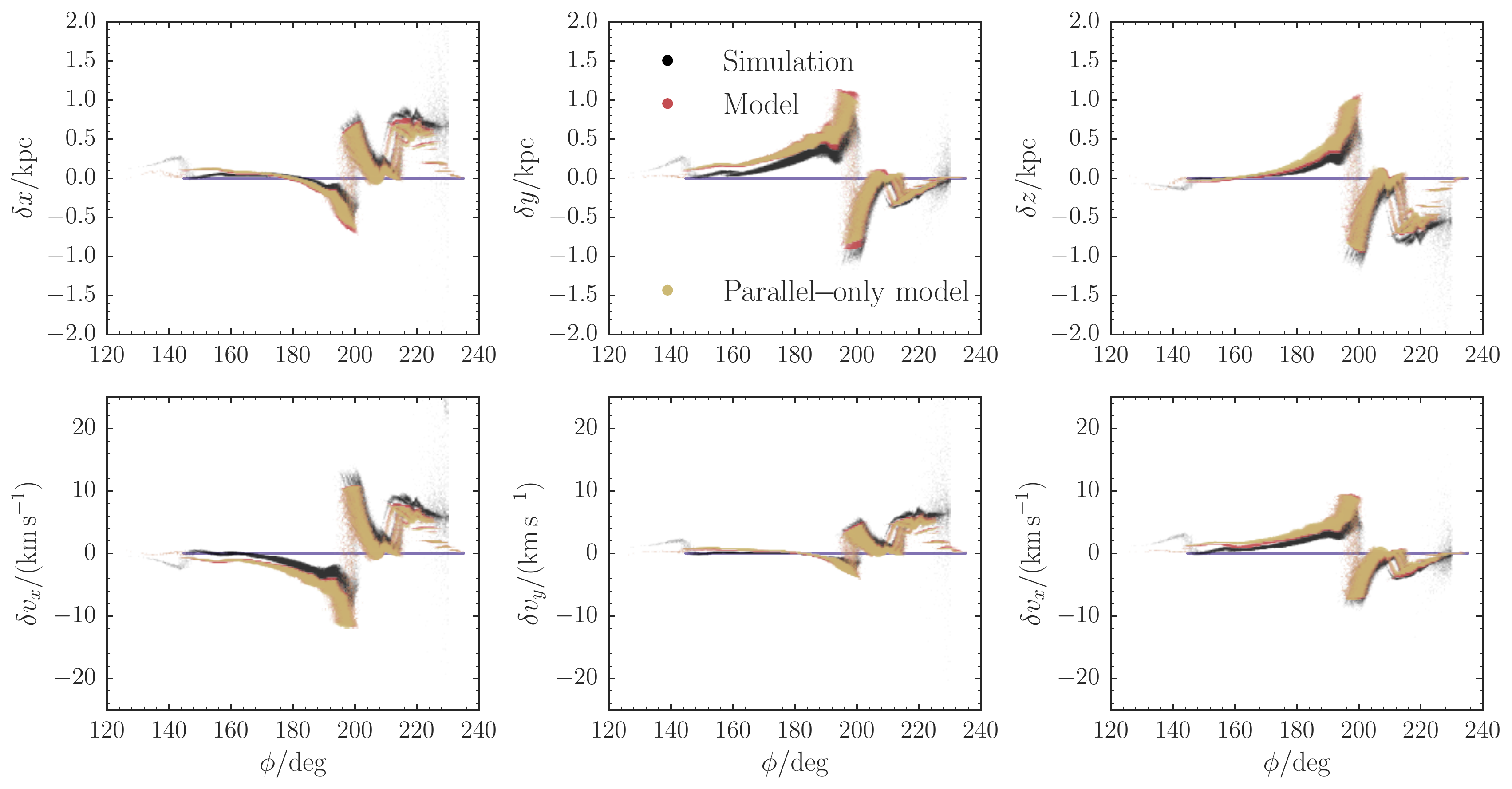}
\caption{Like \figurename~\ref{fig:nbody_1e7}, but for a simulation
  where the stream has been perturbed by 24 subhalos with masses
  between $10^6\msun$ and $10^8\msun$ that impact the stream between
  $10\Gyr$ and $11.75\Gyr$ and that are otherwise sampled using the
  method in \sectionname~\ref{sec:sample}. The phase-space differences
  are shown at $12\Gyr$ as a function of unperturbed azimuth $\phi$
  position all along the trailing arm of the stream. The agreement
  between the $N$-body simulation and the frequency--angle modeling is
  good, with only a small overall offset in $y$, $v_x$, and
  $v_y$.\label{fig:nbody_many}}
\end{figure*}

To further stress-test our framework, we have also run $N$-body
simulations where the same part of the stream gets impacted by two
massive dark--matter subhalos. We impact the same stream as in
\citet{Sanders15a} again at $10\Gyr$ with the same $10^8\msun$ impact
as in \citet{Sanders15a} (except that it has $r_s = 1.05\kpc$ to more
closely follow the expected mass--concentration relation), but follow
it with another impact in almost the same part of the stream at
$11\Gyr$ (the mass of this second impact is either $10^8\msun$ or
$10^7\msun$). The second impact is again a direct impact and it has a
fly-by velocity of $100\kms$ pointed along the $x$ direction; the
$10^7\msun$ subhalo has $r_s = 332\pc$. We compare configuration space
differences in all coordinates $1\Gyr$ after the second impact for
both of these setups. These results are displayed in \figurename
s~\ref{fig:nbody_1e81e8} and \ref{fig:nbody_1e81e7}. While the
agreement between the frequency--angle models and the $N$-body
simulations is not perfect, it is clear that the model overall does a
good job of matching the effects seen in the $N$-body simulation.

This two-impact comparison that we have performed here is in many ways
a worst-case scenario for our framework, because it has multiple large
impacts happen in almost the same part of the stream. Because large
impacts are rare, this should not happen often in reality. As the
($10^8\msun,10^7\msun$) demonstrates, a large direct impact followed
by a smaller direct impact is already much better modeled than that of
two $10^8\msun$ subhalos, and even such combinations are
rare. Therefore, we are confident that our framework correctly
captures the effect of multiple impacts.

As a final test, we sample 24 fly-bys with masses between $10^6\msun$
and $10^8\msun$ that impact the trailing part of the same stream
between $10\Gyr$ and $11.75\Gyr$ and we observe the stream again at
$12\Gyr$. The parameters describing these impacts are sampled using
the statistical procedure of \sectionname~\ref{sec:sample} (the
CDM-like rate for this setup is 21.5 expected impacts). The set of
fly-bys has three impacts with $M\approx3\times10^7\msun$ and many
lower-mass impacts. The phase--space differences at the present time
as a function of Galactocentric azimuth are displayed in
\figurename~\ref{fig:nbody_many}. These differences are dominated by a
single large perturbation due to one of the $M\approx3\times10^7\msun$
fly-bys, but have much structure on smaller scales as well due to the
other 23 fly-bys. The frequency--angle model matches the overall
structure of the phase--space differences and also reproduces most of
the smaller-scales wiggles. There is a slight $\approx$constant offset
in some of the coordinates that may be due an imperfect translation of
our modeling setup to the $N$-body code, a breakdown of the impulse
approximation, or the slight difference in the stripping rate between
the simulation and the model. Close to the progenitor---located around
$\phi=240^\circ$---the simulation and the model also display some
differences, because the details of the stripping rate matter more
there than elsewhere and we only roughly match the stripping rate in
the model.

\begin{figure}
\includegraphics[width=0.49\textwidth]{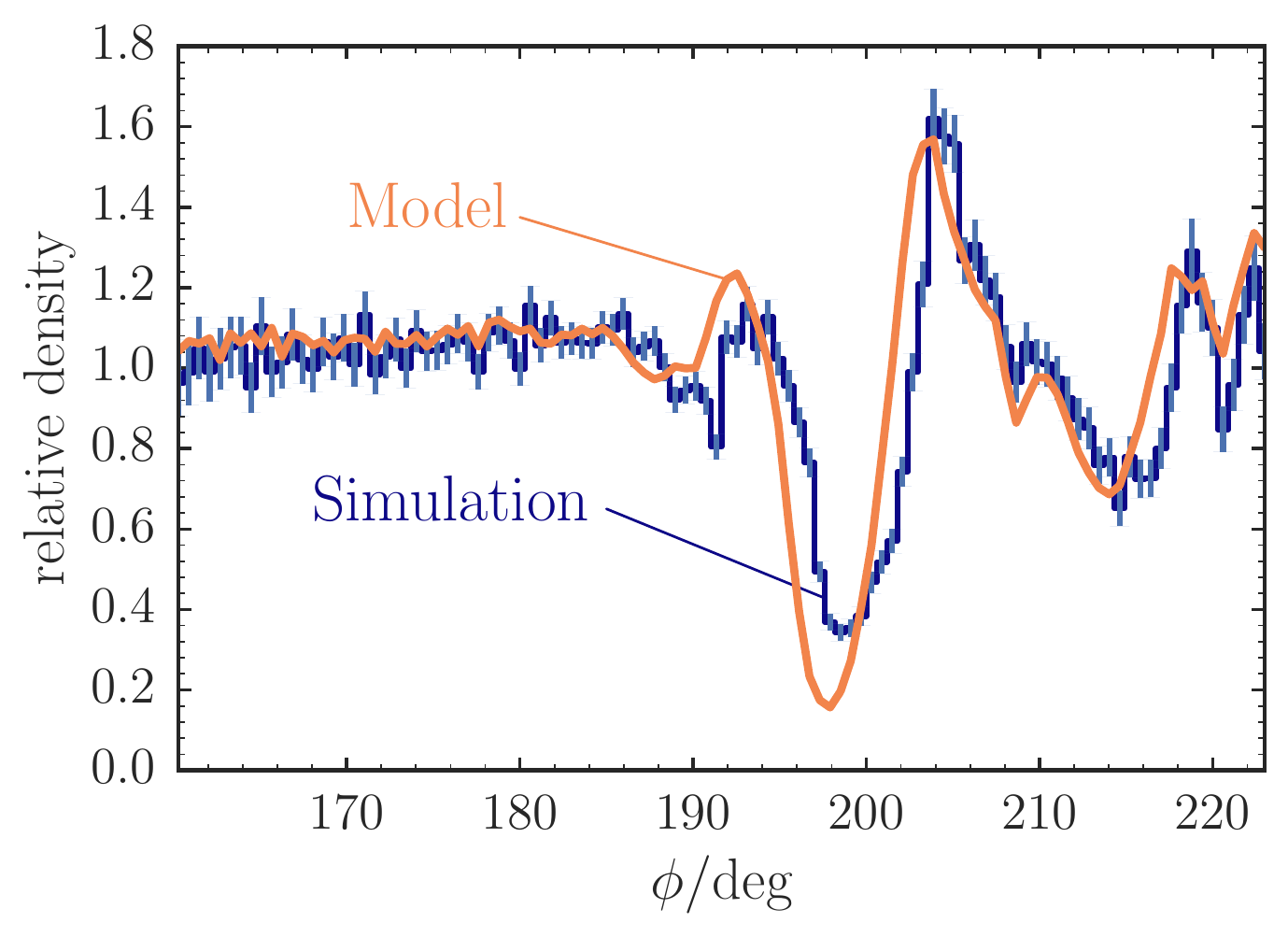}
\caption{Comparison between the relative perturbed/unperturbed
  density for the $N$-body simulation and frequency--angle model in
  \figurename~\ref{fig:nbody_many} (blue and orange curves,
  respectively). The 24 subhalo impacts with masses between
  $10^6\msun$ and $10^8\msun$ create 5 visible gaps that are
  reproduced by the model, albeit imperfectly. The density power
  spectra of both curves (not shown) display good
  agreement.\label{fig:nbody_many_dens}}
\end{figure}

The difference between the relative density (perturbed/unperturbed)
between the $N$-body simulation and the model is displayed in
\figurename~\ref{fig:nbody_many_dens}. The 24 impacts lead to 5
visible gaps in the relative density. All of these are reproduced by
the model. The width of the gaps is very well matched, while the depth
and the exact location are less well modeled, due to the same reasons
that produce the $\approx$constant offset in the phase--space
differences. The power spectra of the relative density of the
simulation and the model are in good agreement; thus we certainly
match the statistical properties of the stream. In all of these
comparisons the model that only applies the $\delta \opar^g$ kicks
produces the same results as the model that applies the full $(\delta
\veco^g,\delta \veca^g)$ kicks, demonstrating once again that $\delta
\opar^g$ kicks are all that is necessary to model the effect of
subhalo perturbations.

\section{Detailed derivation of the line-of-parallel-angle algorithm for multiple impacts}\label{sec:detail}

In this appendix, we go through the derivation of the
line-of-parallel-angle algorithm in \sectionname~\ref{sec:multiple} in
more detail. We denote the piecewise-polynomial representation of kick
$j$ as $\sum_k C_{bk}^j (\dapar-x_b^j)^k$ for segment $b$, with $j=1$
for the final impact that is undone first. For $\dopar \leq
\dopar^{\mathrm{max}}$ we start by undoing the effect of the final
impact as in \sectionname~\ref{sec:single}. As in
\equationname~(\ref{eq:p0single}), the point $(\dopar,\dapar)$ is
located at
\begin{align}\label{eq:anglefinalimpact}
& \left(\dopar-\sum_k C_{bk}^1\,\left(\dapar-\dopar\,t_i^1-x_b^1\right)^k,
  \dapar-\dopar\,t_i^1\right)\,\nonumber\\
  & \qquad \qquad \qquad x_b^1 \leq \dapar-\dopar\,t_i^1 < x_{b+1}^1\,,
\end{align}
before the final impact. This phase--space point can then be evolved
backward until the second-to-last impact at $t_i^2$. Its frequency
remains constant and it arrives at a parallel angle
\begin{align}\label{eq:anglepenultimateimpact}
& \dapar-\dopar\,t_i^1\nonumber\\
 & \qquad -\left(\dopar-\sum_k C_{bk}^1\,\left(\dapar-\dopar\,t_i^1-x_b^1\right)^k\right)\,(t_i^2-t_i^1)\,\nonumber\\
& \qquad \qquad \qquad x_b^1 \leq \dapar-\dopar\,t_i^1 < x_{b+1}^1\,.
\end{align}
Keeping only terms up to linear order in $\dapar-\dopar\,t_i^1-x_b^1$,
we can re-write this as
\begin{align}
& \dapar-(t_i^2-t_i^1)\,\left(-C_{b0}^1-C_{b1}^1\left[\dapar-x_b^1\right]\right)\nonumber\\
& \ \ -\dopar\,\left(t_i^2-[t_i^2-t_i^1]\,C_{b1}^1\,t_i^1\right)\,,\ x_b^1 \leq \dapar-\dopar\,t_i^1 < x_{b+1}^1\,.
\end{align}
If we define
\begin{align}
  \daparb^2 &\equiv \dapar-(t_i^2-t_i^1)\,\left(-C_{b0}^1-C_{b1}^1\left[\dapar-x_b^1\right]\right)\,,\\
  t^2_b & \equiv t_i^2-[t_i^2-t_i^1]\,C_{b1}^1\,t_i^1\,,\label{eq:parsecond}
\end{align}
the parallel angle at the time of the second-to-last impact is simply
\begin{equation}
  \daparb^2 - \dopar\,t^2_b\,.
\end{equation}
This expression is similar to that for the parallel angle at the final
impact (cf. the angle in \equationname~[\ref{eq:anglefinalimpact}]),
except that each segment $b$ now has an individual angle and time
associated with it. Previously these were the same ($\dapar$ and
$t_i^1$) for all segments.

For each segment, we can then undo the kick from the second-to-last
impact by determining which segment $b'$ $[x_{b'}^2,x_{b'+1}^2[$ of
    the second-to-last impact the parallel angle falls in. Undoing the
    effect of the second-to-last impact then changes the frequency to
    (from now on we only consider terms up to linear for each kick)
\begin{align}\label{eq:freqthirdtolast}
 \dopar&-C_{b0}^1-C_{b1}^1\,\left(\dapar-\dopar\,t_i^1-x_b^1\right)\nonumber\\
  & -C_{b'0}^2-C_{b'1}^2\,\left(\daparb^2-\dopar\,t_b^2-x_{b'}^2\right)\nonumber\\
  & \!\!\!\!\!\! x_b^1 \leq \dapar-\dopar\,t_i^1 < x_{b+1}^1\ \mathrm{and}\ 
  x_{b'}^2 \leq \daparb^2-\dopar\,t_b^2 < x_{b'+1}^2\,.
\end{align}
We can then run phase--space points backward to the time of the
third-to-last impact using this frequency, and similar to
\equationname~(\ref{eq:anglepenultimateimpact}) we arrive at
\begin{align}
 \daparb^2 - \dopar\,t^2_b&\nonumber\\
  \qquad -\left(\dopar\right.&-C_{b0}^1-C_{b1}^1\,\left[\dapar-\dopar\,t_i^1-x_b^1\right]\nonumber\\
  &\left.-C_{b'0}^2-C_{b'1}^2\,\left[\daparb^2-\dopar\,t_b^2-x_{b'}^1\right]\right)\,(t_i^3-t_i^2)\,\nonumber\\
  & \!\!\!\!\!\!\!\!\!\!\!\!\!\!\!\!\!\!\!\!\!\!\!\!\!\!\!\!\!\!\!\!\!\!\!\! x_b^1 \leq \dapar-\dopar\,t_i^1 < x_{b+1}^1\ \mathrm{and}\ 
  x_{b'}^2 \leq \daparb^2-\dopar\,t_b^2 < x_{b'+1}^2\,.
\end{align}
If we then define
\begin{align}\label{eq:apdaparb}
  \daparb^3 &\equiv \daparb^2-(t_i^3-t_i^2)\,\left(-C_{b0}^1-C_{b1}^1\left[\dapar-x_b^1\right]\right.\\
& \quad\qquad\qquad\qquad\qquad\left.-C_{b'0}^2-C_{b'1}^2\left[\daparb^2-x_{b'}^2\right]\right)\,,\nonumber\\
  t^3_b & \equiv t_b^2+[t_i^3-t_i^2]\,[1+C_{b1}^1\,t_i^1+C_{b'1}^2\,t_b^2]\,,
\end{align}
we can again write this angle as
\begin{equation}\label{eq:parthird}
  \daparb^3 - \dopar\,t^3_b\,.
\end{equation}
Thus, by performing similar operations for all previous impacts, we
always keep the same form of expression for the parallel angle of each
segment at the previous impact. This is what motivates the updates in
\equationname s~(\ref{eq:update_dapar}) and (\ref{eq:update_t}) and
the definition and updates to $\dd\Omega_b$, which arise from the need
to keep track of \emph{all} previous changes to the frequency (see
\equationname~[\ref{eq:apdaparb}]).

The discussion so far has not described how the piecewise-linear
segments of different impacts mesh throughout the propagation of the
line-of-parallel-angle. In \equationname~(\ref{eq:freqthirdtolast}),
we simply wrote the conditions for the parallel angle to be within
segments $b$ and $b'$ of the last and second-to-last impact. In
practice, we can track the segments by re-writing them as segments in
present-day $\dopar$ rather than $\dapar$, as done for the case of a
single impact in \equationname~(\ref{eq:breakSingle}). Each previous
impact gives rise to a new set of breakpoints in present-day $\dopar$
through equations such as \equationname~(\ref{eq:parsecond}) and
(\ref{eq:parthird}). These are computed using
\equationname~(\ref{eq:newbreak}), where in the notation of this
appendix we would have to write $x_b^j$ instead of $x_b$. Because
after correcting for the final impact, the expression for the new
breakpoints depends on the parameters for each individual segment $b$,
for each segment we only add new breakpoints that are within the
segment $b$. In \figurename~\ref{fig:illustrate_perturbations}, the
new breakpoints for the second-to-last impact are the small dots that
are added to the large dots in the fourth panel.

After correcting for the effect of the third-to-last impact, the
frequency becomes (analogous to
\equationname~[\ref{eq:freqthirdtolast}])
\begin{align}
 \dopar&-C_{b0}^1-C_{b1}^1\,\left(\dapar-\dopar\,t_i^1-x_b^1\right)\nonumber\\
  & -C_{b'0}^2-C_{b'1}^2\,\left(\daparb^2-\dopar\,t_b^2-x_{b'}^2\right)\nonumber\\
  & -C_{b''0}^3-C_{b''1}^3\,\left(\daparb^3-\dopar\,t_b^3-x_{b''}^3\right)\nonumber\\
  & \!\!\!\!\!\! x_b^1 \leq \dapar-\dopar\,t_i^1 < x_{b+1}^1\,, 
 x_{b'}^2 \leq \daparb^2-\dopar\,t_b^2 < x_{b'+1}^2\,,\nonumber\\
 &\mathrm{and}\ x_{b''}^3 \leq \daparb^3-\dopar\,t_b^3 < x_{b''+1}^3\,.
\end{align}
In terms of the breakpoints $\doparb$ defined in terms of $\dopar$, we
can write this as
\begin{align}
 \dopar&-C_{b0}^1-C_{b1}^1\,t_i^1\,\left(\doparb-\dopar\right)\nonumber\\
  & -C_{b'0}^2-C_{b'1}^2\,t_b^2\left(\doparbprime-\dopar\right)\nonumber\\
  & -C_{b''0}^3-C_{b''1}^3\,t_b^3\left(\doparbprimeprime-\dopar\right)\nonumber\\
  & \!\!\!\!\!\! \doparb < \dopar \leq \doparbpone\,,
 \doparbprime < \dopar \leq \doparbprimepone\,,\nonumber\\
 & \mathrm{and}\ \doparbprimeprime < \dopar \leq \doparbprimeprimepone\,,
\end{align}
where $\doparbprime$ with a higher number of primes denotes
breakpoints defined using \equationname~(\ref{eq:newbreak}) at earlier
impacts. Rather than tracking all previous impacts and their
coefficients $C_{bk}^j$ separately, we update a single
piecewise-linear representation of the line-of-parallel angle with
coefficients $c_{b0}$ and $c_{b1}t$. Thus, we specify the equation above
to the narrowest range of the three breakpoint-ranges. Let's say that
this is the final one, $\doparbprimeprime < \dopar \leq
\doparbprimeprimepone$. For that interval, the frequency becomes
\begin{align}
 \dopar&-C_{b0}^1-C_{b1}^1\,t_i^1\,\left(\doparbprimeprime-\doparbprimeprime+\doparb-\dopar\right)\nonumber\\
  & -C_{b'0}^2-C_{b'1}^2\,t_b^2\left(\doparbprimeprime-\doparbprimeprime+\doparbprime-\dopar\right)\nonumber\\
  & -C_{b''0}^3-C_{b''1}^3\,t_b^3\left(\doparbprimeprime-\dopar\right)\nonumber\\
 & \doparbprimeprime < \dopar \leq \doparbprimeprimepone\,,
\end{align}
or
\begin{align}
 &\dopar\nonumber\\
&-C_{b0}^1-C_{b1}^1\,t_i^1\,\left(\doparb-\doparbprimeprime\right)
-C_{b1}^1\,t_i^1\,\left(\doparbprimeprime-\dopar\right)\nonumber\\
  & -C_{b'0}^2-C_{b'1}^2\,t_b^2\left(\doparbprime-\doparbprimeprime\right)
-C_{b'1}^2\,t_b^2\left(\doparbprimeprime-\dopar\right)\nonumber\\
  & -C_{b''0}^3-C_{b''1}^3\,t_b^3\left(\doparbprimeprime-\dopar\right)\nonumber\\
 & \qquad \doparbprimeprime < \dopar \leq \doparbprimeprimepone\,.
\end{align}
Thus, in a single piecewise-linear representation of the frequency
changes due to kicks, after correcting for the third-to-last kick, we
have a constant term
\begin{align}\label{eq:apdefinec0}
c_{b0}= &-C_{b0}^1-C_{b1}^1\,t_i^1\,\left(\doparb-\doparbprimeprime\right)\nonumber\\
  & -C_{b'0}^2-C_{b'1}^2\,t_b^2\left(\doparbprime-\doparbprimeprime\right)\nonumber\\
& -C_{b''0}^3\,,
\end{align}
and a linear factor
\begin{equation}\label{eq:apdefinec1}
c_{b1}t = C_{b1}^1\,t_i^1+C_{b'1}^2\,t_b^2+C_{b''1}^3\,t^3_b\,.
\end{equation}
These equations demonstrate the need for the updates in \equationname
s~(\ref{eq:updatecfirst})-(\ref{eq:updatecfirst2}) and
(\ref{eq:updatecsecond})-(\ref{eq:updatecsecond2}), and in particular,
the need for the auxiliary variable $c_{bx}$. The update to $c_{b0}$ in
\equationname~(\ref{eq:updatecfirst}) takes care of the first term in
parentheses in the first two lines of
\equationname~(\ref{eq:apdefinec0}). The auxiliary variable $c_{bx}$
stores the coefficient of the second term in the parentheses, such
that the second term---which involves the final, finest set of
breakpoints---can be accounted for at the end (in the final update to
$c_{b0}$ in \equationname~[\ref{eq:c0finalupdate}]).

\section{Convergence tests}\label{sec:convtests}

\begin{figure*}
\includegraphics[width=\textwidth]{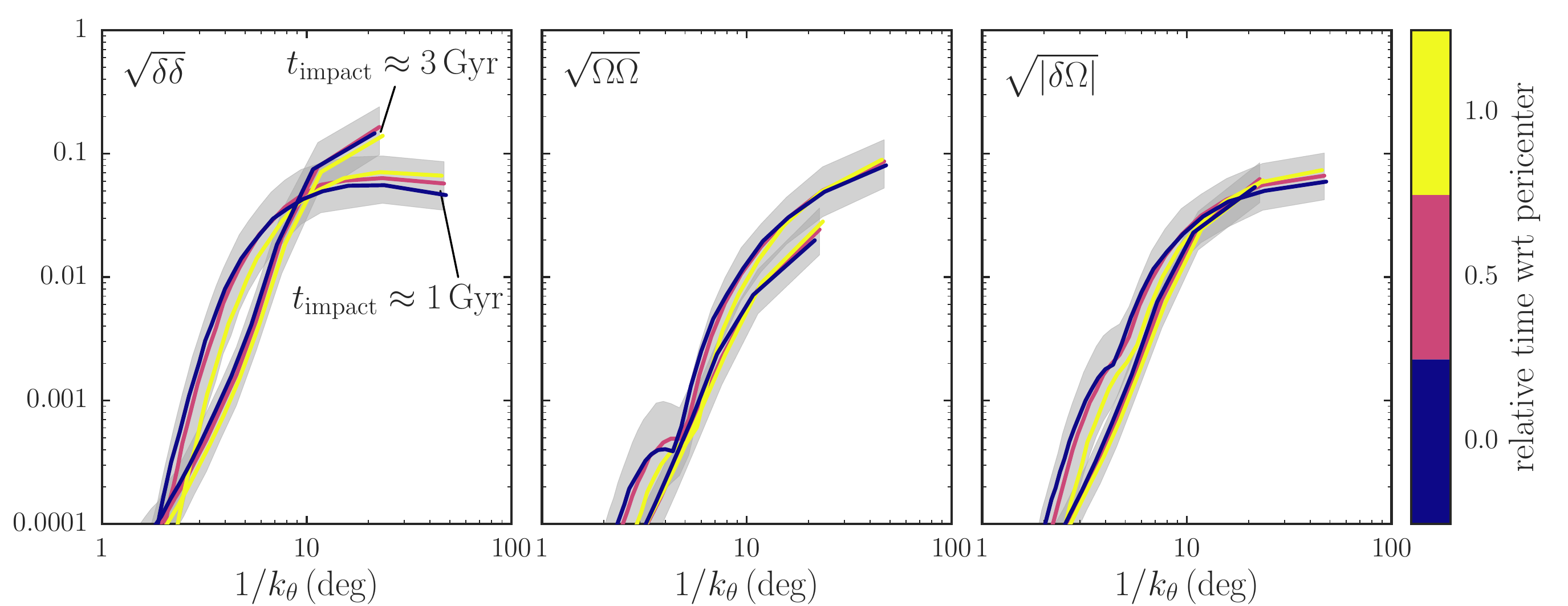}
\caption{Importance of the timing and phase of impacts. This figure
  shows power spectra for simulations of $10^{6.5}\msun$ impacts that
  all occur at the same time, using the rate of impacts for the range
  $10^6\msun$--$10^7\msun$. One group of impacts happens approximately
  $1\Gyr$ ago, the second group $\approx 3\Gyr$ ago (the former
  extends to $1/k_\theta \approx 50^\circ$ in each panel while the
  latter only extends to $1/k\ _\theta \approx 25^\circ$). Each group
  contains impacts at the times corresponding to the pericentric
  passage of the cluster, the apocentric passage, and halfway between
  the two. We only compute the power spectrum over the part of the
  stream that was impacted ($\dapar > 2\Delta
  \Omega^m\,t_{\mathrm{impact}}$, where $\Delta \Omega^m$ is the
  mean-parallel-frequency parameter of the smooth stream), because the
  part of the stream that is dominated by stars stripped after the
  impact time is largely smooth. Impacts that occur further in the
  past give more power on large scales because the effect of the
  impacts has more time to evolve into density fluctuations. The phase
  at which the impacts happen is (statistically at least) not
  important.\label{fig:periapo}}
\end{figure*}

\begin{figure*}
\includegraphics[width=\textwidth]{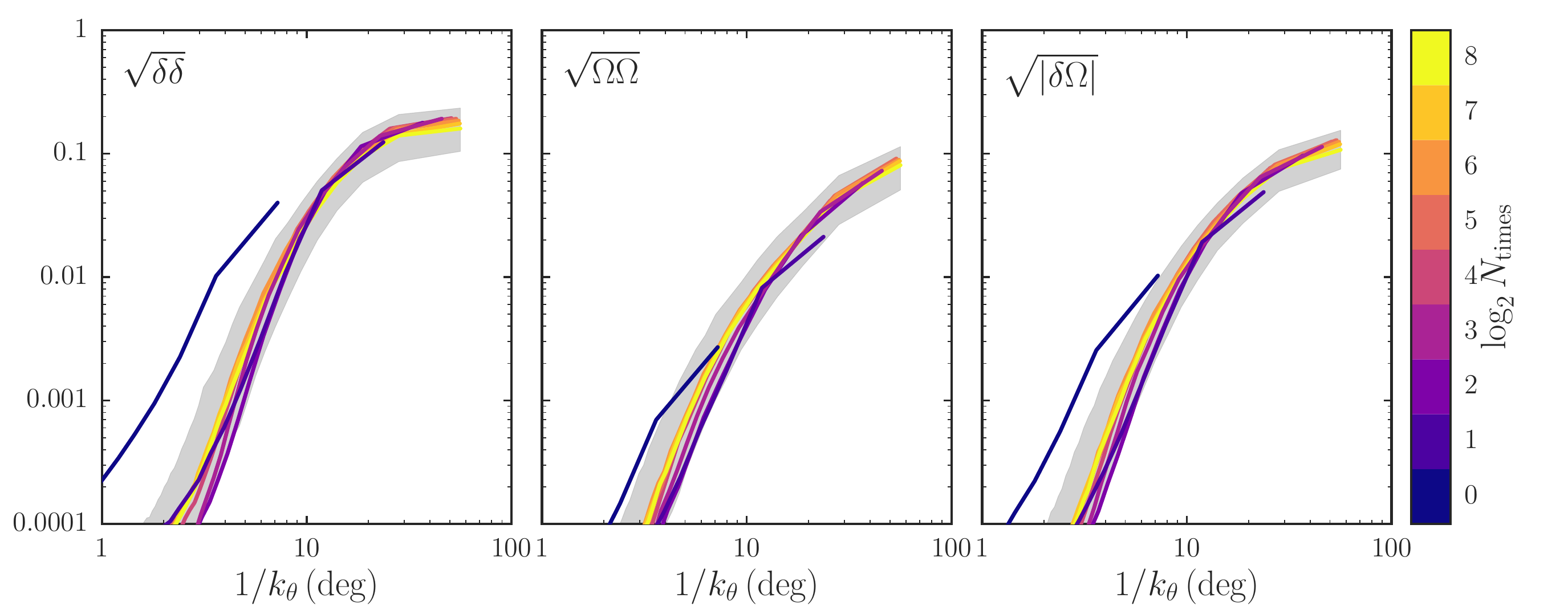}
\caption{Convergence of the power spectrum with the time sampling of
  impacts. This figure displays the power spectra computed when
  allowing impacts at $2^N_{\mathrm{times}}$ different times, varying
  from one (at $4.5\Gyr$) to 256 (every $\approx35\Myr$), for
  simulations of $10^{6.5}\msun$ impacts (similar to those in
  \figurename~\ref{fig:bmaxtime}). The statistical properties quickly
  converge as the number of separate times is increased and we use a
  fiducial value of 64 different times (every $\approx140\Myr$, which
  is less than the radial period of $400\Myr$ of the orbit). The
  convergence for other mass ranges is similar and we use 64 different
  times everywhere.\label{fig:timesampling}}
\end{figure*}

\begin{figure*}
\includegraphics[width=\textwidth]{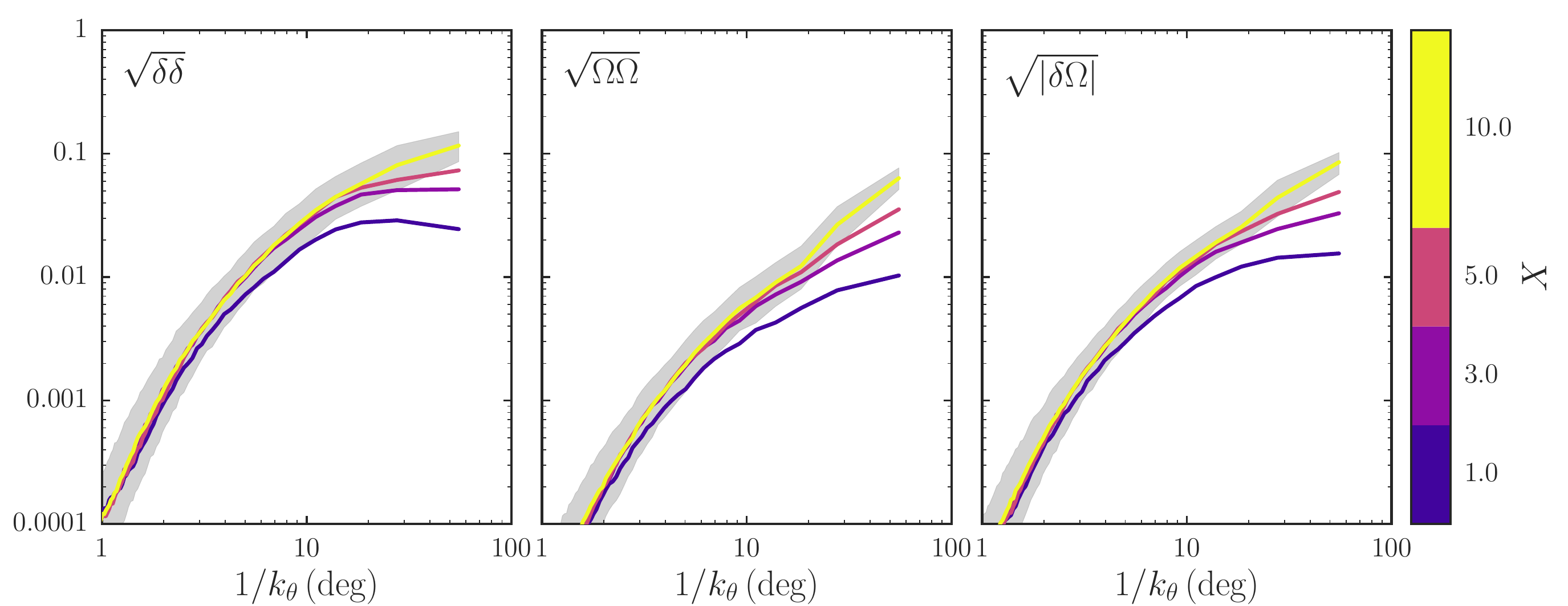}
\caption{Convergence of the power spectrum when considering impacts
  out to $X\,r_s(M)$ for simulations of $10^{5.5}\msun$ impacts
  (similar to those in \figurename~\ref{fig:bmaxtime}). While the
  small-scale structure of the stream quickly converges when $X > 1$,
  the largest scale modes are significantly affected by distant
  encounters. Even for $X = 10$, the largest scale modes do not appear
  to have converged. Higher-mass ranges show similar results (see
  \figurename~\ref{fig:bmaxtime}). Because the rate of impacts scales
  linearly with $X$, we use a fiducial value of $X=5$ to limit the
  computational cost. Thus, the largest-scale modes in the power
  spectra in this paper are not fully converged.\label{fig:bmax}}
\end{figure*}

\begin{figure*}
\includegraphics[width=\textwidth]{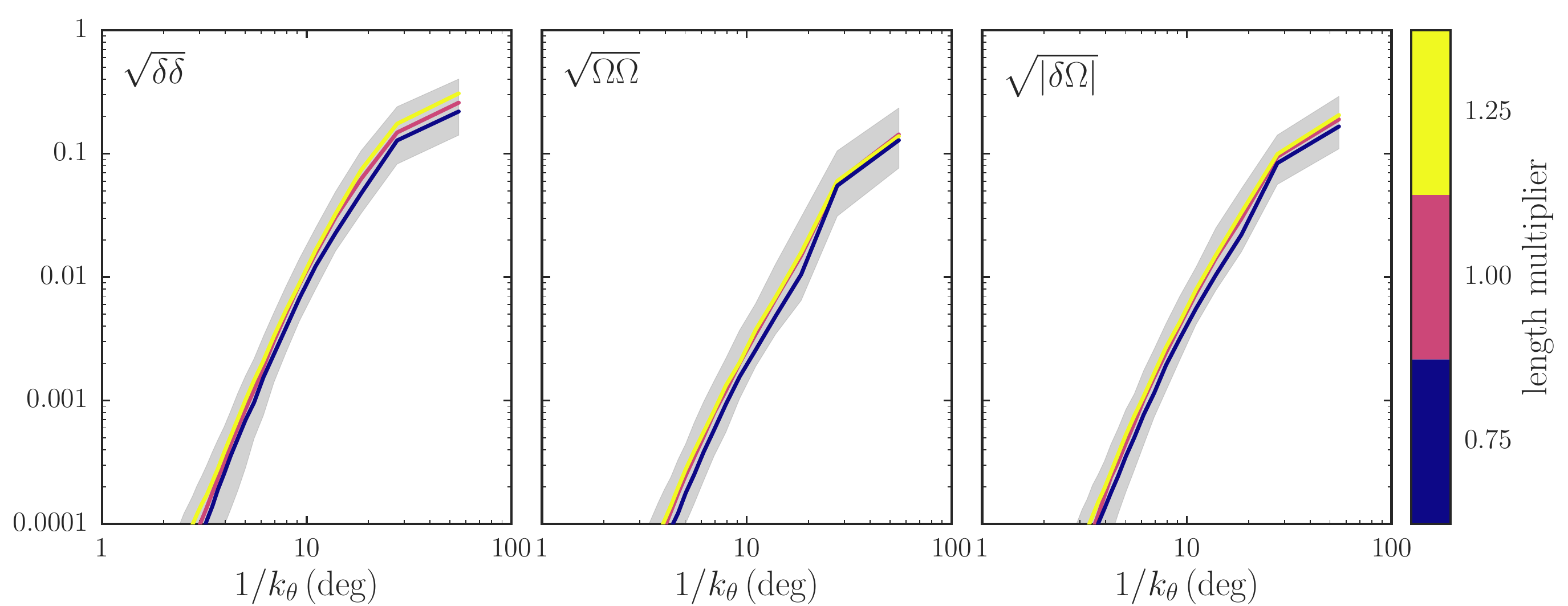}
\caption{Convergence of the power spectrum when considering impacts up
  to a factor times the length of the stream. This figure shows the
  result of simulations of $10^{7.5}\msun$ impacts (similar to those
  in \figurename~\ref{fig:bmaxtime}), where impacts at each impact
  time are considered up to `length multiplier' times the length of
  the stream (as defined at the end of \sectionname~\ref{sec:smooth}
  as the location along the stream where the density drops below
  20\,\% of the peak value). The statistical properties of the stream
  are very similar in all cases. Therefore, we simply consider impacts
  out to the length of the stream in our fiducial
  setup.\label{fig:lenfac}}
\end{figure*}

In this Appendix, we briefly discuss the results from a set of tests
to assess whether the power spectra computed using our default
sampling of subhalo impacts have converged. We test the three most
important approximations in our statistical sampling of the
impacts. These are (a) the resolution of the discrete time sampling,
(b) the maximum possible impact parameter, and (c) the maximum
distance along the stream to consider impacts at each impact
time. While these convergence tests should be repeated for any new
stream modeled using our framework, we derive some rules that should
hold more generally than the specific simulation here.

We sample impacts on a discrete grid of times between the start of
tidal disruption of the stellar stream and the present time. This
saves computational time in the line-of-parallel-angle approach,
because multiple impacts at the same time do not add to the
computational cost. Our default simulation setup for the GD-1-like
stream considers impacts at 64 different times or at a spacing of
$\approx140\Myr$. This is about one-third of the radial period of the
approximate stream orbit, which is $400\Myr$. Sampling impact times on
a finer grid essentially increases the coverage of the orbital phase
of the stream at the impacts.

To assess the importance of sampling the orbital phase, we run
simulations with a single value of the mass, here $10^{6.5}\msun$,
using the rate of impacts corresponding to masses between $10^6\msun$
and $10^7\msun$. Rather than sampling the entire range of times
between the start of disruption and the current time, we apply all
impacts at the same time. This time is chosen to be when the stream is
near pericenter, apocenter, and halfway in between. We do this for two
sets of such times, one $\approx1\Gyr$ in the past and one
$\approx3\Gyr$ ago. The resulting power spectra are displayed in
\figurename~\ref{fig:periapo}. It is clear that the overall time at
which the impact occurs is important: The impacts that happen
$\approx1\Gyr$ ago give rise to less power on large scales than those
that happen $\approx3\Gyr,$ because they have not had enough time to
evolve yet. However, the orbital phase at which the impact happens is
much less important: all different phases near the same time give rise
to the same power spectrum. Thus, the time sampling only needs to be
fine enough to sample the overall history of the stream, but is not
required to be so fine as to densely sample the orbital phase of each
radial oscillation.

In \figurename~\ref{fig:timesampling} we display the power spectrum
computed using different-sized grids of equally-spaced times. We
consider grids between that consisting of a single time at the
mid-point between the start of disruption and the current time up to a
grid with 256 different times (spacing $\approx35\Myr$). This figure
demonstrates that the power spectrum quickly converges, in agreement
with the discussion in the preceding paragraph. The power spectrum has
largely converged when using $16$ different times, which corresponds
to a spacing of $\approx560\Myr$, which is slightly larger than the
radial period. The same convergence happens for other mass ranges
($10^5\msun$--$10^7\msun$ and $10^7\msun$--$10^8\msun$; see
\figurename~\ref{fig:bmaxtime}). In general, we expect from the
behavior in \figurename s~\ref{fig:periapo} and \ref{fig:timesampling}
that sampling impact times somewhat finer than the radial period
should always suffice to obtain a converged power spectrum.

The second simulation parameter that we consider in this Appendix is
the maximum impact parameter. In \sectionname~\ref{sec:sample}, we
describe how we sample impacts up to a maximum impact parameter that
is a function of the mass of the perturber. We sample impact
parameters up to a multiple $X$ of the scale radius of the perturbing
subhalo. \equationname~(\ref{eq:nenc}) shows that the rate of impacts
is linearly dependent on the maximum impact parameter and therefore
also on $X$. Ideally, we would consider impacts out to an impact
parameter of infinity, but this is not advisable both from a practical
and a modeling perspective: Distant encounters are not well described
by the impulse approximation that we use to compute the instantaneous
effect of subhalo impacts. Practically, increasing the rate of
encounters increases the computational cost significantly. Because we
expect distant encounters to be subdominant (velocity kicks go as
$M/b$ at large $b$), we therefore do not want to include unnecessary
distant encounters.

In \figurename~\ref{fig:bmax} we display the $X$ dependence of the
power spectrum for impacts with a single mass of $10^{5.5}\msun$ using
the rate corresponding to the range $10^5\msun$ to $10^6\msun$. While
the power spectrum on small scales quickly converges once $X\approx3$,
the power on large scales keeps increasing as $X$ is increased. The
same behavior occurs for other mass ranges (see
\figurename~\ref{fig:bmaxtime}). To limit the computational cost, we
use $X=5$ in our default setup, but it is clear from
\figurename~\ref{fig:bmax} and \figurename~\ref{fig:bmaxtime} that
this underestimates the power on the largest scales. At these large
distances, the impulse approximation will begin to break down so it is
unclear whether this lack of convergence on large scales is as severe
as it seems from \figurename~\ref{fig:bmax}. If one is interested in
using density and track fluctuations on the largest scales---which is
only be possible for the longest streams with lengths
$\gtrsim30^\circ$---distant encounters are important and need to be
taken into account. However, to compute the statistical properties of
a tidal stream on the smallest scales, $X$ can safely be set to a
value of a few.

Finally, in \figurename~\ref{fig:lenfac} we vary the distance along
the stream where we consider impacts. In our default setup, at each
impact time we only consider impacts that happen closer to the
progenitor than the length of the stream, defined at the end of
\sectionname~\ref{sec:smooth}. However, a tidal stream does not have a
sharp end and impacts that occur beyond this nominal length could push
stream stars back toward the progenitor where they could affect the
current structure of the stream. \figurename~\ref{fig:lenfac}
demonstrates that at least for the GD-1-like stream this definition of
the length is appropriate: the power spectrum is essentially the same
when only considering impacts up to $75\,\%$ of the nominal length at
each time and when considering impacts out to $125\,\%$ of the nominal
length. \figurename~\ref{fig:lenfac} shows this for impacts with a
mass of $10^{7.5}\msun$ using the rate for the mass range $10^7\msun$
to $10^8\msun$, but we find the same for lower mass ranges (see
\figurename~\ref{fig:bmaxtime}). We expect our definition of the
length to work well for most streams, at least for the small expected
CDM-like impact rates.

\section{Mock Pal 5 rate analysis}\label{sec:mockpal5}

\begin{figure}
\includegraphics[width=0.46\textwidth]{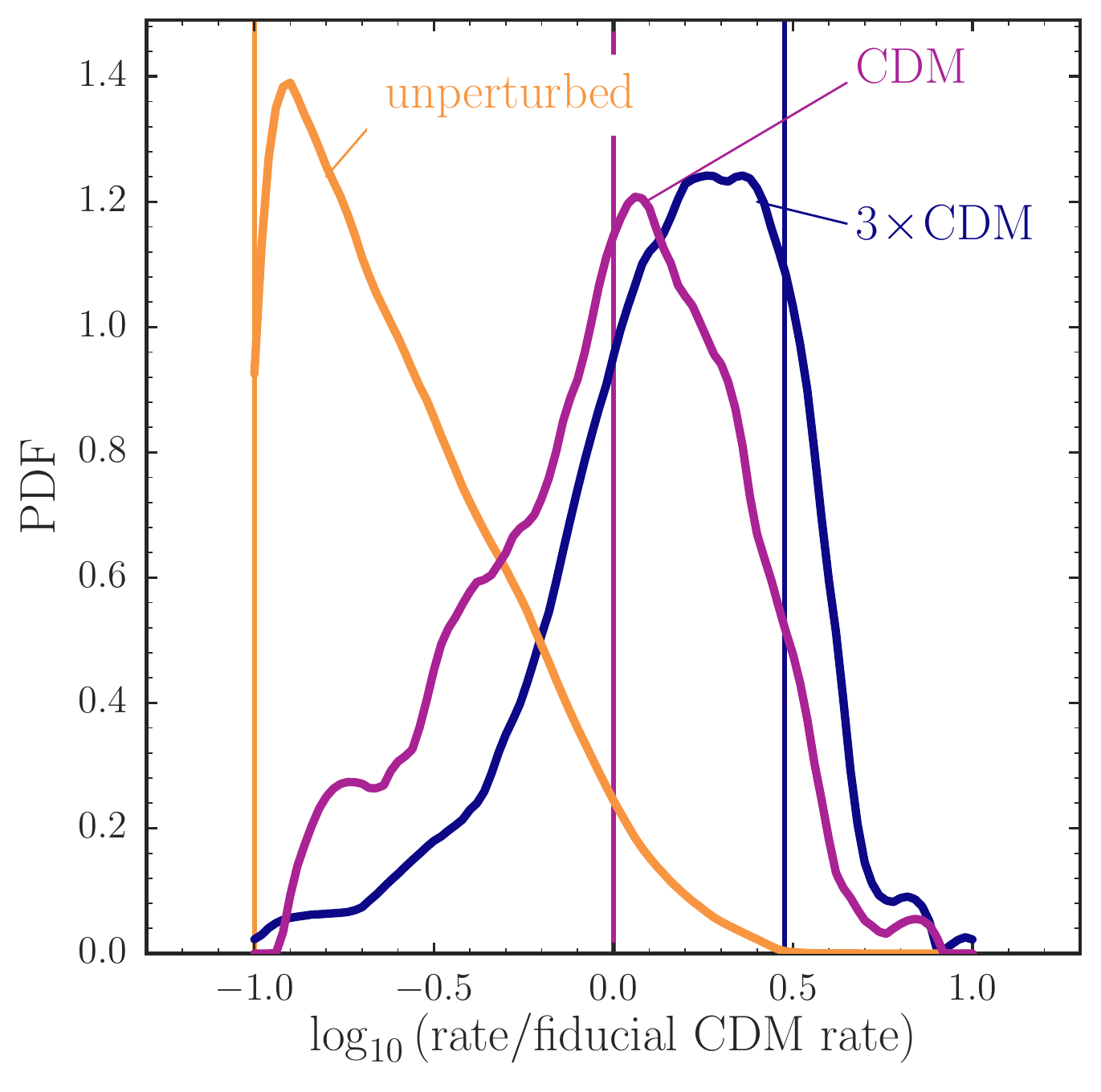}
\caption{Posterior probability distribution functions for the rate of
  dark--matter subhalo encounters based on mock simulations of the Pal
  5 stream in a dark--matter halo with varying levels of substructure
  with masses between $10^6\msun$ and $10^9\msun$. The PDFs are
  obtained in the same manner as those for the real Pal 5 data in
  \figurename~\ref{fig:pal5ratepdf}. Different curves are the PDF for
  different levels of substructure: The orange line is for a
  simulation without any CDM halos, the purple curve for a simulation
  with a CDM-like population of subhalos (without accounting for any
  subhalo destruction by the disk), and the blue curve for a
  simulation with three times the CDM population. The vertical lines
  indicate the truth for each simulation. The downturn of the orange
  curve near $-1$ is an artifact due to the KDE method used to smooth
  the PDF, in reality it peaks at $-1$. All PDFs are consistent with
  the input rate within the width of the PDF, demonstrating that
  matching the observed power using simulations produced with our
  formalism can robustly constrain the amount of substructure in the
  halo.\label{fig:mockpal5}}
\end{figure}

In this Appendix, we repeat the analysis of the Pal 5 data in
\sectionname~\ref{sec:pal5} for a suite of mock-data simulations to
test our procedure. These simulations were run with the $N$-body part
of \textsc{gadget-3} which is similar to \textsc{gadget-2}
\citep{Springel05a}. The code was modified to include external static
potentials as well as the forces from the subhalos which were modeled
as Hernquist profiles. These Hernquist profiles have the same
size-mass relation given in \sectionname~\ref{sec:sample}. The
best-fit phase--space position from \citet{Kuepper15a} is integrated
backward for $5\Gyr$. At this point, a King cluster with a mass of
$2\times 10^4\msun$, central-potential parameter $W_0 = 2$, and $r_c =
15$ pc is instantiated with $100,000$ particles. This cluster is then
integrated forward in time for $5\Gyr$ until the present day, using a
softening of 1 pc. The smooth potential is similar to the
\texttt{MWPotential2014} used in the analysis of the real Pal 5 data,
except that the bulge component has been replaced by a Hernquist
sphere with the same mass and a scale radius of $500\pc$
(\texttt{MWPotential2014} has a power-law bulge with an exponential
cut-off that is computationally more complex). A population of
subhalos is orbiting within this potential with a number density given
by the Einasto fit in \citep{Springel08a} scaled down to a mass of
$10^{12}\msun$. The full gravitational interaction between each
subhalo and the stream is computed whenever the subhalo is within
$30\kpc$ from the galactic center. We perform simulation (a) without
any substructure, (b) with a CDM-like population of subhalos, and (c)
with three times a CDM-like number of subhalos. At the end of the
simulation, these end up at approximately the current position of Pal
5 and we analyze them in the same $(\xi,\eta)$ coordinate system as
for the real data (we shift the simulations such that the surviving
progenitor is located at $(\xi,\eta) = (0,0)$). We have also performed
a simulation with ten times the amount of substructure expected for
CDM. This produces a stream that is very far from the current position
of Pal 5 and that is very significantly perturbed. It does not look
like the observed Pal 5 stream, confirming that a rate as high as ten
times CDM is obviously at odds with the observed Pal 5 stream. We do
not consider this mock simulation further. The expected number of
subhalos within $25\kpc$ for CDM for these simulations is slightly
different than what we have used in the main body of this paper. The
CDM-like number of subhalos within $25\kpc$ with masses between
$10^5\msun$ and $10^9\msun$ in these mock simulations is $525.40$,
obtained by rescaling the fits from the Aquarius simulation
\citep{Springel08a} (see \citealt{Erkal16a} for further details).

We analyze the density data in of these three mock-Pal 5 streams in
the same manner as the real Pal 5 data. That is, we compute the
density in $0.1^\circ$ bins in $\xi$, normalize it using a third-order
polynomial fit, calculate the power spectrum of the part of the stream
between $0.2^\circ < \xi < 14.3^\circ$, and match the power on the
three largest scales and the bispectrum, using largely the same
tolerances as for the real data. For the simulation without any
substructure we change the tolerance for the power on the second- and
third-largest scales to $0.01$, because the power in the mock data on
these scales is very small, thus allowing us to make the tolerance
smaller. For the CDM and $3\times$CDM simulations we relax the
tolerance on the bispectrum slightly. As the uncertainties, we simply
use the Poisson uncertainties on the number of $N$-body particles in
each $\xi$ bin; the stream density is $\approx500\,\mathrm{deg}^{-1}$
for all mock streams. The density uncertainties in the simulations are
therefore typically $\approx14\,\%$.

The resulting PDFs for the rate of impacts are displayed in
\figurename~\ref{fig:mockpal5}. It is clear that we can put a tight
constraint on the incidence of substructure if there is no
substructure (yellow curve). The $95\,\%$ and $99\,\%$ upper limits on
the rate are $0.6\times$CDM and $1.0\times$CDM, respectively, just
from the power on the three largest scales and the bispectrum on a
single scale. We could have used the power on smaller scales to get a
better constraint, but we are here primarily interested in testing the
robustness of our Pal 5 analysis, for which this is not possible. For
the mock data perturbed by a CDM-like or $3\times$CDM-like population
of subhalos, we find relatively broad PDFs similar to the PDF for the
real Pal 5 data in \figurename~\ref{fig:pal5ratepdf}. These are both
consistent with the input value for the rate, which in both cases lies
about $1\sigma$ from the peak of the PDF. The reason that the PDFs are
not narrower than those for the Pal 5 data in
\sectionname~\ref{sec:pal5} even though the mock-data uncertainties
are four times smaller than those for the Pal 5 data is that the power
on the largest scales is much larger than the noise power. The
measurement of this large-scale power is therefore limited by the fact
that we only have a single realization of the density to measure the
power and not by the uncertainties in the density measurement: the
uncertainty in the power is the power itself \citep{Press07a}.

In addition to the results displayed in
\figurename~\ref{fig:pal5ratepdf}, we have performed three more
simulations each of perturbations from a CDM-like and
$3\times$CDM-like population. The resulting PDFs are similar to those
shown here and we find no significant biases in the inferred
rate. Thus, we conclude that our procedure of matching the power on
the largest scales to constrain the number of subhalos with the Pal 5
data is robust.

\end{document}